\documentclass[smallextended]{svjour3}       
\smartqed  
\usepackage{graphicx}
\usepackage{bm}
\usepackage{color}
\usepackage{soul}
\usepackage{ulem}
\usepackage[]{amsmath}
\usepackage[]{dsfont}
\usepackage[]{amsfonts}
\usepackage[]{amssymb}
\usepackage[]{mathrsfs}
\usepackage{enumitem}   
\usepackage{times}
\usepackage{hyperref}
\usepackage{stackengine}
\hypersetup{
  colorlinks=true,        
  linkcolor=blue,         
  citecolor=cyan,         
}
\usepackage{mathtools}
\usepackage{amsbsy}
\usepackage{float}
\usepackage{comment}
\usepackage{lineno}
\usepackage{cite}
\usepackage{xspace} 
\usepackage{multirow}
\usepackage{array}
\newcommand{\beq}{\begin{equation}} 
\newcommand{\eeq}{\end{equation}} 
\newcommand{\beqn}{\begin{eqnarray}} 
\newcommand{\eeqn}{\end{eqnarray}} 
\newcommand{\pa}{\partial}
\newcommand{\na}{\nabla}

\newcommand{\gabu}{g^{\alpha\beta}}
\newcommand{\gabd}{g_{\alpha\beta}}

\newcommand{\gmabu}{\gamma^{ab}}

\newcommand{\tgmabu}{\tilde\gamma^{ab}}

\newcommand{\albe}{{\alpha\beta}}

\newcommand{\Tabu}{T^{\alpha\beta}}

\newcommand{\zD}{{\raise1.0ex\hbox{${}^{\ \circ}$}}\!\!\!\!\!D}
\newcommand{\alone}{{\raise0.5ex\hbox{${}^{\ 1}$}}\!\!\!\!\alpha}

\newcommand{\dl}{\delta}

\newcommand{\Fabd}{F_{\alpha\beta}}

\newcommand{\Fcdu}{F^{\gamma\delta}}
\newcommand{\Fcdd}{F_{\gamma\delta}}
\newcommand{\nalam}{\mathrel{\raise0.9ex\hbox{$^\lambda$}\mkern-14mu
\lower0.0ex\hbox{$\nabla$}}}

\newcommand{\zeroD}{{\raise1.0ex\hbox{${}^{\ \circ}$}}\!\!\!\!\!D}

\newcommand{\zLap}{{\raise1.0ex\hbox{${}^{\ \circ}$}}\!\!\!\!\Delta}
\newcommand{\zna}{{\raise1.0ex\hbox{${}^{\ \circ}$}}\!\!\!\!\!\nabla}
\newcommand{\zS}{{\raise1.0ex\hbox{${}^{\ \circ}$}}\!\!\!\!\!S}

\newcommand{\GA}{\alpha}
\newcommand{\GB}{\beta}
\newcommand{\GG}{\gamma}
\newcommand{\GD}{\delta}
\newcommand{\GE}{\epsilon}
\newcommand{\GK}{\kappa}
\newcommand{\GL}{\lambda}
\newcommand{\GR}{\rho}
\newcommand{\GS}{\sigma}
\newcommand{\GT}{\tau}
\newcommand{\GC}{\psi}

\newcommand{\GO}{\omega}
\newcommand{\GZ}{\zeta}
\newcommand{\GP}{\phi}

\newcommand{\GH}{\eta}
\newcommand{\GU}{\theta}

\newcommand{\Bk}{\mathbf{k}}
\newcommand{\Bn}{\mathbf{n}}

\newcommand{\Bu}{\mathbf{u}}

\newcommand{\ks}{\rm \scriptscriptstyle KS}
\newcommand{\pd}{\partial}
\newcommand{\be}{\begin{equation}}
\newcommand{\ee}{\end{equation}}

\newcommand{\TD}[2]{\tilde{#1}_{#2}}
\newcommand{\TU}[2]{\tilde{#1}^{#2}}
\newcommand{\TUU}[3]{\tilde{#1}^{#2 #3}} 
\newcommand{\TDD}[3]{\tilde{#1}_{#2 #3}}

\newcommand{\TWD}[2]{( \tilde{\mathbb L} \tilde{W} )_{#1 #2}}

\newcommand{\tW}{\tilde{W}}
\newcommand{\tGS}{\tilde{\GS}}

\newcommand{\tGB}{\tilde{\GB}}

\newcommand{\fDu}[1]{\stackon[-0.3ex]{$D^{#1}$}{\kern-1.0ex\scalebox{0.7}{$\circ$}}}

\newcommand{\Sinfty}{S_{_{\infty}}}

\newcommand{\omekep}{\Omega_{\scalebox{0.6}{K}}}

\newcommand{\MK}{M_{\scalebox{0.6}{K}}}
\newcommand{\JK}{J_{\scalebox{0.6}{K}}}
\newcommand{\TEMuab}{T^{\alpha\beta}_{\rm \scalebox{0.6}{EM}}}
\newcommand{\TPFuab}{T^{\alpha\beta}_{\rm \scalebox{0.6}{M}}}

\newcommand{\TFMuab}{T^{\alpha\beta}_{\rm \scalebox{0.6}{FM}}}

\def\QEQ{{%
    \setbox0\hbox{$I$}%
    \rlap{\hbox to \wd0{\hss--\hss}}\box0
}}

\newcommand{\ie}{i.e.,~}

\newcommand{\at}[1]{{\textcolor{red}{ #1}}}

\newcommand{\cocal}{\textsc{cocal}\xspace}

\newcommand{\lorene}{\textsc{lorene}\xspace}
\newcommand{\kadath}{\textsc{kadath}\xspace}
\newcommand{\sgrid}{\textsc{sgrid}\xspace}

\newcommand{\spells}{\textsc{spells}\xspace}

\newcommand{\whiskythc}{\textsc{WhiskyTHC}\xspace}

\begin{document}

\title{Methods for relativistic self-gravitating fluids: 
From binary neutron stars to black hole-disks and magnetized rotating neutron stars. 
\thanks{A.Tsokaros is supported by National Science Foundation Grant No. PHY-1662211 and the 
National Aeronautics and Space Administration (NASA) Grant No. 80NSSC17K0070 to the University 
of Illinois at Urbana-Champaign. K. Ury\=u is supported by JSPS Grant-in-Aid for Scientific 
Research (C) 15K05085 and 18K03624 to the University of Ryukyus.}
}


\titlerunning{Methods for binary neutron stars, black hole-disks, and magnetized rotating neutron stars}    

\author{Antonios Tsokaros \and K{\= o}ji Ury{\= u}  }


\institute{Antonios Tsokaros     \at
              University of Illinois at Urbana-Champaign, Loomis Laboratory, 1110 W Green St, Urbana, IL 61801, USA. \\
              Tel.: +1-217-3004508 \\
              Fax : +1-217-3334898 \\
              \email{tsokaros@illinois.edu}         
           \and
           K{\= o}ji Ury{\= u}   \at
              University of the Ryukyus, Department of Physics, Senbaru, Nishihara, Okinawa 903-0213, Japan.  \\ 
              Tel. \& Fax:  +81988958521  \\
              \email{uryu@sci.u-ryukyu.ac.jp}         
}

\date{Received: date / Accepted: date}

\maketitle

\textit{This article is dedicated to the memory of Yoshiharu Eriguchi, whose
work on self-gravitating fluids has been a source of inspiration.}

\vspace{1cm}

\begin{abstract}
The cataclysmic observations of event GW170817, first as gravitational waves along the inspiral motion
of two neutron stars, then as a short $\GG$-ray burst, and later as a kilonova, launched the era
of multimessenger astronomy, and played a pivotal role in furthering our understanding on a number of 
longstanding questions. 
Numerical modelling of such multimessenger sources is an important tool to understand the physics of compact objects
and, more generally, the physics of matter under extreme conditions. In this review we present a unified
view of various techniques used to obtain equilibrium and quasiequilibrium solutions for three 
astrophysically relevant relativistic, self-gravitating fluid systems: Binary neutron stars, black hole-disks, 
and magnetized rotating neutron stars. 
These solutions are necessary not only for modeling such compact objects, but equally important, for providing 
self-consistent initial data in numerical relativity simulations.
Instead of presenting the full details of the formulations and numerical algorithms, we focus on 
painting the broadbrush picture of the methods developed to address these problems, 
and facilitate future work in the area.
\end{abstract}

\section{Introduction}
\label{sec:intro}

Although the majority of neutron stars are observed as isolated pulsars that emit electromagnetic  
radiation from their magnetic poles \cite{Hewish1968,lyne_graham-smith_2012},
a small percentage of them appear in binary systems. In our own
Galaxy the first binary neutron star (BNS) system, PSR B1913+16, was discovered in 1974
\cite{HulseTaylor74} while currently 19 such binaries are known \cite{Zhu2020}). Despite the
small number of BNS currently observed their strong gravitational and electromagnetic interactions constitutes them as
ideal probes for relativistic astrophysical phenomena.  In particular: 
\begin{enumerate}[label=(\roman*)]
\item During their inspiral and merger they produce gravitational waves an
important feature of the general theory of relativity. Fundamental tests
related to gravity in an otherwise inaccessible strong-field regime are
therefore possible \cite{MillerColeman2016,Abbott2019}. 
\item The late inspiral as well as the merger process can yield important
information about the masses and radii of the component stars and in effect
about their equation of state (EOS) \cite{Yagi2013b,lorimer:lr,Lattimer2016,Ozel2016}. 
\item They can be the progenitors of short duration ($<2\;\rm s$) $\GG$-ray
bursts (sGRBs) \cite{Eichler89,Narayan92,Fong2013}.
\item They constitute premier sites for the production of rare heavy elements like
gold and platinum \cite{Lattimer74,Lattimer76,Symbalisty82}, through a rapid
neutron-capture process (or r-process), in which neutrons are captured by
lighter nuclei like iron in a dense neutron-rich environment
\cite{Burbidge1957,Cameron1957}.
\end{enumerate}
In a single stroke the first detection by LIGO/Virgo of a BNS approximately $40\;\rm
Mpc$ away in the Galaxy NGC 4993, together with follow-up detections by the
Fermi, INTEGRAL spacecrafts and other telescopes, addressed all of the problems
above.  Event GW170817 showed the inspiral of two neutron stars \cite{GW170817prl} which
was followed by a sGRB, named GRB170817A $1.7\;\rm s$ later
\cite{Goldstein2017,Savchenko2017,Abbott2017d}. Lastly a luminous optical
counterpart, named AT2017gfo, was also observed 11 hours after the merger
\cite{Coulter2017,Soares-Santos2017,Arcavi2017,Diaz2017,Abbott2017c}.  This
kilonova \cite{Metzger:2010} transient was powered by the radioactive decay of
heavy neutron-rich elements created in the expanding merger ejecta
\cite{Li:1998} and served as a direct probe of the astrophysical origin of the
heaviest elements in the Universe (see recent review by Metzger
\cite{Metzger2019}).  Despite the scarcity of detections like event GW170817,
it is apparent that the synthesis of gravity, electromagnetism, and microphysical processes
in binary black hole-neutron star (BHNS) or BNS mergers 
(binary black holes are not expected to produce electromagnetic radiation due
to the absence of matter) is astrophysically very fruitful
\cite{ShibataTaniguchilrr-2011-6,Berger2013b,Baiotti2016,Paschalidis2016,Shibata2019a,Meszaros2019,Friedman2020,Foucart2020,Ciolfi2020,Radice2020}.

The background spacetime in which many of the complex phenomena mentioned above
develop, is the strong gravity regime.  Depending on the question at
hand, or the timescale, one can consider two categories of problems. In the
first category there is a dominant compact object (either single or composite) that sets up the spacetime
and the rest of the system is evolving either without affecting the
gravitational sector or by perturbing it. An example of such system is a black hole
surrounded by a non self-gravitating disk \cite{Abramowicz2011}. A widely
separated BNS where the neutron stars are treated as point masses,
also falls within this class of problems \cite{Blanchet06}.  The
complement of the first category constitutes of all systems where the
gravitational interactions between the various components are equally
significant, cannot be isolated, and vary with time. In this category one necessarily ends up
with a dynamical spacetime where the self-gravity of every component is
important. Analytical work for these kind of systems is limited, and full 3+1
(spatial and temporal) computational calculations are the main tool of
investigation. In this endeavor the Einstein's equations are playing the
leading role since both electromagnetic and nuclear interactions are acting within the
unknown spacetime.  Therefore numerical simulations that try to explain complex
phenomena behind BHNS or BNS mergers inevitably have to be able to integrate in
a stable way the equations of general relativity, a second order system of
partial differential equations. In addition in order to incorporate
magnetic fields one needs Maxwell equations, and finally the
energy-momentum and radiation transport equations for the evolution of
matter and radiation. 

Any differential equation needs initial values in order to be integrated and
the aforementioned dynamical system is no exception. Even more, due to the
complexity of the problem there are two main difficulties.  The first one is
mathematical and is associated with the fact that the Einstein dynamical system
itself, therefore its initial values too, is not trivially posed. Afterall the
Einstein equations for the pair $(M,g_{\GA\GB})$, where $M$ is a 4-dimensional manifold,
and $g_{\GA\GB}$ a Lorentzian metric, are invariant under diffeomorphisms of $M$
and the associated isometries of $g_{\GA\GB}$.  In addition even if a
formulation is mathematically well posed it does not mean that it will be
numerically stable. The second difficulty is rooted into the fact that the
initial value functions must represent the physical system under consideration
which is a complicated task in the general theory of relativity.  Basic
concepts like the mass, angular momentum or center of mass are not trivially
defined as they are in Newtonian mechanics.  In other words it is not always
clear what kind of assumptions one has to make in order to get a snapshot of
the system under consideration which will be physically meaningful. The subject
of numerical relativity and relativistic hydrodynamics has grown considerably
since the pioneering binary black hole simulations
\cite{Pretorius:2005gq,Campanelli06,Baker:2005vv} and details can be found in a
number of textbooks
\cite{Wilson2003,Alcubierre:2008,Bona2009,Baumgarte2010,gourgoulhon20123+1,Friedman2012,Rezzolla_book:2013},
while recent reviews \cite{Lehner2014,Tichy2016rev,Paschalidis2017b,Duez2019,Radice2020}
present different aspects of it.

In this review we will touch upon the initial value problem and its
numerical implementation of three astrophysically relevant self-gravitating systems that include matter:
BNSs, black hole-disks (BHD), and magnetized rotating neutron stars (MRNSs).  For reasons mentioned
above the modeling of a neutron star in a binary or single setting as well as its
magnetic field is very important not only for gravitational wave astronomy but
also multimessenger astrophysics.  In addition since the most promising
scenario for the existence of a sGRB is a black hole surrounded by a massive disk
\cite{Eichler89,Narayan92,Shibata99d,Baiotti08,Anderson2007,Liu:2008xy,Bernuzzi2011,Paschalidis2014,Ruiz2016}
the study of such systems is also well-motivated. Furthermore BHD are
ubiquitous in the Universe and self-gravity can be important at certain times
during their evolution. Although BHD have been intensely studied in the
past \cite{Frank2002,Kato2008,Abramowicz2011,Rezzolla_book:2013}, here 
we will focus only on general relativistic self-gravitating
disks that are not covered in the bibliography, where a fixed Kerr black hole is assumed.
Apart from the BNS initial value problem which is discussed extensively in the
bibliography, BHD and MRNS are still missing a concrete exposition,
therefore our effort here is to close this gap and offer a unified approach to
the subject.  Also we do not discuss the initial value problem of binary black holes, BHNSs or
rotating neutron stars (RNS) since it is sufficiently covered in the textbooks and the reviews mentioned
above.  Due to limitations in space, our discussion will focus only on the
salient features and methods of obtaining self-gravitating solutions of
generic BNS, BHD, and MRNS (quasi)equilibria, without getting into the
details of those calculations. We will narrow our exposition only in full general relativistic
methods.  Neither we will address the important subject of stability and
evolution in general for these compact objects that inevitably will make this
review grow manyfold. Our main goal is to offer a bird's eye view of the
subject, while at the same time motivate further research in the field
of numerical general relativistic solutions.

In this review Greek indices are taken to run from 0 to 3 while Latin indices from 1 to 3.
We use a signature $(-,+,+,+)$ for the spacetime line element, and a system of units 
in which $c=G=M_\odot =1$ (unless explicitly shown). The list of acronyms adopted in
this paper are listed below:
\begin{table}[h!]
\label{tab:acro}
\begin{tabular}{lllll}
BNS: & Binary neutron star               &  $\qquad$ & sGRB: & Short $\gamma$-ray burst   \\ 
RNS: & Rotating neutron star             &  $\qquad$ & EOS:  & Equation of state   \\            
MRNS:& Magnetized rotating neutron star  &  $\qquad$ & ADM:  & Arnowitt-Deser-Misner \\             
BHD: & Black hole-disk                   &  $\qquad$ & KEH:  & Komatsu-Eriguchi-Hachisu    \\   
MHD: & Magnetohydrodynamics              &  $\qquad$ & IWM:  & Isenberg-Wilson-Mathews    \\     
IMHD:& Ideal magnetohydrodynamics        &  $\qquad$ &       &
\end{tabular}
\end{table}


\section{Binary neutron stars}
\label{sec:bns}

In general relativity contrary to Newtonian gravity binary compact objects evolve by the
emission of gravitational waves. When two neutron stars are separated by a distance much
larger than their radii one can approximate them as point particles
\cite{Blanchet2013}. At that stage the neutron stars evolve in an adiabatic manner with
the gravitational wave timescale being larger than the orbital period $P_{\rm orb}$, 
\be 
\frac{t_{\rm gw}}{P_{\rm orb}} \approx \left(\frac{r}{6M}\right)^{5/2}\left(\frac{M}{4\mu}\right) .
\label{eq:timescale} 
\ee 
Here $M$ is the total mass of the binary and $\mu$ its
reduced mass ($\mu=M/4$ for equal mass binaries).  When the separation becomes
smaller, the gravitational wave timescale decreases, as does the orbital period.
When the two become comparable, the radial velocity of the neutron stars increases
significantly and the adiabatic approximation breaks down. For a typical system
this happens at $35\;\rm km$ or $1\;\rm KHz$ gravitational wave frequency with
the orbital period being approximately $2\;\rm ms$. The neutron stars merge shortly
afterwards. At the intermediate stage when on one hand the two neutron stars are not
too far (distances less than $60\; \rm km$), but on the other not too close,
(distances greater than $35\;\rm km$) the system can be approximated as
stationary in the corotating frame and finite size effects are important. The
combined Eistein-Euler system needs to be resolved for an accurate
representation of the system. The methods described below aim at that stage.
The solutions obtained can be used as accurate initial values for performing
full general relativistic simulations and study the late inspiral, merger and postmerger at the
nonlinear regime, but they can also be used on their own for gaining important
information regarding this stage in the evolution of the binary.

\subsection{Isenberg-Wilson-Mathews formulation}
\label{sec:IWM}

One of the pillars in constructing binary neutron star initial data is the so-called
Isenberg-Wilson-Mathews (IWM) formulation \cite{Isenberg08,Wilson89,Wilson95,Wilson96}. 
Three waveless formulations were proposed by Isenberg in 1978 \cite{Isenberg08} 
which simplify computations of astrophysical systems of compact objects by decoupling 
the gravitational wave part and an ``induced'' part of strong gravity (which may be associated 
with the matter source terms). Isenberg never implemented the formulation
into a numerical code, but later Wilson and Mathews \cite{Wilson89,Wilson95,Wilson96} 
implemented one of the waveless formulations, 
in which the spatial metric is assumed to be conformally flat, for the evolution of BNS.  

Since a binary system is inherently nonaxisymmetric, all 10 metric components in the spacetime line 
element are necessary. A convenient way to express that is by the use of the 3+1 form  
\beq
ds^2 =  -\GA^2 dt^2 \,+\, \GG_{ij}(dx^i+\GB^i dt)(dx^j+\GB^j dt)\, ,
\label{eq:dstpo}
\eeq
where $\GA, \GB^i$ are the lapse function and shift vector respectively, and $\GG_{ij}$ the
spatial metric. A conformal 3-metric, $\TDD{\gamma}{i}{j}$, and a conformal 
traceless extrinsic curvature, $\TDD{A}{i}{j}$, are introduced through \cite{Lichnerowicz44,Nakamura1994}
\be
\GG_{ij} = \psi^4 \TDD{\gamma}{i}{j} , \qquad\mbox{and}\qquad
A_{ij} = \GC^4 \TDD{A}{i}{j} ,
\label{eq:baumgij}
\ee
where $\GC$ is the confomal factor. In the IWM formulation, the spatial metric is conformally flat
$\TDD{\gamma}{i}{j}=f_{ij}$ ($f_{ij}=\GD_{ij}$ in Cartesian coordinates), the slicing is maximal
$K=0$, and the time derivatives of the conformal metric vanish 
$\pd_t \TDD{\GG}{i}{j}=0$. Under such assumptions, part of the Einstein system reduces
to 5 elliptic equations for the conformal factor $\GC$, the lapse $\GA$ and the shift $\GB^i$.
These are the Hamiltonian constraint, the spatial trace of $\pd_t K_{ij}$, 
and the momentum constraint. In the IWM formulation the tracefree part of
the extrinsic curvature is written in terms of the lapse, the shift, and the conformal factor as
\be
A^{ij} = K^{ij} = \frac{\GC^{-4}}{2\GA}\left( \zD^i\GB^j + \zD^j\GB^i - \frac{2}{3} f^{ij} \zD_k \GB^k \right) \, ,
\label{eq:IWMaij}
\ee
where $\zD$ is the covariant derivative with respect to the flat metric, $\zD_c f_{ab}=0$
(in Cartesian coordinates $\zD_i=\pd_i$).
This scheme was first used in an evolutionary study of BNS by
Wilson {\it et~al.} \cite{Wilson95,Wilson96}. Soon after, it was realized that
the IWM formulation was even more useful for the construction of accurate
initial data sets for full numerical relativity simulations. The IWM
formulation is still used in simulations of binary neutron stars or binary white
dwarf mergers in order to incorporate (part of) relativistic gravity by
replacing the Newtonian gravitational potential with the above metric
potentials \cite{Bauswein2010,Bauswein2014}.  The connection between gravity and matter can be
accomplished in many ways and the IWM formalism is one of them. A better
formulation will be presented later in Section \ref{sec:ncfid} \cite{Bonazzola:2003dm,Shibata04a}.

\subsection{Mass, angular momentum and the first law for binary systems}
\label{sec:mj}

Two important characteristics of a BNS system are its mass-energy content, and its angular momentum.
In an asymptotically flat spacetime like the one representing an isolated BNS 
a definition for a global mass was presented by Arnowitt-Deser-Misner (ADM) \cite{Arnowitt1960,Arnowitt62unfindable}. 
It is now called the ADM mass 
\be
M= \frac{1}{16\pi}\oint_{\Sinfty} (f^{ai} f^{bk} - f^{ki} f^{ab})\zD_k \GG_{ab} dS_i  ,
\label{eq:Madm}
\ee
where the integral is performed on a sphere whose radius tends at infinity.
Similarly the ADM angular momentum associated with a rotational Killing vector $\GP^i$ is
\be
J= -\frac{1}{8\pi}\oint_{\Sinfty} \pi^i_{\ j}\GP^j dS_i  ,
\label{eq:Jadm}
\ee
were $\pi^{ij}=-(K^{ij}-\GG^{ij}K)\sqrt{\GG}$ the momentum conjugate to $\GG_{ij}$. Having 
a spacetime metric in the form of Eq. (\ref{eq:dstpo}), one can rewrite 
Eqs. (\ref{eq:Madm}) and (\ref{eq:Jadm}) in terms of $\GA,\GC,\GB^i,K_{ij}$
(see e.g. \cite{gourgoulhon20123+1})
and therefore have a measure of the mass and angular momentum in terms of the 3+1 variables.

According to the first law of thermodynamics for binary systems by Friedman {\it et~al.}  \cite{Friedman01a},
if one assumes a spatial geometry $\Sigma_t$ that is conformally flat, neighboring equilibria of asymptotically flat 
spacetimes with a helical Killing vector satisfy
\be
\GD M  =  \Omega \delta J + \int_{\Sigma_t} [\bar{T}\Delta dS+\bar{\mu}\Delta dM_0 + V^\GA\Delta d \mathcal{C}_\GA] 
      +  \sum_i \frac{1}{8\pi}\GK_i \GD A_i .  
\label{eq:1stlaw}
\ee
Here $M$ and $J$ are the ADM mass and angular momentum of the spacetime
while $\Omega$ is the orbital angular velocity; $\bar{T}$ 
and $\bar{\mu}$ are the redshifted temperature and chemical potential; $dM_0$ is the baryon mass of a fluid 
element; $d \mathcal{C}_\GA$ is related to the circulation of a fluid element, and $V^\GA$ is the velocity with respect to the
corotating frame; $\GK_i,\ A_i$ are the surface gravity, and the areas of black holes.
For isentropic fluids, dynamical evolution conserves the baryon mass, entropy, and vorticity of each fluid element, 
and thus the first law yields 
\be
\dl M = \Omega \dl J  .
\label{eq:dmdj}
\ee
Eq. (\ref{eq:1stlaw}) implies that a natural measure
to characterize the spin of a neutron star in a binary setting is its circulation in a similar manner to the way 
rest mass characterizes the mass.

\subsection{Equations for perfect fluids}
\label{sec:pf}

In this review the  stress-energy tensor for the matter  will be described 
by a perfect fluid with 4-velocity $u^\GA$, pressure $p$ 
and total energy density $\GE$,
\be
\TPFuab = (\epsilon +p) u^\alpha u^\beta + p g^{\alpha\beta}\, .
\label{eq:pfset}
\ee
Its conservation leads to
\be
\nabla_\GB \TPFuab = \GR[u^\GB\nabla_\GB(hu^\GA)+\nabla^\GA h] + 
hu^\GA\nabla_\GB (\GR u^\GB) -\GR T \nabla^\GA s = 0 , 
\label{eq:coset}
\ee
where the first law of thermodynamics $dh=Tds+dp/\GR$ has been used for the 
pressure gradient. Here $\GR,\, h,\, s$ are the rest-mass density, specific 
enthalpy ($h:=(\GE+p)/\rho$), and specific entropy respectively.

The inspiral of two cold neutron stars can be considered as an
isentropic process therefore the last term in Eq. (\ref{eq:coset})
can be set to zero. In addition the system conserves rest mass, $\nabla_\GA(\GR u^\GA)=0$, 
therefore one arrives at the relativistic Euler equation, 
$u^\GB\nabla_\GB(hu_\GA)+\nabla_\GA h=0$.

Decomposing the 4-velocity with respect to the corotating observer as
\be
u^\GA=u^t (k^\GA+V^\GA), 
\label{eq:4velocity}
\ee
with 
\be
k^\GA =t^\alpha +\Omega\phi^\alpha
\label{eq:baumhkv}
\ee
being the helical vector, one can rewrite the conservation of rest mass as
\be
\mathcal{L}_{\bf k} (\rho u^t) + \na_\alpha (\rho u^t V^\alpha)=0, 
\label{eq:rmcons}
\ee
and the Euler equation as, 
\be
\mathcal{L}_{\bf k}(hu_\GA) + \nabla_\GA\left(\frac{h}{u^t}+hu_\GB V^\GB\right)
+V^\GB \GO_{\GA\GB} = 0 \, ,
\label{eq:ree}
\ee
where $\GO_{\GA\GB} = \nabla_\GB(hu_\GA)-\nabla_\GA(hu_\GB)$ is the vorticity tensor.
Here $\mathcal{L}_{\bf k}$ is the Lie derivative along $k^\alpha$, and vector
$\GP^\GA$ generates rotations in the $\GP$ direction. Eqs. (\ref{eq:rmcons}) and (\ref{eq:ree})
are valid for any type of flow field. In the next sections we will describe 3 types of flows:
corotating, irrotational, and arbitrary spinning.

\subsection{Corotating binary neutron star initial data}
\label{sec:corot}

The first BNS quasiequilibrium initial data have been constructed by 
Baumgarte \textit{et~al.} \cite{Baumgarte97,Baumgarte98b} using the IWM formalism.
Neglecting deviations from a strictly periodic circular orbit and 
assuming the two stars to be corotating,
the fluid 4-velocity is proportional to the helical vector  $k^\GA$,
as in a single rotating neutron star, which is assumed to be a Killing vector -- 
a time translation symmetry in a rotating frame. 
Here $\Omega$ in Eq. (\ref{eq:baumhkv}) is the orbital angular velocity of the binary system
which is also the spin angular velocity of the corotating component stars.
Similarly to the single rotating star case the Euler equations can be integrated 
to yield a first integral. Equation (\ref{eq:ree}) with $V^\alpha =0$ and $\mathcal{L}_{\bf k} (hu_\GA)=0$
results to 
\be
\frac{h}{u^t}= C \, ,
\label{eq:baumbi}
\ee
where $C$ is a constant. This first integral  
can be used to compute the single unknown thermodynamic variable (for example $h$).
The component $u^t$ can be found from the normalization of the 4-velocity and
the gravitational variables, $u^t=1/\sqrt{\GA^2-\GO^i\GO_i}$, where 
$\GO^i=\GB^i+\Omega\GP^i$ is the so-called corotating shift vector.
There are two constants that appear in the equations, which are the orbital angular velocity
$\Omega$ and the constant $C$ that has the meaning of the injection energy \cite{Thorne1967}. 
These are evaluated from two conditions, one is to fix $\rho$ at the center of the star, 
and the other to fix the separation.
\footnote{ Sometimes the number of constants is augmented to 3 or 4, and additional conditions 
are needed. This results to a system of 3 or 4 non-linear equations that needs 
to be solved at every iteration.}
The whole procedure above is repeated until 
the differences in all computed quantities between two subsequent iteration steps drop
below a certain threshold error.

Baumgarte \textit{et~al.} \cite{Baumgarte97,Baumgarte98b} built quasiequilibrium binary
sequences, i.e. sequences of solutions of constant rest mass and decreasing separation, 
which approximate evolutionary trajectories of neutron star binaries undergoing slow inspiral via the
generation of gravitational radiation. By construction these sequences
maintained corotation (i.e. the spin frequency of the neutron stars increases as they become closer)
which is not realistic, since tidal torques due to realistic viscosity mechanisms are not strong enough 
to synchronize the neutron star spin. The conserved quantity in
a BNS system besides the rest mass is circulation \cite{Kochanek92,Bildsten92,Friedman01a}.  
Nevertheless, constructing corotating sequences was a very important first step in understanding close binary dynamics. 
They found that the maximum density of the component stars decreases as they approach each other and become 
tidally deformed 
\footnote{The behavior of maximum density as the binary system merges was the subject of intense debate which stemmed
from relativistic numerical simulations \cite{Wilson95,Wilson96,Mathews97}
in which it has been noted that as the stars approach each other their interior maximum density \textit{increases}. 
As a consequence, depending on the EOS, BNS would collapse individually 
toward black holes prior to merger. At that time many authors have disputed the finding of density increase 
\cite{Lai96,Wiseman97,Brady97,Shibata98b,Baumgarte98c}, while Flanagan \cite{Flanagan:1998c} pointed to 
an inconsistency in the solution of the shift vector employed in \cite{Wilson95,Wilson96,Mathews97}
and was responsible for this erroneous behavior \cite{Mathews00}. }.
At the same time, the mass that a given maximum
density can support increases as the stars approach each other. The authors found 
that the maximum allowed mass of neutron stars in quasiequilibrium
binaries increases with decreasing separation. These effects are larger for a
smaller polytropic index and hence a stiffer EOS.
In this review we use the term stiff in two ways. If the EOS is a simple polytrope, then a stiff EOS has a higher value
of polytropic exponent $\Gamma$. On the other hand if we have a realistic EOS (as those represented in piecewise poytropic 
form \cite{Read:2009a}) a stiff EOS signifies that a more massive neutron star can be supported\footnote{A lot 
of times this means that for a given mass, stiff EOSs result to larger radius. Confusion may arise since for simple
polytropes (in terms of normalized mass and radius), a higher $\Gamma$ result to neutron stars with smaller 
radius (for a given mass), or to a smaller maximum mass, although the maximum compactness $M/R$ becomes higher.}. 

Baumgarte \textit{et~al.}  computed the binding energy of the system $E_b=M-M_\infty$, where $M$
is the ADM mass at a given separation, and $M_\infty=M_1+M_2$ the sum of the two gravitational 
masses of the two component stars at infinite separation.  In a corotating binary as the 
stars approach each other, the angular velocity increases while the binding 
energy decreases as in Newtonian gravity. When the separation becomes sufficiently
small, finite size effects start playing an important role, and in particular
the (positive) rotational energy of the stars reduces the negative binding energy. If the 
EOS is sufficiently stiff ($\Gamma\gtrapprox 2$ so that the moment of inertia
is sufficiently large) the binding energy goes through a minimum and then 
increases again prior to contact. This minimum in energy which coincides with the 
minimum in the angular momentum of the system (see Eq.(\ref{eq:dmdj}) \cite{Friedman01a}),
approximately signifies the innermost stable circular orbit (ISCO) beyond which 
a rapid plunge and merger occurs. By locating the turning points in their
total energy versus separation curves, the authors identified the onset of orbital
instability, and the orbital parameters at that critical radius finding no evidence 
of destabilization \cite{Wilson95,Wilson96,Mathews98}.
For stiff EOSs the stars are more extended and such minimum in the energy is absent 
i.e. the binding energy is monotonically decreasing and merger happens at some 
minimum distance at the inner Lagrange point. 


Corotating BNSs have also been calculated by Marronetti {\it et~al.} \cite{Marronetti98} 
using the IWM scheme similarly to \cite{Baumgarte97}, where the focus was in the direct determination of the 
ISCO. The authors presented equal mass corotating binaries from a larger separation, where the 
Newtonian approximation is valid, to a smaller separation close to ISCO.
Further analysis of conformally flat corotating configurations has been presented by Miller 
{\it et~al.} \cite{Miller:2003vc}, who explicitly showed that if one takes into account the spin 
energy of the neutron stars, the effective binding energy (i.e. the binding energy of an effective irrotational BNS) 
no longer exhibits a minimum (this was in agreement with irrotational quasiequilibrium BNS sequences as we describe in Sec \ref{sec:irrot}).
The authors made a complete comparison of the quasiequilibrium corotating sequences with full general relativistic simulations and
analyzed the conformal flatness assumption, and the assumption of the 4-velocity being proportional to the Killing vector 
field $k^\GA$ in  a BNS evolution.
It turned out that the violation of the assumption regarding the timelike helical Killing vector field was
an order of magnitude larger than the violation of the assumption of conformal flatness. In particular, for
corotating BNSs at separations less than $47M_0$ (which was slightly more than 6 neutron star radii),
where $M_0$ is the neutron star rest mass, the 
quasiequilibrium solutions violate the Einstein field equations at the order of $10\%$.
On the other hand the conformal flatness assumption was violated at the order of $1\%$. 


\subsection{Irrotational binary neutron star initial data}
\label{sec:irrot}

As mentioned earlier corotation is not maintained during the evolution of a BNS 
due to their small viscosity \cite{Bildsten92,Kochanek92,Miller:2003vc}. Instead the
conserved quantity is circulation, and it is expected that most BNSs can
be approximated as irrotational, i.e. having no vorticity in the inertial frame.
In this case the stars are counterotating in the rotating frame and therefore the 
fluid velocity is nonzero there. This means that in contrast to the corotating 
case, the continuity equation is nontrivial and leads to an equation for the 
divergence of the fluid velocity. A first formulation that treated the case of
nonsynchronized binaries was presented by Bonazzola {\it et~al.}
\cite{Bonazzola97} and Asada \cite{Asada1998}. 
Later Shibata \cite{Shibata98} and Teukolsky \cite{Teukolsky98}
presented a simplified formulation introducing the relativistic velocity potential. 
Since for irrotational BNSs the three formalisms are equivalent (see Appendix A in \cite{Gourgoulhon01}), 
with the latter two being much simpler, and becoming the workhorse in all subsequent numerical implementations,
we will limit our discussion to those.
As in the corotating case, we assume that
the system preserves its properties under the action of the helical Killing vector 
(\ref{eq:baumhkv}) (stationarity property), i.e.
\be
\mathcal{L}_{\bf k}(u_\GA) = \mathcal{L}_{\bf k}(h) = 0 .
\label{eq:hkvsym}
\ee
For an irrotational flow the vorticity tensor $\GO_{\GA\GB}$ is by definition zero
which implies that the  4-velocity can be derived from a potential $\Phi$
\be
h u_\GA = \nabla_\GA \Phi  ,
\label{eq:velpot}
\ee
which together with the conservation of rest mass, Eq. (\ref{eq:rmcons}), 
will yield an elliptic equation for $\Phi$, $\nabla_\GA (\frac{\GR}{h}\nabla^\GA\Phi) = 0$, 
or in 3+1 form under helical symmetry (\ref{eq:hkvsym}), (\ref{eq:rmcons})
\be
D_i(\GA\GR u^t V^i) = 0, \qquad \mbox{with}\qquad     V^i = \frac{D^i\Phi}{hu^t} - \GO^i \, .
\label{eq:poissonphi}
\ee
Here $D_i$ is the 3D covariant derivative associated with $\gamma_{ij}$.
A boundary condition for this Poisson type of equation 
can be derived by setting the fluid velocity in the corotating frame $V^\GA$
at the stellar surface, $r=R_s$, to be tangent to the surface itself, 
$V^\alpha \na_\alpha p |_{r=R_s} = 0$,   which yields 
\be
(D^i\Phi - hu^t\GO^i)D_i\rho|_{r=R_s} = 0 .
\label{eq:irrbc}
\ee

Having determined the enthalpy vector (canonical momentum), $hu_\GA$, one needs to 
calculate the enthalpy $h$ in order to have a complete solution of the fluid equations for a barotropic EOS. 
Equation (\ref{eq:ree}) under the assumptions of Eqs. (\ref{eq:hkvsym})  yields the first integral
\be
\frac{h}{u^t}+V^\GB \nabla_\GB\Phi= {\rm const} .
\label{eq:irrfi}
\ee
We emphasize here that in order to  
arrive to the first integral Eq. (\ref{eq:irrfi}), and the elliptic equation for $\Phi$, 
Eq. (\ref{eq:poissonphi}), both the first (in particular $\mathcal{L}_{\bf k}u^t=0$)
and the second parts of Eq. (\ref{eq:hkvsym}) are used. 
In other words both the thermodynamic profile and the velocity field need to respect the 
approximation of stationarity in the corotating frame.


The irrotational formulation by Shibata \cite{Shibata98} and Teukolsky
\cite{Teukolsky98} together with the IWM formalism
\cite{Isenberg08,Wilson95,Wilson96} has been used by many groups around the
world in order to obtain realistic BNS initial data.  The first such
calculation has been done by Bonazzola, {\it et~al.}
\cite{Bonazzola98b} which explicitly showed that the maximum density of
constant rest-mass irrotational binaries \textit{decreases} with respect to its
value at infinite separation, as the binaries are approaching each other.  As 
expected the decrease is not as pronounced as the corresponding decrease of
the corotating binaries. This can be explained by the fact that the latter have 
considerable spin at close distances which leads to a larger decrease in their
central density, as in a sequence of single rotating neutron stars that starts
at the spherical limit and evolves towards the mass-shedding limit. 
These first results have soon been confirmed by
Marronetti {\it et~al.} \cite{Marronetti99}
as well as by Ury\=u \& Eriguchi \cite{Uryu00}.  The picture that emerged from these early
BNS initial data studies
\cite{Bonazzola98b,Gourgoulhon-etal-2000:2ns-initial-data,Marronetti99,Uryu00,Uryu00a,
 Usui2000,Taniguchi02b,Faber:2002zn,Taniguchi03,Usui2002,Bejger04}
can be summarized as follows: 1) Irrotational binary neutron star systems have been found
to be dynamically stable during the inspiraling stage until Roche lobe overflow
starts at the L1 point as in the Newtonian cases. The quasiequilibrium
sequences showed no increase of the maximum energy density during
the inspiraling phase as a result of gravitational wave emission.  2) In general, equal mass corotating
BNS sequences terminate at contact point, while all
the other types of sequences (corotating nonequal mass BNSs, irrotational
equal mass, and irrotational nonequal mass binaries) end at the mass shedding point.
3) Turning points (defining the ISCO) in the binding energy and angular
momentum of constant rest-mass sequences are found for corotating binary
systems for $\Gamma\geq 2$ and for irrotational ones for $\Gamma\geq 2.5$.  It
was conjectured, that if the turning points are found for the Newtonian
binaries for some adiabatic index $\Gamma$, they should exist in the general
relativistic computations for the same value of $\Gamma$.  4) For different mass
binary systems a turning point may not always exist.
5) The deformation of the star is determined by
the orbital separation and the mass ratio and is not affected much by its
compactness. 6) The decrease of the central energy density depends on the
compactness of the star and not on that of its companion.

The first three groups that computed irrotational BNS quasiequlibrium initial data 
used the same formalism for the gravitational field and hydrodynamics but completely 
different numerical techniques to solve these equations. In particular Bonazzola {\it et~al.} 
\cite{Bonazzola98b,Gourgoulhon-etal-2000:2ns-initial-data} used a  multi-domain
pseudo spectral method with several coordinate systems one of which
is fitted to the stellar surface. The implementation has been done within the
\lorene code \cite{lorene_web} and produced solutions with excellent accuracy with only limited
amount of resources. 
Marronetti {\it et~al.} \cite{Marronetti99} on the other hand, 
used a Cartesian coordinate system and a finite difference method. They used domain decomposition 
techniques which provided a natural way of code parallelization, that reduced the processing time 
enormously. For the conservation of rest-mass density equation, they splitted the elliptic equation
for the potential field into homogeneous and inhomogeneous parts, with
the homogeneous field satisfying the boundary condition. While formally
correct, this solution was difficult to implement accurately and the
results strongly depended on grid resolutions (see Sec. \ref{sec:other} for recent work on this issue).
In the works by Ury\=u \& Eriguchi \cite{Uryu00} spherical coordinates and finite differences
were used while the Poisson equations were inverted using a Green's function approach. This was the
first application of the Komatsu-Eriguchi-Hachisu (KEH) method to BNSs in general relativity. Separate coordinates
for the gravitational and fluid equations have been employed (surface fitted coordinates) that enabled the accurate
determination of the neutron star surface.


A systematic study of quasiequilibrium BNSs has been performed by Taniguchi \& Shibata \cite{Taniguchi2010}
where a large number of systems with different mass ratios $q=M_1/M_2$
($0.71\leq q \leq 1$), total masses, and EOSs has been studied under
the IWM formulation using the \lorene code. By constructing a large number of BNS sequences or constant
rest mass, the authors investigated the behavior of the binding energy and total angular momentum, 
the endpoint of such sequences, and the orbital angular velocity as a
function of time. They found that for piecewise polytropic EOSs
the change in stellar radius at fixed core stiffness makes
the orbital angular velocity at the mass-shedding limit vary widely, while the
change in stiffness of the core EOS at fixed stellar radius does not change
the orbital angular velocity at mass-shedding significantly.
Since the less massive star in an unequal-mass binary is tidally deformed by the
companion more massive star and starts shedding mass at larger separation than
that for the equal-mass case,  the orbital angular velocity at the closest separation
decreases as we decrease the mass ratio. On the other hand the orbital angular velocity at
the mass-shedding limit increases as the neutron star mass increases, since a more massive 
star becomes more compact and more difficult to be tidally disrupted for the same EOS. This implies that BNS
with massive stars need to come closer than those with less massive stars for
reaching the mass-shedding limit. The authors derive an empirical formula for the orbital angular velocity at the
mass-shedding limit using a Newtonian argument
\be
M \Omega_{\rm ms} = 0.27 C_1^{3/2}\left(1+\frac{1}{q}\right)^{3/2} q^{1/2}
\label{eq:omegams}
\ee
where $C_1=M_1/R_1$ the compactness of the first neutron star.

In all the initial data above, the IWM formulation as well as helical symmetry for the matter (Eq.(\ref{eq:hkvsym})) 
is assumed. Any sequence constructed from such initial data, satisfies Eq. (\ref{eq:dmdj}). Stated differently,
a sequence whose entropy and rest mass is kept constant, and the flow is either 
corotational or circulation conserving, approximately inspirals as a result of 
gravitational wave emission.  However, because such solutions represent circular orbits of the
IWM spacetime, their sequences deviate from the realistic inspiraling orbits in full general relativity.  
When such initial data are used for precise numerical relativity 
simulations of inspiraling BNS (typically for highly accurate gravitational wave extraction), an approaching velocity for the 
component stars is added to minimize the deviation from the inspiraling orbit (see section \ref{sec:ecc}).


\subsection{Spinning binary neutron star initial data}
\label{sec:spinid}

One of the most important characteristic of a neutron star is its rotational frequency, which in isolation
has been observed to be as high as $716\;{\rm Hz}$, corresponding to a period of 
 $1.4\;{\rm ms}$ for PSR J1748-2446ad \cite{Hessels2006}. For
the BNS systems currently known in the Galaxy \cite{Tauris2017,ZhuX2018} the rotational frequencies are 
typically smaller. The neutron star in the system J1807-2500B has a period of $4.2\;{\rm ms}$ while
systems J1946+2052 \cite{Stovall2018}, J1757-1854 \cite{Cameron2018}, J0737-3039A \cite{Kramer2006}
have periods $16.96,\ 21.50,\ 22.70\;{\rm ms}$ respectively.
Although the majority of the BNS simulations are based on formulations that assume the neutron stars to be irrotational 
this is no longer true when the spins are as high as approximately 10 times the orbital period at merger 
($\sim 3\;\rm ms$). In other words for accurate gravitational wave analysis we cannot ignore neutron star spins that are
$30\;\rm ms$ or less. According to  \cite{ZhuX2018} this will be the case for binaries J1946+2052, J1757-1854, 
J0737-3039A which at merger will have periods of $18.23,\ 27.09,\ 27.17\;{\rm ms}$, respectively.
Note that typical spin down rate observed for neutron stars is around $10^{-13} - 10^{-17}\mbox{s s}^{-1}$, 
where $10^{-15}$s s${}^{-1}=30$ ms$/10^6$ yr.
Also, let's not forget that event GW170817 \cite{GW170817prl} was unable to rule out high spin priors and thus two sets
of data (for low and high spins) were consistent with the observations.

\subsubsection{Earlier formulations and calculations}
\label{sec:earlyspin}

The first attempt to address the neutron star spin in a general relativistic binary setting was by 
Marronetti \& Shapiro \cite{Marronetti03}.
Despite the problems with their approach, it laid the foundations for the more recent advances in this topic
by Tichy \cite{Tichy11} that we will discuss in the next section. 
The central points of \cite{Marronetti03} are the use of the generalized Bernoulli law
\be
\mathcal{L}_{\Bu}(hu_\GA k^\GA) = 0
\label{eq:gbl}
\ee
together with a decomposition 
\be
v^i = v_{\rm \scriptscriptstyle RS}^i + v_{\rm \scriptscriptstyle RI}^i  + 
\GE^i_{\ jk}\Omega^j x^k   
\label{eq:msvel}
\ee
for the fluid velocity. Here $v^i=u^i/u^t$ and Eq. (\ref{eq:msvel}) decomposes the velocity with respect
to the inertial frame $v^i$, into an orbital velocity (last term) and 
a velocity with respect to the rotating frame (other two terms). The latter is further written as the sum of 
a solenoidal $v_{\rm \scriptscriptstyle RS}^i$, and an irrotational part $v_{\rm \scriptscriptstyle RI}^i$, i.e.
\be
\pd_i v_{\rm \scriptscriptstyle RS}^i = 0 ,\qquad 
\GE_{kji} \pd^j v_{\rm \scriptscriptstyle RI}^i = 0.
\label{eq:msvel1}
\ee
Here $\GE_{kji}$ is the three-dimensional Levi-Civita tensor. Equation (\ref{eq:gbl}) holds if $k^\GA$ is a Killing
vector, and a symmetry vector for the stress-energy tensor. It generalizes the relativistic Bernoulli law
for stationary flows, i.e. that the enthalpy per unit rest mass is constant along the flow lines, $hu_t=\rm const$,
and one can show that in the cases of corotating or irrotational flows the more stringent relation
\be
hu_\GA k^\GA = {\rm const}\qquad \mbox{everywhere inside the star}
\label{eq:firstint}
\ee
holds. The authors make their first assumption at this point by taking Eq. (\ref{eq:firstint}) to hold for the
arbitrary spinning binaries too. For the components of the fluid velocity in Eq. (\ref{eq:msvel}) they assume 
\be
v_{\rm \scriptscriptstyle RS}^i = (a-1) \GE^i_{\ jk} \Omega^j (x^k-x_0^k),\qquad
v_{\rm \scriptscriptstyle RI}^i = \pd^i\Phi  .
\label{eq:msvel2}
\ee
The first equation above assigns a uniform angular velocity of magnitude $(a-1)\Omega$ (a fraction of the orbital
angular velocity) to the neutron star companions in the rotating frame ($x^k,\, x^k_0$ are
the coordinate position vector and the position of maximum baryonic density respectively), while the second expresses
the fact that every irrotational vector field is a gradient of a scalar. The free parameter $a$ 
controls the approximate spin of the neutron star $\Omega_s^i = a \Omega^i$ as seen by a distant observer in the inertial frame.
Since neither $a$ nor $\Omega_s^i$ have a strict physical meaning, in order to measure the spin of each neutron star
the authors introduce the concept of circulation,
\be
\mathcal{C} = \oint_c h u_\GA dx^\GA ,
\label{eq:mscirc}
\ee
which according to the Kelvin-Helmholtz theorem \cite{Carter1979_conservation_laws} is conserved for isentropic flows
along any closed path $c$ on hypersurfaces of constant proper time. Carter \cite{Carter1979_conservation_laws} has 
shown that conservation laws like Eq. (\ref{eq:mscirc}) can hold beyond hypersurfaces of constant proper time, and
Marronetti \& Shapiro \cite{Marronetti03} assume constant circulation along a sequence of orbits. This 
assumption is consistent with neglecting the radial velocity of the fluid as the binary moves to closer
separations. In reality both the radial velocity is nonzero and the circulation will slightly change
as the stars get closer together. For an irrotational flow the closed integral (\ref{eq:mscirc}) over
a gradient yields zero circulation, while for a corotating binary $\mathcal{C}$ is monotonically increasing from
zero at infinite separation to a finite value (but relatively small with respect to the maximum possible, 
see Sec \ref{sec:other}) all the way to merger. 
The irrotational part of the velocity, $v_{\rm \scriptscriptstyle RI}^i$ in (\ref{eq:msvel2}), leads to a Poisson-type of 
equation, similar to Eq. (\ref{eq:poissonphi}), which the authors solved following the methods of 
Marronetti {\it et~al.} \cite{Marronetti99}.  
Sequences of different values of circulation have been constructed and the authors reported on a spin-up effect
for all cases examined, except for the sequence with the largest value of compactness.


A different approach to construct spinning BNS initial data has been introduced by Baumgarte \& Shapiro 
\cite{Baumgarte:2009,Baumgarte:2009e}. In an effort to reduce the Euler equations to an elliptic problem they
took the divergence of Eq. (\ref{eq:ree}) and 
derived a Poisson-type of equation for the auxiliary variable
\be
H = \frac{h}{u^t} + V^i \hat{u}_i , 
\label{eq:bsH}
\ee
where $\hat{u}_i= \GG_i^{\ \GA} h u_\GA$ the projected enthalpy current. 
Two generalizations for the enthalpy current have been introduced:
\begin{eqnarray}
U1)\quad  \hat{u}_i & = & D_i\Phi + \GH \Omega\GP_i ,   \\
U2)\quad  \hat{u}_i & = & D_i\Phi + \GH h u^t (\GB_i+\Omega\GP_i) 
\end{eqnarray}
where $\GH$ is a parameter. Decomposition ($U1$) does not lead to a
corotating flow when $\GH=1$, while ($U2$) does. For both decompositions when $\GH=0$ one gets
an irrotational flow. Although the introduction of decompositions $(U1)$ or $(U2)$ were in
the correct dirrection (see \cite{Tichy11} and \cite{Tsokaros2018}) the main problem of the 
approach in \cite{Baumgarte:2009} was the effort to reduce the Euler equation into an elliptic problem.
As it was identified by Gourgoulhon \cite{Baumgarte:2009e}, by doing so,
one solves only a subset of the equilibrium equations in general, but not all of
them. In particular for the Euler equation to be satisfied, both its divergence and its curl has to vanish. 
For the limiting cases of corotational or irrotational fluid flow the curl of these equations does vanish
identically, but this is not necessarily the case for intermediate circulation.


\subsubsection{Tichy's formulation for spinning binary neutron star initial data}
\label{eq:Tichy}

A formulation that remedies the problems of \cite{Marronetti03,Baumgarte:2009} has been proposed by Tichy 
\cite{Tichy11}. The main ingredients of his approach was the decomposition of the spatialy projected enthalpy 
current 
\be
\hat{u}_i = D_i\Phi + s_i,
\label{eq:hatui}
\ee
as in the decomposition $(U1)$ of \cite{Baumgarte:2009}. Here $s^i$ is the neutron star spin, which in principle
can have any form. The second main ingredient in Tichy's formulation is a first integral of the Euler equation 
as in Eq. (\ref{eq:irrfi}). Although this first integral has the same 
form as the one in the irrotational case in reality it is different because the velocity with respect to the 
corotating observer is now 
\be
V^i = \frac{D^i\Phi +s^i}{hu^t} - \GO^i  ,
\label{eq:spinvel}
\ee
i.e. it depends on the choice of the input spin $s^i$. Tichy derived this first integral by 
using the following assumptions: 
(A1) Helical symmetry for the metric $\mathcal{L}_{\Bk}g_{\GA\GB}=0$. 
(A2) Helical symmetry for the fluid thermodynamic variables like $h$ or $\GR$. 
(A3) Helical symmetry for the irrotational part of the velocity $\GG^\nu_{\ i}\mathcal{L}_{\Bk}(\nabla_\nu \Phi)=0$.
(A4) Spin is constant along $\nabla\Phi/(hu^t)$ which is parallel to the worldline of the star center,
$\GG^\nu_{\ i}\mathcal{L}_{\nabla\Phi/(hu^t)}(s_\nu)=0$.
(A5) Second order terms on spin are neglected $\GG^\nu_{\ i}\mathcal{L}_{{\bf s}/(hu^t)}(s_i)=0$.
In particular, Tichy argued that for the spinning case one cannot assume helical symmetry on the 4-velocity 
$\mathcal{L}_{\Bk} u_\GA=0$, but only on the irrotational part, assumption (A3) \cite{Tichy2016rev}. 
In such case, from the relation
$u^t = g^{t\GA}u_\GA$, and assumption (A1), one has that $\mathcal{L}_{\Bk} u^t \neq 0$, which implies that
the conservation of rest mass in the form of  Eq. (\ref{eq:poissonphi}) (first part) will not hold. Fortunately
the assumption $\mathcal{L}_{\Bk} u_\GA \neq 0$ is not necessary, and the same set of equations can still be
derived. We will come back to this point in the following section.

Equation (\ref{eq:hatui}) generalizes
the irrotational condition Eq. (\ref{eq:velpot}) with the inclusion of a spin vector $s^i$. Although this vector in
principle can be chosen arbitrarily, it was shown \cite{Tichy11,Tichy12} that a choice 
\be
s^i=\Omega_s^a \GP_{s(a)}^i
\label{eq:spin}
\ee
minimizes differential rotation and leads to a negligible shear.
Here $\Omega_s^a$ are the three components of the spin angular velocity and $\GP_{s(a)}^i$ are the rotation vectors
along the neutron star's three axes (the subscript $(a)$ denotes a different vector). 
One important difference between the works of Tichy \cite{Tichy11} and Marronetti \& Shapiro \cite{Marronetti03}
is the fact that the latter decomposes $u^i/u^t=v^i$ in irrotational and rotational parts while the former
performs this decomposition on the spatial enthalpy current $\hat{u}_i$. This choice is better first because $v^i$ 
is not a spatial vector while $\hat{u}_i$ is, and second because it leads to equations that have the correct
limit in the irrotational case while the ones in \cite{Marronetti03} do not.

With the help of Eqs. (\ref{eq:irrfi}), (\ref{eq:hatui}), (\ref{eq:spinvel}), the
conservation of rest-mass density, Eq. (\ref{eq:poissonphi}), as well as the rest of the IWM equations for the gravitational
fields, one can obtain a quasiequilibrium solution for spinning BNS following the same steps as in the irrotational case.
Spinning equilibria with their spins aligned with the orbital angular momentum have been presented in 
\cite{Tichy12} using a polytropic EOS ($\Gamma=2$), 
and the \sgrid code \cite{Tichy:2009} that employs pseudospectral methods. Those sequences
conserve the rest-mass of the neutron star while keeping the spin parameter $\Omega_s$ the same. Although a better description 
for BNS quasiequilibria conserve rest mass and circulation \cite{Marronetti03,Tsokaros2018}, the two descriptions
give similar results \cite{Tsokaros2018}. 
A large suite of results using the \sgrid code have been presented in \cite{Dietrich:2015b}, where
different spins, eccentricities, mass ratios, compactions, and EOSs have been explored.
The authors produced highly eccentric and eccentricity-reduced initial data (see Section \ref{sec:ecc}), as well as
unequal-mass binaries with mass ratios $q\approx 2$. In addition they constructed binaries with arbitrary
spins, misaligned with the orbital angular momentum, and studied precession. In all cases the dimensionless spins
of the neutron stars were $\lesssim 0.16$. Recent upgrades of the \sgrid code \cite{Tichy2019} that involved a new grid structure, 
the use of different coordinates, as well as a reformulation of the equations for the conformal factor and the velocity 
potential, enabled the same group to reach spins all the way to the mass-shedding limit ($\sim 0.59$ for a $\Gamma=2$ 
polytrope) and compactions up to $C\sim 0.28$.

\subsection{Other developments}
\label{sec:other}

A completely different method to compute BNS initial data with arbitrary spin has also been presented by 
Tsatsin and Marronetti \cite{Tsatsin2013}. Their method does not look for solutions of the Hamiltonian and
momentum constraints: their satisfaction is only asymptotic with binary separation. Instead it consists of a
variant of metric superposition that addresses two common problems, large
stellar shape oscillations and orbital eccentricities. It reduces the former to
variations of the order of $1\%$ and offers great control over orbital
eccentricities. The authors also found that these initial data sets
possess less junk radiation than that found in standard quasiequilibria. 
Techniques based on superposition and the conformal thin sandwich formulation of the constraint equations
have been used for generic initial data calculations in \cite{East2012d}, while spinning initial data based on 
superposition of irrotational and corotating flows have been employed in \cite{Kastaun2013}.


Another approach to BNS initial data has been presented in \cite{Tsokaros2015,Tsokaros2018} 
which employs the \cocal library that has been used in the past to succesfully 
compute a great variety of quasiequilibria, from RNS and quark stars (axisymmetric or triaxial) 
\cite{Huang08,Uryu2016a,Uryu2016b,Uryu2017,Zhou2017xhf,Zhou2019hyy},
binary black holes  \cite{Uryu2012,Tsokaros2012,Uryu:2012b},
MRNSs with mixed poloidal and toroidal magnetic fields \cite{Uryu2014,Uryu2019} (see section \ref{sec:mrns}),
as well as self-gravitating BHDs \cite{Tsokaros2018a} (see section \ref{sec:bht}).
The main characteristics of the \cocal code is the use of finite differences and a Green's function approach as 
first developed in \cite{Uryu00} for neutron stars and in \cite{Tsokaros2007} for black holes to achieve a convergent solution
through a Picard type of iteration. 
The field equations are solved in spherical coordinates in multiple patches and a smooth solution is obtained
everywhere through boundary surface integrals. For the fluid equations surface fitted coordinates are
being implemented that allow accurate representation of the neutron star surface which is important in order to impose
boundary conditions.
Comparison both with the pseudospecral code \kadath \cite{Uryu:2012b} as well as with \lorene \cite{Tsokaros2015,Tsokaros2016} 
showed excellent agreement, with the small initial differences being leveled out in the first couple of 
iterations of an evolution. The \cocal code can produce accurate initial data for a variety of neutron or quark star EOSs in a 
piecewise or tabulated form. It has been employed for the construction of the highest compactness BNS system
($C=0.34$ with a total mass of $M=7.90 M_\odot$) in quasiequilibrium to date, using a causal and compressible EOS \cite{Tsokaros:2019lnx}.

The first spinning quasiequilibrium sequences with a nuclear EOS have been presented by Tsokaros {\it et~al.} \cite{Tsokaros2015}.
The authors used the formulation by Tichy \cite{Tichy11} but started from different assumptions. In particular using 
Eqs. (\ref{eq:hatui}) and (\ref{eq:spinvel}) one can rewrite the Euler Eq. (\ref{eq:ree}) as 
\be
\GG_i^{\ \GA}[\mathcal{L}_{\bf k}(hu_\GA) + \mathcal{L}_{\bf V}(s_\GA)] + D_i\left(\frac{h}{u^t} + V^j D_j\Phi\right) = 0 \, .
\label{eq:spinree}
\ee
Instead of using assumptions (A1)-(A5) \cite{Tichy11}, they assumed (A1), (A2), and
\be
{\rm (B1)}:\ \mathcal{L}_{\Bk}\hat{u}_i=0,\qquad\mbox{and}\qquad
{\rm (B2)}:\ \mathcal{L}_{\bf V} s_i=0.
\label{eq:spassum}
\ee
Assumption (B1) enforces helical symmetry on the fluid velocity, consistent with the helical symmetry 
assumption on the spacetime (A1), and more importantly, consistent with the helical symmetry on the thermodynamic variables (A2).
Assumption (B1) is also necessary for taking $\mathcal{L}_{\Bk} u^t = 0$ and deriving the elliptic equation for the fluid potential
$\Phi$. On the other hand, assumption (B2) roughly expresses the fact that the spin does not change significantly 
with respect to the  corotating observer. 
Although helical symmetry is increasingly more accurate as the binary separation becomes larger (irrespective of the spin),
closer binaries with rapidly spinning neutron stars (with ms or smaller periods) will not satisfy the helical symmetry 
assumptions, and in fact will not satisfy any of (A1)-(A5), (B1),(B2).


BNS initial data within the IWM formulation have also been presented by Tacik {\it et~al.} 
\cite{Tacik15,Tacik16} based on the \spells elliptic solver libraries developed by
Pfeiffer {\it et~al.} \cite{Pfeiffer:2002wt,Foucart2008}. These methods employ pseudospectral techniques
that have been applied from the same group with great success in the binary black hole problem 
\cite{Cook:2004kt,Pfeiffer:2004qz,Caudill:2006hw,Boyle:2006ne}. The authors implement the spinning formulation
by Tichy \cite{Tichy11} and introduce a new diagnostic for measuring the neutron star spin in a BNS setting which is 
based on their work on binary black holes \cite{Lovelace2008c}. In particular they construct the quasilocal spin angular
momentum  \cite{Brown1993,Ashtekar01a,Ashtekar03a}
\be
J_{\rm ql} = \frac{1}{8\pi} \oint_S K_i^{\ j}\GZ^i dS_j ,
\label{eq:qls}
\ee
where $\GZ^i$ is an approximate Killing vector of the spacetime\footnote{Note that this expression is essentially 
identical to the ADM angular momentum calculated at infinity Eq. (\ref{eq:Jadm}).}. 
For spacetimes with axisymmetry, as for example a rotating black hole,
$\GZ^i$ should be chosen as the rotational Killing vector $\GP^i$. For binary systems where no such symmetry exists one needs to construct 
$\GZ^i$ through a minimization principle that results to an eigenvalue problem \cite{Lovelace2008c}.
The surface of integration $S$ is the apparent horizon for the case of a black hole, while for neutron stars although there is no clear 
choice, the stellar surface is the most natural one.
The authors performed this integration both on the stellar surface (which is calculated when a convergent solution is 
obtained) as well as on spheres of increased radius outside the star. They showed that the calculated spin
is independent of the precise choice of $S$ within an accuracy of $1\%$. For irrotational BNSs they found
a quasilocal spin residue of $\sim 10^{-4}$. Highly spinning BNS initial data with $J_{\rm ql}/M_1^2 \sim 0.43$,
where $M_1$ is the ADM mass of a single star at infinite separation, were presented, as well as precessing binaries. 
The spin and orbital precession of the stars were well described by 
the post-Newtonian approximation. In addition the authors implemented
eccentricity control algorithms \cite{Pfeiffer2007} for its reduction targeted for highly accurate waveform
production (see section \ref{sec:ecc}). 


Different spin measures for BNSs were examined in detail in Tsokaros {\it et~al.} \cite{Tsokaros2018}. The authors
examined the relationship between the circulation $\mathcal{C}$, dimensionless spin $J_{\rm ql}/M_1^2$, and spin 
parameter $\Omega_s$ in Eq. (\ref{eq:spin}). For the quasilocal spin, Eq. (\ref{eq:qls}), the authors assumed $\GZ^i$ 
to be the rotational vector around each star's center $\GP_{s}^i$. Realistic spinning sequences that conserve both 
rest mass and circulation were presented and compared with corotating and irrotational (zero circulation) ones.
Regarding the corotating sequence, it was found that the circulation and the dimensionless spin of the 
neutron stars close to the ISCO (turning point of the $E_b(\Omega)$ curve) are much smaller than the possible maximums 
at the mass-shedding limit. Practically this means that corotating neutron stars at the ISCO can be considered as slowly 
rotating. Similarly to isolated slowly uniformly rotating neutron stars, the circulation and the dimensionless spin of the 
component stars increase linearly with the angular velocity as the binary proceeds to closer separations.

For a spinning sequence of constant rest mass and circulation, the binding energy (or angular momentum) as a function
of angular velocity typically follows a curve parallel to the irrotational sequence, but shifted to higher energy
(or angular momentum). This is expected since now the system contains the spin rotational energy which remains
approximately constant along the sequence. For moderate spins (i.e. spins smaller than the corotating value at the ISCO),
$E_b(\Omega)$ or $J(\Omega)$ curves of a spinning sequence exhibit an intersection with the corotating curve at some
angular velocity $\Omega_i$. For $\Omega < \Omega_i$ the binding energy (or angular momentum) of the spinning sequence
is larger than the corresponding one from the corotating and irrotational binaries (with the same rest mass) due to
the excess of rotational energy. On the other hand for $\Omega> \Omega_i$ (binaries in closer separations) 
the binding energy (angular momentum) of the spinning sequence although larger than the corresponding irrotational one,
it is smaller than the corotating. 
The dimensionless spin of each star, $J_{\rm ql}/M_1^2$, is nearly constant for
larger separations and exhibits an increase of the order of $10-15\%$ as ones moves closer to the ISCO. For closer
separations, the choice $\GZ^i=\GP^i_s$ is less accurate and one needs to compute the approximate Killing vector for 
the quasilocal spin \cite{Lovelace2008c}. In addition the rotational parameter $\Omega_s$ was found to be approximately 
constant along a constant rest mass and circulation sequence.

Motivated by the circulation expression for single stars and corotating binaries
\be
\mathcal{C}=\oint_c hu^t \GG_{ij}(\GB^i+\Omega\GP^i)dx^j  \, ,
\label{eq:starcirc}
\ee
the authors proposed a modification for the fluid velocity decomposition Eq. (\ref{eq:hatui}) according to 
\be
\hat{u}_i = D_i\Phi + h u^t s_i\, ,  
\label{eq:uihnew}
\ee
which apart from the shift term in Eq. (\ref{eq:starcirc}), resembles the circulation of a single rotating star. 
This decomposition is similar to the choice $(U2)$ in Baumgarte \& Shapiro \cite{Baumgarte:2009}.
The velocity with respect to the corotating observer, Eq. (\ref{eq:spinvel}), is now modified as
\be
V^i = \frac{D^i\Phi}{hu^t} - (\GO^i - s^i)  \, ,
\label{eq:spinvelnew}
\ee
and new equations of hydrostatic equilibrium are derived. Notice that Eq. (\ref{eq:spinvelnew}) resembles the corresponding
velocity of irrotational binaries Eq. (\ref{eq:poissonphi}), with the spin vector $s^i$ modifying the corotating shift $\GO^i$.
Similar modification is present in the boundary condition, Eq. (\ref{eq:irrbc}). Using Eq. (\ref{eq:uihnew})
the authors computed new sequences of constant rest mass and circulation and found that for a fixed angular velocity, 
the binding energy (and angular momentum) is now larger than the corresponding one using the decomposition of Eq. (\ref{eq:hatui}). 
This shows that the way the 4-velocity is decomposed can influence important quantities like the energy of the system or its angular 
momentum.


The irrotational and spinning formulations involve the solution of an elliptic equation for the velocity potential $\Phi$,
which typically is performed on the so-called surface fitted coordinates for greater accuracy. 
These are fluid coordinates that track the position
of the surface of the star at every iteration and are used to impose the boundary condition Eq. (\ref{eq:irrbc}) for the
irrotational case (or a similar one for the spinning case). 
In order to avoid such complication, Tsao {\it et~al.} \cite{Tsao2020} developed a new technique 
by employing the source term method proposed by Towers \cite{Towers2018}, where the boundary condition is treated as a jump
condition and is incorporated as additional source terms in the Poisson equation. 
If the domain of the star is denoted by $Q^+$,  its exterior by $Q^-$, and its boundary by $\partial Q^+$,
Tsao {\it et~al.} considered the boundary value problem 
\begin{align}  
\nabla^2 \Phi & = S^+(x),\quad x \in Q^+,    & \nabla^2 \Phi  = S^-(x),\quad  x \in Q^-,  \\
\left[\frac{\pd\Phi}{\pd n}\right]  & = a(x),\quad x \in \pd Q^+,   & [\Phi] = b(x),\quad  x \in \pd Q^+,  
\label{eq:gbvp}
\end{align}      
which is a generalization of the irrotational/spinning one for the velocity potential $\Phi$. Here
we denote by $[\Phi] \equiv \Phi^+ - \Phi^- = b(x)$ and 
$\left[\frac{\pd\Phi}{\pd n}\right] \equiv \left(\frac{\pd\Phi}{\pd n}\right)^+ - \left(\frac{\pd\Phi}{\pd n}\right)^- = a(x)$,
with $\Phi^+, \left(\frac{\pd\Phi}{\pd n}\right)^+$ being the solution $\Phi$ and its normal derivative evaluated at the interior
of $Q^+$, while $\Phi^-, \left(\frac{\pd\Phi}{\pd n}\right)^-$ the same quantities at the exterior $Q^-$.

The source term method converts the boundary conditions on  $\partial Q^+$ to jump conditions that can be absorbed into the sources. 
The generalized Poisson equation that is solved in the extended domain $Q=Q^+ \cup Q^-$ becomes
\begin{equation}
\label{eq:source_term}
\nabla^2 \Phi = \nabla^2(bH) - H\nabla^2 b - \left(a -\frac{\partial b}{\partial n}\right) |\nabla\GR | \delta(\GR) + S,
\end{equation}
where $\GR$ is the rest-mass density, $S(x)=S^+ H(\GR(x)) + S^-(1-H(\GR(x)))$ with 
$H(\GR)$ the Heaviside function, and $\delta(\GR)$ the Dirac delta function.
The authors presented a comparison between the solution of Eq. (\ref{eq:source_term}) on a Cartesian grid, with the one obtained
by \cocal on surface fitted coordinates, and found excellent agreement with a maximum difference of $1.4\%$. The source term method 
can be used in other problems where nonsmooth solutions are present, as for example in MRNSs.


More recently spinning BNS solutions have been presented by Papenfort {\it et~al.} \cite{Papenfort2021} using the KADATH
spectral library \cite{kadath,Grandclement09}. The authors used Tichy's formulation \cite{Tichy11} to produce a large suite
of highly spinning, and asymmetric systems including one with mass ratio $q=0.455$ with the primary having dimensionless 
spin of $\sim 0.6$. Eccentricity reduced techniques (as described in the next section) have also been implemented.


\subsection{Eccentricity in binary neutron stars}
\label{sec:ecc}

All BNS initial data computations that we have mentioned are based on the IWM formulation, whose
equations hold even without the existence of a helical Killing vector. Since gravitational radiation 
reaction and the accompanying approaching velocity are neglected, quasicircular binary initial data 
of this kind exhibit a small but nonzero deviation from strict circularity of the order of $\sim 0.01$
Despite its small value, the residual eccentricity becomes problematic in the evolution of a binary 
when accurate gravitational wave analysis is needed in order to determine various parameters such as the tidal deformability 
\cite{Favata2014,Yagi2014c,Wade2014}, which is crucial for constraining the neutron star EOS. 

Kyutoku {\it et~al.} \cite{Kyutoku2014} presented a method to further reduce the orbital 
eccentricity by using a similar methodology employed for binary black holes \cite{Pfeiffer2007}. The main
idea is to incorporate an approaching velocity term in the formulation that can be adjusted through 
sequential evolutions until the eccentricity is reduced. The difference between binary black holes and 
BNSs is that this approaching velocity is incorporated on the apparent horizon boundary 
conditions in the black hole case, while for neutron stars it is through the helical Killing vector that controls the 
hydrodynamic equations. The procedure starts with the computation of quasicircular initial data 
and a modified symmetry vector for the hydrodynamical fields as
\be
k^\GA = t^\GA + \Omega \GP^\GA + v \frac{r}{r_0}(\pd_r)^\GA \, ,
\label{eq:ecck}
\ee
instead of Eq. (\ref{eq:baumhkv}). 
Here $v$ is the radial velocity, and $r_0$ the separation from the coordinate origin.
Note that the extra approaching velocity term is a 
conformal Killing vector of the flat metric, and hence does not affect the gravitational field equations. 
The quasicircular initial data ($v=0$) are then evolved for a sufficient time interval
whose duration on one hand has to be long enough to include more than one modulation eccentricity cycles,
and on the other, short enough to avoid strong influence of a long-term secular evolution. The 
authors assumed an interval of $[0.5P,3P]$, where $P=2\pi/\Omega$ is the initial orbital period.
To measure the eccentricity they
locate the star center $x^i_{_{\rm NS}}=(x_{_{\rm NS}},y_{_{\rm NS}},0)$ through the maximum of the conserved rest-mass
density, $\GR\GA u^t\sqrt{\GG}$, and compute the coordinate orbital separation and orbital phase as
\begin{eqnarray}
d(t)  &=& \sqrt{(x_{_{\rm NS,1}}-x_{_{\rm NS,2}})^2 + (y_{_{\rm NS,1}}-y_{_{\rm NS,2}})^2} , \label{eq:eccd}  \\
\GP(t)&=& \tan^{-1}\left(\frac{y_{_{\rm NS,1}}-y_{_{\rm NS,2}}}{x_{_{\rm NS,1}}-x_{_{\rm NS,2}}}\right) + 2\pi N ,  \label{eq:eccphi}
\end{eqnarray}
where $N$ is the number of orbits (in order to make $\GP$ continuous) and indices $1,2$ refer to the two neutron stars.
Following \cite{Pfeiffer2007,Boyle:2007ft}, Kyutoku \textit{et~al.} assume 
\be
\frac{d\Omega}{dt} = A_0 + A_1 t + B \cos(\GO t + \GP_0)  \, ,
\label{eq:eccfp}
\ee
where $\{A_0,A_1,B,\GO,\GP_0\}$ are fitting parameters. The term $A_0+A_1 t$ is due to radiation reaction, while
the term $B \cos(\GO t + \GP_0)$ represents the modulation from the nonzero eccentricity $e$. For a perfectly circular
orbit it should be $B=0$. For a Newtonian orbit with small eccentricity ($e<<1$), it is 
$\Omega(t)\approx\Omega [1+2e\sin(\GO t+\GP_0)]$ which leads to the estimate $e \approx \frac{|B|}{2\GO\Omega} $.
Following Newtonian considerations \cite{Buonanno2011} the adjustments in the orbital velocity and the approaching velocity
turn out to be
\be
\Omega \rightarrow \Omega -\frac{B\GO\sin\GP_0}{4\Omega^2},\qquad\mbox{and}\qquad
v \rightarrow v + \frac{Bd\cos\GP_0}{4\Omega} .
\label{eq:eccadj}
\ee
Using the updated values of $\Omega$ and $v$, the elliptic initial value equations are then solved again, and the 
whole procedure is repeated until an acceptable value of eccentricity ($e\lesssim 10^{-3}$) is obtained. Typically this
requires 3 such iterations. By calculating the gravitational waves in every iteration the authors prove that this eccentricity reduction
procedure leads to gauge invariant results, i.e. the considerations based on the coordinate orbital separation 
are not gauge artifacts.

The authors provide an alternative way to measure the eccentricity based on the gravitational wave angular velocity $\Omega_{\rm gw}(t)$,
\be
e_{\rm gw}(t) = \frac{\Omega_{\rm gw}(t) - \Omega_{\rm gw,fit}(t)}{2\Omega_{\rm gw,fit}(t)}
\label{eq:eccgwfit}
\ee
where $\Omega_{\rm gw, fit}(t)$ a fourth order polynomial (5 fitting constants). They showed that the results obtained 
with this method are similar to the ones based on the orbital motion adding confidence about the reliability of both methods.

Similar algorithms to remove the eccentricity were presented by Tacik {\it et~al.} \cite{Tacik15,Tacik16},
Dietrich {\it et~al.} \cite{Dietrich:2015b}, and Papenfort {\it et~al.} \cite{Papenfort2021}. 
Dietrich {\it et~al.} \cite{Dietrich:2015b} compared two merger simulations
with the SLy EOS one from quasicircular initial data having $e=1.241 \times 10^{-2}$, and the other from eccentricity
reduced initial data having $e=8.7\times 10^{-4}$. They found that the phase difference $\GD\GP_{22}$ oscillates in
the range of $[-0.06,0.06]$ rad, i.e. it produces an approximate dephasing of $\sim 0.12$ rad. 
The amplitude of the non-eccentricity-reduced data 
oscillates around the eccentricity-reduced ones by $5\%$ at early times and decrease as the system approaches merger. 
Note that in this comparison the initial data are at a large distance of $\sim 10$ orbits before merger. At much
closer distances of $\sim 3$ orbits, Tsokaros {\it et~al.} \cite{Tsokaros2016} found that the dephasing that comes by evolving
quasicircular initial data from the \cocal and \lorene initial value codes and evolved by the 
\whiskythc{} \cite{Radice2013b,Radice2013c} evolution code, can be 10 times larger. Therefore for accurate
gravitational wave analysis one has to take multiple factors into consideration including the truncation errors
at various resolutions.

When BNSs merge they will follow highly circular trajectories, since gravitational radiation reaction 
circularizes the orbit \cite{Peters:1964}. Despite of that, BNS mergers in eccentric orbits are still possible, either
by dynamical interactions in dense stellar regions, such as globular clusters \cite{Oleary2009,Lee2010,Tsang2013},
or by exciting their eccentricity by, e.g., the Kozai mechanism in a hierarchical triple \cite{Antonini2012,Naoz2012,Seto2013,Antognini2014}.
Although eccentric BNSs have been constructed by Gold {\it et~al.} \cite{Gold2012}, as well by East \& Pretorius \cite{East2012c}
using different approximations, the first consistent method to construct such initial data has been presented by 
Moldenhauer {\it et~al.} \cite{Moldenhauer2014}. Their method generalizes the
approximate helical Killing vector approach that is used to solve the Euler equation in quasicircular binaries,
to a pair of inscribed helical symmetry vectors (one for each star), 
and allows them to provide initial data for binary neutron stars with arbitrary eccentricity.
The authors assume each star center moves along a segment of an elliptic orbit at apoapsis
(which they take to be on the x-axis), and approximate this small orbital segment by the circle inscribed into the elliptical orbit there. 
The radii of the inscribed circles are 
$r_{c1,2}=(1-e)d_{1,2}$, where $e$ is the eccentricity of the elliptical orbit, and 
$d_{1,2}=|x_{1,2}-x_{\rm cm}|$ are the distances of the neutron stars
from the center of mass. The centers of the inscribed circles are at $x_{c1,2}=x_{\rm cm}+e(x_{1,2}-x_{\rm cm})$, 
thus the approximate Killing vector (near each star) for the elliptical orbit is
\be
k_{\rm ecc1,2}^\GA = t^\GA + \Omega[(x-x_{c1,2})y^\GA-yx^\GA] \, .
\label{eq:kellipt}
\ee
In addition in order to accommodate for the energy loss due to gravitational wave emission, a radial velocity can be added to Eq. (\ref{eq:kellipt})
similar in spirit to Eq. (\ref{eq:ecck}) \cite{Kyutoku2014}. 
Consistent initial data with eccentricities as large as $e=0.5$ have been presented in \cite{Dietrich:2015b}.


\subsection{Non-conformally flat binary neutron star initial data}
\label{sec:ncfid}

In an effort to correct the error coming from the choice of 
the conformally flat three-geometry, 
two groups presented an improved formulation for initial data
starting from different viewpoints. Bonazzola {\it et al.} \cite{Bonazzola:2003dm} aimed to reformulate 
the whole 3+1 numerical relativity system (the Einstein system), and 
instead of using a free-evolution scheme (like the Baumgarte-Shapiro-Shibata-Nakamura formulation \cite{Shibata95,Baumgarte99}) 
that involves hyperbolic
equations, they proposed to use a fully constrained evolution method.
In that formulation one only solves the maximum number of elliptic equations and the minimum number of 
hyperbolic equations: the two wave equations corresponding to the two degrees of freedom of the gravitational field.
To achieve this they assumed a decomposition of the conformal spatial metric as
\be
\TDD{\GG}{i}{j} = f_{ij} + h_{ij} ,
\label{eq:confmetric}
\ee
where $f_{ij}$ is the flat metric in the chart of the 3-dim hypersurface, and $h_{ij}$ the 
components that need to be evaluated. 
The authors imposed a condition for the determinants, $\det\TDD{\gamma}{i}{j}=\det f_{ij}$,
and used maximal slicing and a generalization of the Dirac gauge to curvilinear coordinates
\be
\zD_b\tgmabu=0 .
\label{eq:dirac}
\ee
This gauge fixes the spatial coordinates $x^i$ in each hypersurface $\Sigma_t$, and has been introduced by Dirac
\cite{Dirac59a} as a way to fix the coordinates in the Hamiltonian formulation of general relativity. The non-conformal
flat part of the metric $h_{ij}$ satisfies a wave-like equation 
which in spherical coordinates and in the presence 
of the Dirac gauge can be reduced to two scalar wave equations. This new evolution scheme in effect determines
a new initial data formulation also, where the quantities $\{\GC,\GA,\GB^i,h_{ij}\}$ must be determined from a set
of elliptic equations.

At the same time Shibata {\it et~al.} \cite{Shibata04a} proposed a formulation for BNSs
where the full Einstein system is solved similar to \cite{Bonazzola:2003dm}. 
They also used the Dirac gauge to derive elliptic equations for the non-conformal part of 
the methric $h_{ij}$, but not in spherical coordinates.  
The authors provide asymptotic falloff
conditions for the lapse, the shift, the spatial metric, and the extrinsic curvature 
which ensure the equality of ADM and Komar masses, $\MK=M$, therefore generalizing
the results by Beig \cite{Beig1978}, and Ashtekar \& Magnon-Ashtekar \cite{Ashtekar79a} 
for stationary systems.
The equality of the ADM and Komar masses is related to a virial relation for equilibrium,
\be
\int_\Sigma x^a\GG_a^{\ \GA}\nabla_\GB T_\GA^{\ \GB}\sqrt{-g}d^3x = 0  ,
\label{eq:rvr}
\ee
and the first law, Eq. (\ref{eq:1stlaw}), that are used for evaluating the accuracy of the numerical solutions 
including non-axisymmetric ones computed from the above formulation.

The first quasiequilibrium sequences of irrotational BNS under the formulation
described above were calculated by Ury\=u {\it et~al.} \cite{Uryu2006}. Together with the assumptions of maximal
slicing and the Dirac gauge (or spatially transverse condition), the authors restrict the time derivative
terms so that all components of the field equations are elliptic, and
hence that all metric components, including the spatial metric have Coulomb-type falloff.
In particular for the conformal metric a condition $\pd_t \TDD{\GG}{i}{j}=0$ is imposed while for the
extrinsic curvature and the fluid variables helical symmetry is assumed $\mathcal{L}_{\bf k}K_{ij}=0$.

In order to impose conditions (\ref{eq:dirac}) and have a self-consistent iteration scheme, 
an adjustment is necessary for the $h_{ij}$. This is accomplished by introducing new
gauge vector potentials $\xi^a$ as in \cite{Uryu:2009ye}, (or \cite{Uryu2016a} Eq. (29)-(32))
through the transformation
\be
\dl \gmabu \rightarrow \dl \gmabu \,-\, \zD^a\xi^b \,-\, \zD^b\xi^a  ,
\label{eq:gauge_transf}
\ee
which when combined with Eq. (\ref{eq:dirac}), yield another set of elliptic equations for
$\xi^a$. The augmented system for $\{\psi,\alpha\psi,\beta^a, h_{ij},\xi^a\}$ is then solved
using a self-consistent method.
The authors named this new formulation as ``waveless'' formulation due to the absence of gravitational waves 
in the constructed spacetimes. The work of Ury\=u {\it et al.} \cite{Uryu2006} 
has two main conclusions. First when one computes the binding energy of the system
it is found that although at large separations the results agree with the ones coming from post-Newtonian
or from using the IWM formulation, this is not true any more for close orbits. In particular it was shown
that the IWM formalism overestimates $|E_b|$ by neglecting  the nonconformally flat part
of the conformal geometry. The second important conclusion of \cite{Uryu2006} is that the IWM formulation
does not enforce circularity as accurately as the waveless formulation even for large separations. In order 
to prove that, the authors calculated the norm of $\mathcal{L}_{\bf k}K_{ij}$ which should be zero
in exact helical symmetry. They found that in the waveless formulation the number is at least one order
of magnitude less than in the IWM formulation and the discrepancy is larger at larger separations. 

These results have been extended in the sequel work of Ury\=u {\it et~al.} \cite{Uryu:2009ye}
where two formulations for nonconformally flat initial data are examined: waveless
and near-zone helically symmetric \cite{Yoshida06a}. In each formulation, the
Einstein-Euler system, written in 3+1 form on an asymptotically flat
spacelike hypersurface, is exactly solved for all metric components, including
the spatially nonconformally flat potentials, and assuming an irrotational flow. 
In the helically symmetric approach helical symmetry is imposed in the near
zone, from the center of mass to a radius $\sim \GL=\pi/\Omega$, 
and either the waveless formulation is applied outside, or the computational
domain is truncated at that radius. Here $\GL$
is the wavelength of the dominant mode (primarily $\ell=m=2$ quadrupole) of
the gravitational waves expected to be radiated from the system.
As the compactness of the component neutron stars
increases, the behavior of the binding energy and angular momentum of binary
sequences more closely approximates that of point masses.
By using a variety of realistic EOSs in a piecewise representation the authors show that 
this is true for the IWM and waveless/helically symmetric sequences, but the effect
is more pronounced for the IWM models than in the waveless/helically symmetric ones.
The behavior of the IWM sequence was interpreted as the effacing of the tidal
effects due to the spatially conformally flat approximation. 
In effect this correction reflects an overestimation 
in the IWM formulation by 1 cycle in the gravitational wave phase during the last several orbits.


\section{Self-gravitating black hole-disks.}
\label{sec:bht}

BHD systems are omnipresent in astrophysics, from the core
collapse of massive stars \cite{Woosley1993,MacFadyen1999}, and the cores of active galactic nuclei
\cite{LyndenBell1969,Shakura1973,Paczynski1978}, to the merger of two compact objects 
at least one of which is not a black hole
\cite{Shibata99d,Shibata06d,Etienne2007b,Liu:2008xy,Baiotti08,Anderson2007,Rezzolla:2010,Lovelace2013}.
In addition, stellar-mass binaries in active galactic nuclei and quasars or massive black hole binaries in extended disks 
are also possible scenarios with high astrophysical interest.
Such systems constitute  excellent candidates of ``multimessenger astronomy'' since they
will produce copious amounts of electromagnetic and gravitational radiation, detectable by the future Laser 
Interferometer Space Antenna (LISA) \cite{Amaro2017}.

In many of the cases above the spacetime can be approximated by the Kerr solution, and one assumes that the
fluid motion is not backreacting on the gravitational sector. However it is possible that a massive black hole
(or a binary black hole system) is immersed in an exteded disk with mass comparable or even greater 
than the black hole itself. In other words, there may be a timeframe where such BHD spacetimes cannot be described 
by the Kerr metric and the self-gravity of the disk needs to be taken into account. The disk geometry, and density
profile can instigate a number of instabilities \cite{Papaloizou84,Papaloizou85,Blaes1985,Goldreich1986,Narayan1987,
Abramowicz1980} that can affect the gravitational wave signal detected by LISA \cite{Korobkin2011,Korobkin2013,Mewes2016,
Kiuchi2011b,Wessel2020}. Sufficiently massive disks can cause spin precession or even spin flips \cite{Gergely:07}.
At the same time gas accretion can appreciably change the mass and the spin of the black hole(s) \cite{Gammie:2003qi}. 
If a binary black hole is found within a massive disk its gravitational pull, hydrodynamical drag, and planetary migration
can change the trajectories of the companions leaving an imprint on the inspiral rate and the GW phase evolution
\cite{Petrich89,Barausse2007,Barausse2007a,Yunes2011}.
More recently, the announcement of the high-mass binary black hole, GW190521 by aLIGO and VIRGO \cite{Abbott2020},
initiated a debate regarding the several possibilities for forming black holes in the mass gap $\sim 50-120\ M_\odot$.
One of the intriguing proposals \cite{Shibata2021} was that the source of GW190521 is a stellar collapse of a very massive star
leading temporarily to a black hole with a massive disk that is dynamically unstable to the one-armed spiral-shape 
deformation. 

For all these reasons simulating self-gravitating BHDs is important in order to understand the plethora of current and 
future observations. Only by including self-gravity in full general relativity and tracking the nonaxisymmetric perturbations that
it may trigger can gravitational waves from the disk be calculated reliably. The methods that we will describe below aim 
towards calculating self-gravitating BHD (quasi)equilibria, that can serve as self-consistent initial data for various 
evolutionary scenarios.

\subsection{Formulation}
\label{sec:BHdiskFormulation}

Most of the work in BHDs as well as in MRNSs has been done under the assumption of a circular, stationary, and 
axisymmetric spacetime. In these spacetimes (used extensively in the construction of
RNSs \cite{Friedman2012,Paschalidis2017b}), the line element, Eq. (\ref{eq:dstpo}) can be 
constrained to a smaller number of unknown metric components. In particular a spacetime is stationary 
and axisymmetric if there exist a timelike Killing vector $\xi^\GA$ with integral curves $b(t),\, t\in \mathbb{R}$, 
and a spacelike Killing vector $\chi^\GA$ with closed integral curves $c(\GP),\, \GP\in [0,2\pi]$.
Here $t,\,\GP$ are arbitrary parameters. 
Carter has shown \cite{Carter1970} that there is no loss of generality if one further assumes that these 
two vector fields commute, $\xi^\GA\nabla_\GA\chi^\GB = \chi^\GA\nabla_\GA \xi^\GB$. The commutation property 
can be used to promote parameters $t,\,\GP$ into coordinates of the spacetime manifold. In particular
one can choose coordinates $\{t,x^1,x^2,\GP\}$ such that $\xi^\GA:=(\pd_t)^\GA$ and $\chi^\GA:=(\pd_\GP)^\GA$.
In that case the 10 spacetime metric components $g_{\mu\nu}$ will depend only on coordinates $x^1,x^2$
and not on $t,\,\GP$.

If in addition the spacetime is circular, then 
\cite{Papapetrou66,Kundt66,Wald84}
\be
\xi^\mu  R_\mu^{\ [\GA} \xi^\GB \chi^{\GG]} = 0, \qquad \chi^\mu R_\mu^{\ [\GA} \xi^\GB \chi^{\GG]} = 0,
\label{eq:intcon}
\ee
where $R_{\GA\GB}$ is the Ricci tensor, and square brackets denote antisymmetrization. Eq. (\ref{eq:intcon})
guarantee that the 2-dimensional planes orthogonal to $\xi^\GA$ and $\chi^\GA$  are integrable (we have tacitly assumed 
that the spacetime is asymptotically flat, therefore a rotational axis where $\chi^\GA$ vanishes, must exist). 
Under such assumptions the spacetime metric can be further simplified as
\be
ds^2 = g_{tt} dt^2 + 2 g_{t\GP} dt d\GP + g_{\GP\GP} d\GP^2 + 
g_{\scriptscriptstyle AB} dx^{\scriptscriptstyle A} dx^{\scriptscriptstyle B} ,
\label{eq:circular}
\ee
where $A,B\in \{x^1,x^2\}$, with only 6 nonzero components.
A proper choice of $x^1,x^2$ will make the 2-dimensional metric $g_{\scriptscriptstyle AB}$ diagonal
(for example spherical coordinates $\{x^1=r,x^2=\GU\}$) and since all 2-dimensional metrics are conformally 
related, one more component can be eliminated. At the end, the form of a circular, 
stationary and axisymmetric spacetime in the so-called quasi-isotropic form is \cite{Bardeen71,Shibata2007} 
\be
ds^2 = -\GA^2 dt^2 +\GC^4[e^{2q}(dr^2+r^2d\GU^2)+r^2\sin^2\GU(d\GP+\GB dt)^2]
\label{eq:shibatads}
\ee
where all four functions $\GA,\GC,q,\GB$ depend on $r,\,\GU$ only. From now on, in accordance with the coordinate
notation of the metric, we will denote the Killing vectors $\xi^\GA$ and $\chi^\GA$  as $t^\GA$ and $\GP^\GA$.

In the presence of nonzero sources in the Einstein's equations, the Ricci 
tensor in conditions (\ref{eq:intcon}) can be replaced by 
the stress-energy tensor, which result into imposing the circularity property on them. 
In particular for a perfect fluid, conditions (\ref{eq:intcon}) imply that
\be
u^{[\GA} t^\GB \GP^{\GG]} = 0, 
\label{eq:cpf} 
\ee
i.e. the fluid 4-velocity belongs to the hyperplane spanned by $t^\GA$ and $\GP^\GA$, or
\be
u^\GA = u^t (t^\GA + \Omega \GP^\GA),\qquad\mbox{with}\qquad \Omega:=\frac{u^\GP}{u^t}=\frac{d\GP}{dt}.
\label{eq:c4vel}
\ee
Such kind of fluid flow, which is typically assumed in RNSs, will be employed for the BHDs considered below.
In this case, the rest mass conservation equation, 
$\na_\alpha(\rho u^\alpha)=0$, is identically satisfied, and
the equations of motion for the fluid 
$(g_{\GA\GB}+u_\GA u_\GB)\nabla_\GG T^{\GB\GG}=0$ yield 
\be
\frac{\nabla_i p}{\GE+p} = -\nabla_i \ln(u_t) + \frac{\Omega \nabla_i \ell}{1-\Omega\ell} ,
\label{eq:feom}
\ee
where 
\beqn
&\displaystyle
\ell :=-\frac{u_\GP}{u_t}=-\frac{g_{t\GP}+\Omega g_{\phi\phi}}{g_{tt}+\Omega g_{t\GP}}\quad ,
\qquad
\Omega = \frac{u^\phi}{u^t}=-\frac{g_{t\GP}+\ell g_{tt}}{g_{\GP\GP}+\ell g_{t\GP}} ,
&
\label{eq:ellOmega}
\\
&\displaystyle
u_t=\sqrt{\frac{g_{t\GP}^2-g_{tt}g_{\GP\GP}}{g_{\GP\GP}+2\ell g_{t\GP}+\ell^2 g_{tt}}}.  
&
\label{eq:ut}
\eeqn 
Eq.(\ref{eq:feom}) is integrable if a one-parameter EOS\footnote{We often use a barotropic EOS, 
$p=p(\epsilon)$, or $p=p(\rho)$ and $\epsilon=\epsilon(\rho)$.} and a rotation law, 
$\ell=\ell(\Omega)$ (or $\Omega=\Omega(\ell)$), are prescribed.
The same equation  can be written alternatively as
\be
\na_\alpha\ln\frac{h}{u^t}+u^t u_\phi\na_\alpha\Omega=0, 
\label{eq:feom2}
\ee
where $h$ is the relativistic specific enthalpy. Similarly Eq.(\ref{eq:feom2}) is integrable if a one-parameter EOS 
and the rotation law $j:=u^t u_\phi = j(\Omega)$ (or $\Omega = \Omega(j)$) are prescribed.  
The first integral of Eq.(\ref{eq:feom}) or Eq. (\ref{eq:feom2}) defines the surfaces of 
constant pressure (which correspond to the equipotential surfaces in Newtonian theory). 
When the self-gravity of the disk is negligible, these surfaces are derived from 
the Kerr metric components and the rotation law $\ell=\ell(\Omega)$. Such
equilibria, the so-called Polish doughnuts, have been computed by Fishbone \& Moncrief \cite{Fishbone76},
Abramowicz {\it et~al.} \cite{Abramowicz78}, and Kozlowski {\it et~al.} \cite{Kozlowski1978}. They
constitute the workhorse of most BHD studies where disk's gravity is neglected. They are also taken as
the starting point for the calculation of a self-gravitating disk around a black hole. 

For the case where  $\ell=\mbox{constant}$, and  assuming a polytropic EOS $p=k\GR^\Gamma$, Eq.(\ref{eq:feom}) 
can be integrated as
\be
\GR = \left[\left(\frac{\Gamma-1}{\Gamma k}\right) \left(\frac{u_{\rm in} - u_t}{u_t}\right)\right]^{\frac{1}{\Gamma -1}} .
\label{eq:lanzarho}
\ee
Here $u_{\rm in}=u_t(r_{\rm in},\GU=\pi/2)$ is the specific
energy at the inner point of the disk on the equatorial plane. For a radiation dominated disk, $\Gamma=4/3$.
Ignoring the self-gravity of the disk and assuming a Kerr metric results in the determination of $u_t$ through
Eq. (\ref{eq:ut}), 
and therefore the rest-mass density $\GR$. On the other hand if one assumes a general differential rotation 
law of the form $\ell=\ell(\Omega)$, either of Eqs. (\ref{eq:ellOmega}) 
is solved pointwise to calculate $\ell=\ell(x,y,z)$ (or $\Omega=\Omega(x,y,z)$), which in turn is used to compute
$u_t$ from Eq. (\ref{eq:ut}), and finally the rest-mass density is obtained from the first integral of 
Eq. (\ref{eq:feom}) or (\ref{eq:feom2}).

The inclusion of disk's self-gravity is achieved by solving simultaneously the Einstein equations 
in conjunction with Eq. (\ref{eq:feom}) or Eq. (\ref{eq:feom2}) through an iterative method. 
An assumption about the EOS, and the disk's differential rotation law as mentioned above, is required.
Thus the procedure described in the paragraph above for a
non self-gravitating disk, is now repeated at every iteration step, with the difference being the metric components that 
determine $u_t$ are now the solutions of the Einstein system, and not the ones coming from the Kerr spacetime.

\subsection{Mass, angular momentum, and Smarr formula}
\label{sec:mj}

In the case of axisymmetric systems the angular momentum 
can be expressed covariantly (and therefore in a gauge invariant way) as a Komar integral
\be
\JK = \frac{1}{8\pi}\oint_{\Sinfty} \nabla^\GA\GP^\GB dS_{\GA\GB} = 
    \underbrace{-\frac{1}{8\pi}\oint_H \nabla^\GA\GP^\GB dS_{\GA\GB} }_{J_h}
  + \underbrace{\int_{\Sigma_t} T^\GA_{\ \GB}\GP^\GB dS_\GA }_{J_t} ,
\label{eq:JKomar}
\ee
where $\Sinfty$ denotes a sphere whose radius tends at infinity. From Eq.
(\ref{eq:Jadm}) one can show that the Komar angular momentum coincides with the ADM angular momentum,
$J=\JK$. Using Stokes theorem, the surface integral at infinity  can be converted 
to a surface integral on the black hole horizon $H$, and a volume integral on the spatial slice $\Sigma_t$. 
The former can be identified with the black hole angular
momentum $J_h$, while the latter with the disk angular momentum $J_t$
\cite{Bardeen73b,Bardeen73_Stars_Disks_Black_Holes,Carter73_BH_Equilibrium_States},
which with the help of $\nabla_\GB\nabla_\GA\GP^\GB=R_{\GA\GB}\GP^\GB$ can be expressed in terms of 
the stress-energy tensor $T_{\GA\GB}$, Eq. (\ref{eq:JKomar}). 

In an analogous manner to the Komar angular momentum one can define in the presence of the timelike
Killing vector field $t^\GA$ (stationary spacetimes) the Komar mass as a surface integral at infinity
\be
\MK = -\frac{1}{4\pi}\oint_{\Sinfty} \nabla^\GA t^\GB dS_{\GA\GB} = 
    \underbrace{\frac{1}{4\pi}\oint_H \nabla^\GA t^\GB dS_{\GA\GB} }_{M_h}
   +\underbrace{\int_{\Sigma_t} \left(T\GD^\GA_{\ \GB}-2T^\GA_{\ \GB}\right)t^\GB dS_\GA }_{M_t} ,
\label{eq:MKomar}
\ee
where $T=T^\mu_{\ \mu}$. 
Beig \cite{Beig1978}, Ashtekar \&  Magnon-Ashtekar \cite{Ashtekar79a}, and Shibata, {\it et~al.~} \cite{Shibata04a} 
have proved that the Komar and ADM masses are identical, $M=\MK$, for stationary systems.
The integral over the horizon can be identified as the black hole mass $M_h$, while the volume integral
as the gravitational mass of the disk $M_t$. 
In the vacuum case the volume integral over the stress-energy tensor $M_t=0$, and 
$M_h$ can be evaluated through the Smarr formula \cite{Smarr73a}
\be
M_h = \frac{1}{4\pi} \GK_h A_h + 2 \GO_h J_h  \, .
\label{eq:bhmass}
\ee
Here $\GK_h$ is the surface gravity of the black hole, $A_h$ the area of the horizon, and $\GO_h$
the frame-dragging at the horizon defined by $\omega_h = -g_{t\phi}/g_{\phi\phi}$.
For the Kerr black hole, $\GK_h=\sqrt{M^2-a^2}/(2Mr_+)$, and $A_h=8\pi M r_+$, with $r_+=M+\sqrt{M^2-a^2}$
the radial coordinate of the event horizon in Boyer-Lindquist coordinates.
One important point already discussed in \cite{Bardeen73_Stars_Disks_Black_Holes} is that in the presence 
of matter around the black hole, $M_h$ is not a good choice for the gravitational mass of the black hole and can lead to 
erroneous results (for example that $J_h/M_h^2>1$). 
On the contrary, $J_h$ is always a good measure of the black hole angular momentum.

\subsection{Self-gravitating thin disk around black hole}
\label{sec:lanza}

The first general relativistic computation of a BHD system in which the self-gravity of the disk 
was  taken into account has been computed by Lanza \cite{Lanza1992a} 
for a thin pressureless (dust) disk whose stress energy tensor is
$T_{\GA\GB}=\GS u_\GA u_\GB$, $\GS$ being its surface energy density. 
In his method $\sigma$ was prescribed, 
and then the metric and the rotation law were iterated 
until the field equations and the equation of motion were satisfied.  For computing the metric, 
the author used the Bardeen \& Wagoner formulation \cite{Bardeen71} in which 
the self-gravity of a massive infinitesimally thin disk was incorporated as boundary 
conditions of the Einstein's equations at the equatorial plane. 
Although Lanza's black hole-disk model involves a significant simplification in computing the thin disk, 
its qualitative behavior agrees well with that of the self-gravitating equilibrium 
thick disk (having nonzero pressure) around a black hole.

The starting point of \cite{Lanza1992a} was the determination of a non self-gravitating density profile $\GR$,
Eq. (\ref{eq:lanzarho}), around a rotating black hole \cite{Abramowicz78,Kozlowski1978} under the assumption of 
an $\ell={\rm const}$ law and the polytropic EOS for a radiation dominated gas ($\Gamma=4/3$). 
The free quantities to be chosen are the specific angular momentum $\ell$, and the inner
point of the disk $r_{\rm in}$. During the iteration procedure to incorporate the disk's  
self-gravity, the author fixes the surface density which is 
calculated as a quadrature of $\rho$ in the $\GU$ direction, $\GS(r)=\int_0^{\pi}\GR r\GC^2 e^q d\GU$, 
taken from the above non-self-gravitating thick disk. 
The projection of the divergence of the stress-energy tensor orthogonal to the 4-velocity (Euler equation), when 
$p=0$, leads to the geodesic equation 
\be
\pd_\GA g_{tt}+2\Omega\pd_\GA g_{t\phi}+\Omega^2\pd_\GA g_{\GP\GP}=0  \,  ,
\label{eq:geome}
\ee
which the author employs to calculate $\Omega$ on the equatorial plane. The specific angular momentum profile,
and the 4-velocity are  then calculated from Eqs. (\ref{eq:ellOmega}), (\ref{eq:ut}), while the rest mass of 
the disk, $M_0$, is computed as a volume integral over the fluid support. 

Since the reference model of non-self-gravitating thick disk is composed of an ideal gas plus radiation, the polytropic
constant $k$ in Eq. (\ref{eq:lanzarho}) is a function of the ratio $\GD=p_{\rm gas}/p$, $p$ being the total pressure.
By varying $\GD$, and thus $k$, solutions of BHDs are computed where 
it was found that for fixed specific angular momentum $\ell$, the disk rest mass $M_0$ increases with $\GD$, as
the central density becomes larger. Also for fixed ratio $\GD$, increasing $\ell$ results to larger disks and 
therefore $M_0$ also increases. 

Lanza computes BHD sequences of increasing rest mass, where the disk has the same inner radius $r_{\rm in}$ and 
specific angular momentum $\ell$, while the black hole has fixed angular momentum $J_h$ (aligned with the disk
angular momentum), and area $A_h$.
He found that along this sequence:
\begin{enumerate}
\item The ADM mass and angular momentum of the system are increasing.
\item The mass of the black hole, given by the diagnostic $M_h$, is decreasing.
\item The black hole surface gravity $\GK_h$ is decreasing.
\item The black hole horizon radius $r_h$ is decreasing.
\item The black hole horizon angular velocity $\Omega_h$ is decreasing.
\end{enumerate}
Finding (1) is not surprising; it is due to the increase of the gravitational energy and angular momentum caused by
the disk's self-gravity. Finding (2) is also expected since $M_h$ contains part of the binding energy of the system
\cite{Bardeen73_Stars_Disks_Black_Holes}, that becomes increasingly negative as the mass of the disk increases. In fact 
$M_h$ is not a good diagnostic for the black hole mass, as clearly shown by Shibata \cite{Shibata2007}.

The fact that the surface gravity of the black hole decreases (finding (3)) can be explained as follows: 
As the gravitational field of the disk becomes significant, a zero angular momentum
observer located between the black hole  and the disk will feel the outward pull of the disk decreasing his physical acceleration
with respect to infinity, i.e. decreasing the effective gravity. In the limit as one goes to the horizon this is decribed
by $\GK_h$\footnote{A particle between the black hole and the disk will need less angular momentum to stay in a Keplerian orbit
\cite{Abramowicz1984,Chakrabarti1988}.}. Because the surface gravity and the radius of the horizon are related, 
$\GK_h=4\pi r_h/A_h$, and since $A_h$ is kept constant, the radius of the horizon behaves analogously (finding (4)). 

Will \cite{Will1974,Will1975} has pointed that in the presence of self-gravitating matter one cannot make a clear 
distiction for the individual contributions of the black hole and the disk. This makes possible for the black hole
to have zero angular momentum but non zero angular velocity or zero angular velocity and negative angular momentum.
This effect is due to the dragging of the inertial frames by the external self-gravitating disk.
In the case of a slowly rotating black hole the metric function $-\GB$ has no longer a maximum at the horizon but 
at the center of the disk and this maximum increases with $M_0$. This explains finding (5).

Lanza also computes sequences of increasing disk mass with fixed $\Omega_h$, and horizon radius $r_h$, for a slowly rotating
black hole (small and positive $\Omega_h$). Now the black hole's angular momentum is decreasing and becomes increasingly 
more negative (in order to keep $\Omega_h={\rm const}$)
consistent with finding (5). On the other hand, the apparent horizon area (or the irreducible mass) is increasing, consistent
with finding (3), and the fact that  $\GK_h=4\pi r_h/A_h$.

\subsection{Black hole-toroid in equilibrium}
\label{sec:nishida}

The first full calculation of a self-gravitating thick disk in general relativity has been performed by Nishida \& Eriguchi
\cite{NishidaEriguchi94:ToroidAroundBH}. The starting point of their work was the Bardeen formalism and its 
horizon boundary conditions. The new ingredients were: 1)
The stress energy tensor was assumed to be a perfect fluid, Eq. (\ref{eq:pfset}), with a 
polytropic EOS $p=k\GE^\Gamma$\footnote{Notice that in most works reported in this review the polytropic EOS is
$p=k\GR^\Gamma$.}. Contrary to Lanza \cite{Lanza1992a}, 
the authors assumed $p\neq 0$ and computed the hydrostatic equilibrium of
geometrically thick disks. 2) The differential rotation law for the disk was assumed
to be $u^tu_\GP=A^2 (\Omega_c-\Omega)$ folllowing the KEH \cite{Komatsu89,Komatsu89b} 
works on rotating stars. Here $A$ is an input constant parameter that determines the degree of differential rotation 
and $\Omega_c$ a constant that is evaluated during the iteration scheme. 3) The numerical solution of the elliptic 
equations was performed using the KEH method where the second order operator (a Laplacian) 
is inverted by employing a Green's function approach. 

In general, each equilibrium solution of a BHD is obtained by specifying the following: 
i) the black hole mass, 
ii) the black hole spin, 
iii) the EOS of the toroid, 
iv) the rotation law of the toroid, 
v) the rest mass of the toroid, 
vi) the position of the toroid relative to the black hole,
vii) the total angular momentum of the toroid.  
For a strict equilibrium, the directions of black hole spin and the disk's angular momentum should be 
aligned or antialigned. Properties i), ii)  are free parameters, and similarly functional relations iii) and iv) 
can be chosen freely. On the other hand properties v)-vii) determine a unique BHD model.
In actual numerical computations different authors use different ways to specify v)-vii). For example property
v) is controlled by specifying the maximum density $\rho_{\rm max}$ (or $\epsilon_{\rm max}$).
For properties vi) and vii) Nishida \& Eriguchi choose to fix the inner and the outer point of the disk.

The authors constructed sequences of BHDs around nonspinning as well as 
mildly rotating black holes ($J_h/M_h^2 \lessapprox 0.6$) for $\Gamma=2$, and $\Gamma=5/3$ polytropes. 
In general they found that  the disk plays the role of an ``anchor'' since a larger 
angular momentum is needed to keep the same angular velocity as the corresponding Kerr black hole. 
They paid special attention to the cases with $\GO_h=0$, and confirmed previous results
\cite{Lanza1992a,Will1974,Will1975} that even when $\ell\neq\rm const$ the self-gravity of the torus induces 
negative angular momentum $J_h$ on the black hole. Alternatively one can have a black hole with $J_h=0$, while $\GO_h\neq 0$. 
The authors find that black holes with $J_h=0$ have ratio of polar to
equatorial proper circumference $C_p/C_e$  always equal to unity irrespective of the mass of the disk.
This motivates them to define a ``nonrotating'' black hole as one that has $J_h=0$ or $C_p/C_e=1$. Since those
equilibria will have $\GO_h\neq 0$, they will exhibit ergoregions (by definition) despite being spherical
in shape. In other words in the presence of self-gravitating disks one can imagine energy extraction even 
from ``nonrotating'' black holes.


Ansorg \& Petroff \cite{Ansorg05} using their highly accurate multidomain pseudospectral code \cite{Ansorg:2003br},
and the Bardeen formalism \cite{Bardeen73_Stars_Disks_Black_Holes}, 
constructed a wide range of uniformly rotating and constant density self-gravitating rings. They were able
to investigate various properties up to machine precision (for example the authors claim that when $J_h=0$ the
ratio $C_p/C_e$ is close but not exactly one). They presented a BHD with $J_h/M_h^2>1$, while
in a follow-up work \cite{Ansorg2006} they calculate a model with $M_h<0$, lending more evidence that the 
Komar mass for the black hole in the presence of matter is not a good diagnostic.


More recently, the method of Nishida \& Eriguchi was employed by Stergioulas
\cite{Stergioulas2011,Stergioulas2011c} in order to produce more accurate solutions using
a finite difference code. In particular he applied a compactification in the numerical
domain, as in RNS calculations \cite{Cook94b}, using a redefinition of the radial coordinate. 
Self-gravitating sequences of BHD were produced with vanishing
horizon angular velocity and it was found that self-gravitating heavy tori with constant specific 
angular momentum can fill their Roche lobe only if $\ell<4M_{\rm bh}$, similar to
massless disks around a Schwarzschild black hole.


\subsection{Black hole-toroid in the puncture framework}
\label{sec:bhtshibata}

New ideas for the construction of self-gravitating BHDs have been introduced by Shibata \cite{Shibata2007}, 
where methods from binary black hole initial data calculations were employed. In particular the author uses the 3+1 formalism
and the puncture framework with inversion symmetric boundary conditions at the black hole throat, 
together with assumptions on stationarity and axisymmetry to derive a new set of equations 
(4 elliptic plus one first order) that solve the Einstein system. His starting point was the formulation of Krivan
\& Price \cite{Krivan:1998td} according to which in axisymmetric spacetimes around rotating black holes, a nonconformally 
flat form of the 3-geometry can be chosen which allows a simple superposition of Kerr black holes with arbitrary mass and 
spin. For a spacetime element in quasi-isotropic form, Eq. (\ref{eq:shibatads}), the nontrivial
components of the extrinsic curvature are  
\be
K_{i\varphi} = \frac{\GC^4}{2\GA}r^2 \sin^2\GU \pd_i \GB, \qquad\mbox{where}\qquad i\in\{r,\GU\},
\label{eq:shibataKij}
\ee
and the slicing is maximal, i.e. $K=0$.
As a consequence there is only one component for the momentum constraint (the $\varphi$ component) that needs to be 
satisfied and results to 
\be
\frac{1}{r^2}\pd_r(r^2 \GC^2 K_{r\varphi}) +
\frac{1}{r^2\sin\GU} \pd_\GU(\sin\GU \GC^2 K_{\GU\varphi}) = 8\pi
T^t_{\ \varphi}\GA\GC^6 e^{2q}  \, .
\label{eq:shibatamc}
\ee
The equation above is linear in the conformal frame, 
\ie in  $\GC^2 K_{i\varphi}$ \cite{Krivan:1998td}. Thus one can make a decomposition that separates the 
contribution of the Kerr black hole from the torus as 
\be
\GC^2 K_{i\varphi} := K^{\scriptscriptstyle  K}_{i\varphi} + K^{\scriptscriptstyle T}_{i\varphi} ,
\qquad\mbox{and}\qquad
\GB := \GB_{\scriptscriptstyle K} + \GB_{\scriptscriptstyle T}
\label{eq:shibataKijdeco}
\ee
with 
\be
K^{\scriptscriptstyle  K}_{i\varphi} := \frac{H_i\sin\GU}{r} , \qquad\mbox{and}\qquad
K^{\scriptscriptstyle T}_{i\varphi} := \frac{\GC^6}{2\GA} r^2\sin^2\GU\pd_i \GB_{\scriptscriptstyle T}.
\label{eq:shibataKijdeco1}
\ee
with $H_i=(H_r,H_\GU)=(H_E\sin\GU/r, H_F)$ and $H_E$, $H_F$ having well-known expressions in Kerr spacetime
\cite{Krivan:1998td}. The ``Kerr'' contribution to the shift is calculated from  
\be
\pd_r \GB_{\scriptscriptstyle K} := \frac{2\GA H_E}{\GC^6 r^4}, 
\label{eq:shibatabk}
\ee
although this does not mean that $\GB_{\scriptscriptstyle K}$ will be the Kerr shift,
since the conformal factor $\GC$ in Eq. (\ref{eq:shibatabk}) will be a solution of the Hamiltonian
constraint and will have a contribution from the torus. Equation (\ref{eq:shibataKijdeco}) and the momentum constraint 
will then yield an elliptic equation for the torus shift $\GB_{\scriptscriptstyle T}$. 
Elliptic equations for the lapse $\GA$, and the conformal factor $\GC$, are derived from the $\pd_tK=0$ equation,
and the Hamiltonian constraint respectively. On the other hand a combination of the $\pd_t K_{ij}=0$ equations,
\be
\GG^{rr}\pd_t K_{rr} + \GG^{\GU\GU}\pd_t K_{\GU\GU} -3\GG^{\GP\GP} \pd_t K_{\GP\GP} - \pd_t K = 0,
\label{eq:shibataq}
\ee
will yield an elliptic equation for $q$. Together with Eq. (\ref{eq:shibatabk}), this new system in 3+1 and
axisymmetry has one more elliptic equation (in the absence of a torus one has 4 elliptic equations
for $\GA,\GB,\GC,q$) than the Bardeen-Wagoner system
\cite{Bardeen71,Bardeen73_Stars_Disks_Black_Holes,Butterworth1975}  and
completely determines the Einstein equations. 

Shibata introduces the puncture by a transformation for the conformal factor and the lapse to a new set of 
variables $s,\, B$ 
\be
\GC := \left(1+\frac{r_h}{r}\right) e^s,\qquad \GA\GC:= \left(1-\frac{r_h}{r}\right) e^{-s} B ,
\label{eq:shibatapun}
\ee
where $r_h=\sqrt{m^2-a^2}/2$ is the black hole horizon in quasi-isotropic coordinates, 
and $m$, $a$ the mass and spin parameter of the
initial Kerr black hole. This transformation leads to a new elliptic system in terms of $\{s,B,\GB_{\scriptscriptstyle T},q\}$.
Because the vacuum equations are inversion symmetric with respect to
the $r=r_h$ surface, Shibata extends that symmetry in the presence of the fluid and imposes 
\be
\pd_r s = \pd_r B = \pd_r \GB_{\scriptscriptstyle T} = \pd_r q  = 0, \qquad\mbox{at}\qquad r=r_h  .
\label{eq:shibatahbc}
\ee
With the above boundary conditions the 2-surface $r=r_h$ becomes a marginally outer trapped surface,
or an apparent horizon, which in stationary spacetimes agrees with the event horizon \cite{Hawking73a}.

Another important contribution of this work is the careful study of the
diagnostics for the BHD that corrected many of the problems that
plagued previous works, and in particular the diagnostic for the black hole mass.
Contrary to the angular momentum which can unambiguously be separated into the
black hole and the torus components, Eq. (\ref{eq:JKomar}), the
case of mass needs more care. A similar separation for the mass, 
Eq. (\ref{eq:MKomar}), does not lead to a good measure of the black
hole mass $M_h$ in the presence of a massive disk. This point is already being
discussed by Bardeen \cite{Bardeen73_Stars_Disks_Black_Holes} who notes that
$M_h$ defined as such is smaller than the true value of the black hole mass, by
an amount first order in the mass of the torus, since it includes the binding
energy of the system. Numerically this was already reported in the self-gravitating
BHD computed by Lanza \cite{Lanza1992a} who found $M_h$ to be decreasing as the
rest mass of the torus increased. Assigning $M_h$ to the black hole mass is problematic and Shibata
showed explicitly the following erroneous conclusions that such a diagnostic
entails: i) For heavy tori $M_h$ becomes smaller than the irreducible mass of
the black hole (which is impossible by definition of $M_{\rm irr}$) and ii) The sum of the 
rest mass of the torus and $M_h$ is smaller than the ADM mass of the system (which is
is impossible for a bound system).  In order to remedy this problem the
author proposed two new diagnostics for the black hole mass both of which are
inspired from the isolated Kerr spacetime. In the first one the black hole mass is 
estimated from $M_{\scriptscriptstyle C}:=C_e/(4\pi)$,
where $C_e$ the proper equatorial circuference of the black hole horizon. In the
second one, since the angular momentum of the black hole is $J_h$ (Eq. (\ref{eq:JKomar}))
and the irreducible mass can be found from the apparent horizon area
$M_{\rm irr}:=\sqrt{A_h/(16\pi)}$, Shibata uses the Christodoulou formula
\cite{Christodoulou70}, $M_{\rm bh}:=M_{\rm irr}\sqrt{1+J_h^2/(4M_{\rm irr}^4)}$,
to get another measure for the black hole mass. For disk masses up to twice the black hole mass
the two diagnostics, $M_{\scriptscriptstyle C}$ and $M_{\rm bh}$ agree to $\sim 3\%$.

\renewcommand{\arraystretch}{2}
\begin{table} 
\label{tab:diag}     
\caption{Dependence of various black hole diagnostics as a function of the disk rest mass, $M_0$,
along a sequence of $j=\rm const$ self-gravitating BHDs.}                                                
\begin{tabular}{|l|l|l|l|l|l|l|l|}           
\hline           
$M_{\rm bh}\uparrow$  & $M_{\scriptscriptstyle C}\uparrow$     & $M_h\downarrow$ & 
$M_{\rm irr}\uparrow$ & $\frac{|J_h|}{M_{\rm bh}^2}\downarrow$ & $\frac{C_p}{C_e}\uparrow$ & 
$|M_{\rm bh}\Omega_h|\downarrow$ & $M_{\rm bh}\GK_h\downarrow$  \\[5pt]
\hline 
\end{tabular}                                                               
\end{table}

Using a $\Gamma=4/3$ polytropic EOS and a $j=hu_\GP={\rm const}$ differential 
rotating law the author calculates self-gravitating sequences of disks around highly spinning black holes.
The author fixes the mass $m$, and the spin parameter $a$ of the Kerr black hole, therefore the radius of the 
throat is also fixed. As the mass of the toroid increases, the mass and spin of the black hole diverge from
the fixed Kerr quantities $m$ and $a$. Also because of the boundary condition on the shift $\GB$ 
at the throat, $J_h=ma$ always, irrespective of the mass and the spin of the torus. 
In Table \ref{tab:diag} we summarize the behavior of various diagnostics, as the mass of the
disk increases. In particular: 
\begin{enumerate}
\item The angular velocity of the black hole
horizon, $\Omega_h$, decreases with increasing disk mass, due to the fact that the black hole frame-dragging is 
reduced by the slower rotation of the disk (the disk plays the role of an ``anchor'' \cite{Lanza1992a,NishidaEriguchi94:ToroidAroundBH}).
\item The strength of gravity on the event horizon and the magnitude of its surface gravity $\GK_h$ is weakened by the tidal 
force of the torus (less angular momentum is needed to stay in a Keplerian orbit) \cite{Lanza1992a,NishidaEriguchi94:ToroidAroundBH}).
\item The area of the black hole horizon and therefore its irreducible mass is increasing. This results to an actual
increase in the black hole mass as measured by the diagnostic $M_{\rm bh}$, since $J_h$ is constant. At the same
time the dimensioless spin decreases with the disk mass, consistent with finding (1).
\item Also consistent with the behavior of the dimensioless spin is the fact that the ratio $C_p/C_e$ increases
as the torus becomes heavier, and the black hole becomes more spherical.
\end{enumerate}


More recently the method of Shibata has been employed by the Krak\'ow group in a series of works
that investigate the properties of rotation around self-gravitating axisymmetric disks as well as to construct magnetized
equilibria using a second order finite difference code
\cite{Karkowski2017,Karkowski2018,Mach2019,Kulczycki2019,Dyba2019,Kulczycki2021}. 
In particular Karkowski {\it et~al.~} \cite{Karkowski2017,Karkowski2018} proposed a new Keplerian rotation 
law that holds for massless as well as massive disks according to 
\be
u^tu_\GP = -\frac{3}{2\GL}\frac{d}{d\Omega}\ln\left[1-
\frac{\GL}{3}[a^2\Omega^2+3w^{\frac{4}{3}}\Omega^{\frac{2}{3}}(1-a\Omega)^{\frac{4}{3}}]\right] .
\label{eq:karkowskij}
\ee
Here $\GL$ is a parameter that takes values around $\GL_{\rm Kerr}=3$, which corresponds to the exact formula that
characterizes the motion of circular geodesics at the equatorial plane of a Kerr black hole with spin parameter
$a$ and mass $m=w^2$. For self-gravitating tori $w^2\neq m$ in general. In the Newtonian limit, Eq. (\ref{eq:karkowskij})
yields the Keplerian angular velocity $\Omega=w/(r\sin\GU)^{3/2}$. Using the rotation law Eq. (\ref{eq:karkowskij}), 
and the Shibata formulation, the authors constructed sequences of 
stationary and axisymmetric equilibria around nonrotating as well as spinning black holes. 
In Newtonian gravity von Zeipel's theorem \cite{vonZeipel1924} states that for a barotropic fluid the surfaces of 
constant $\Omega$ coincide with the surfaces of constant $\ell_N=\Omega \varpi^2$, where $\varpi$ the distance from 
the axis of rotation. This implies that $\ell_N=\ell_N(\varpi)$ or $\Omega=\Omega(\varpi)$, i.e. the angular 
velocity depends only on the distance from the axis of rotation 
(Poincar\'e-Wavre \cite{Tassoul-1978:theory-of-rotating-stars}). In general relativity Abramowicz \cite{Abramowicz1974} 
showed that for massless disks the surfaces of constant $\Omega$ have cylindrical topology, therefore they depend not 
only on the distance from the rotation axis but also on the distance from the equatorial plane of symmetry. As expected 
this is also true for self-gravitating tori \cite{Karkowski2018}.

Many properties of self-gravitating BHDs have been investigated by Dyba {\it et~al.~} \cite{Dyba2019} using 
the infrastructure developed in previous works. In particular the authors find that by fixing the black hole parameters
($a$ and $m$), 
the polytropic exponent $\Gamma$, and the inner and outer coordinate equatorial radii of the torus, as well as its maximal 
rest-mass density, there exist two solutions differing in the ADM mass. In other words, it is possible 
to obtain a sequence of tori with a decreasing maximum rest-mass density and increasing mass, simply because their size is 
also growing. Applying Seguin's \cite{Seguin1975} criterion for linear stability against axially symmetric 
perturbations the authors found that the massive branch must be dynamically unstable. The location of the ISCO
for a variety of self-gravitating equilibria showed non-negligible differences from the corresponding Kerr value, 
even for light toroids. A typical behavior is the increase of the circumferential radius of the ISCO relative to
that of the Kerr black hole. On the other hand Dyba {\it et~al.~} found that for sufficiently massive disks the effective 
potential $\mathcal{V}_{\rm eff}$ due to its nonmonotonic behavior further from the black hole, can exhibit a region outside 
the ISCO in which circular geodesics can be unstable ($\mathcal{V}^{''}_{\rm eff}(r)<0$). This provides yet another 
reason why very massive disks are dynamically unstable.


\subsection{Black hole-toroid with magnetic fields}
\label{sec:mbht}

Mach {\it et~al.~} \cite{Mach2019} constructed self-gravitating BHDs with toroidal magnetic fields in the general relativistic 
ideal magnetohydrodynamics (IMHD) regime \cite{Komissarov2006a,Shibata2007} under stationarity and axisymmetry.
The main difference here is the integral of the Euler equation, since for the gravitational field equations
the analysis follows previous works \cite{Shibata2007,Karkowski2017,Karkowski2018} (apart from the modification 
of the sources in the Einstein equations). Mach {\it et~al.~} assume that $T^{\GA\GB} = \TPFuab + \TEMuab$ 
with  
\be
    \TEMuab = b^2 (u^\GA u^\GB + \frac{1}{2}g^{\GA\GB})-b^\GA b^\GB ,
\label{eq:imhdset}
\ee
the IMHD stress-energy tensor, and $\TPFuab$  the perfect fluid stress-energy tensor Eq. (\ref{eq:pfset}). 
Here $b^\GA=B^\GA/\sqrt{4\pi}$ is the magnetic field measured by an observer with 4-velocity $u^\GA$ 
(at rest with respect to the fluid), and $b^2=b_\mu b^\mu$ the magnetic pressure $p_{\rm mag}=b^2/2$. Since
\be
b_\mu u^\mu = 0 \, ,
\label{eq:imhd4v}
\ee
assuming a toroidal magnetic field  with $b^r=b^\GU=0$ yields the following realation:
\be
b_t = -\Omega b_\GP .
\label{eq:Bu0}
\ee
In this set up, the continuity equation, the induction equation, as well as the $t$ and $\GP$ components of 
$\nabla_\GA T^{\GA\GB}=0$ are trivially satisfied, while the $r$ and $\GU$ components satisfy
\be
\na_\mu \ln\frac{h}{u^t} + u^t u_\GP \na_\mu\Omega + \frac{1}{2\GR h\mathcal{A}}\na_\mu (b^2 \mathcal{A}) = 0 ,
\label{eq:mei}
\ee
where $\mathcal{A}=g_{\GP\GP}g_{tt}-g_{t\GP}^2$. The above equation can be integrated if and only if
$u^t u_\GP$ is a function of $\Omega$, and $b^2 \mathcal{A}$ a function of $\GR h\mathcal{A}$. When the magnetic 
field is zero one recovers the integral of the Euler Eq. (\ref{eq:feom2}). The 
authors assumed $b^2|\mathcal{A}| = f(\GR h|\mathcal{A}|)$ with $f'(x)=2nC_1 x/(1+C_1x)$, so that 
\be
\int\frac{d(b^2\mathcal{A})}{2\GR h\mathcal{A}} = \ln(1-C_1\GR h\mathcal{A})^n \, .
\ee
Constants $n,\, C_1$ characterize the topology of the magnetic field and the authors construct a series of solutions
with variable $C_1$ while fixing $n=1$. They found that the larger the $C_1$, the smaller the thermal pressure, even 
in cases in which the maximum of the baryonic density is fixed. In general they also observe an increase of the value
of the magnetic pressure.
For the rotation law ($u^t u_\GP$ term) they employed Eq. (\ref{eq:karkowskij}) with $\GL=3$. 
Each solution is specified by the black hole parameters ($m$ and $a$), the inner $r_{\rm in}$ and outer $r_{\rm out}$
radii of the disk,
the polytropic exponent of the EOS, the maximum rest-mass density, and the magnetic field constants $C_1$ and $n$.
On the other hand, the constant $w$ that appears in the differential rotation law, Eq. (\ref{eq:karkowskij}), the 
constant that appears in the integral of the Euler Eq. (\ref{eq:mei}), and the two angular velocities at the inner
and outer points are determined during the iteration scheme. As in Dyba {\it et~al.~} \cite{Dyba2019} 
fixing $a,\, m,\, r_{\rm in},\, r_{\rm out},\, \Gamma,\, \GR_{\rm max}$ does not lead to unique solutions. Two solutions 
exist with different ADM masses, with one of them being larger than the mass of the central black hole.
The authors measure the 
magnetization of their solutions by the parameter $\GB_{\rm mag}=2p/b^2$ and they construct a large set of
solutions with $\GB_{\rm mag}$ as small as $\sim 6\times 10^{-4}$. The main characteristic of the toroidal 
magnetic field is to shift the location of the maximum rest-mass density towards the black hole.


\subsection{Arbitrary spinning black hole-toroid}
\label{sec:sbht}

In an effort to go beyond axisymmetry or stationarity Tsokaros {\it et~al.~} \cite{Tsokaros2018a} presented a 
new formalism for 3-dimensional self-gravitating BHD solutions. This method goes beyond the minimal construction of 
initial data (solution of constraints in binary black hole calculations) 
\cite{York71,York72,Pfeiffer:2002iy,Brandt97b,York99,Isenberg08,Wilson95,Marronetti99} and solves the {\it full} Einstein 
system. In particular the 5 equations related to the conformal geometry which are associated with the true dynamical 
degrees of freedom of the gravitational field \cite{York72} are resolved. The following suite of benchmarks have been 
used to assess the new formalism: 1) In the absence of matter it can reproduce the exact Kerr-Schild solution, even for
extreme spins, which makes the method suitable for a broad range of \textit{nonaxisymmetric} problems, such as tilted 
disks or binary systems. 2) The domain of the solution extends inside the apparent horizon, which is well-suited for 
evolution simulations. 3) In the presence of massless disks around the black hole our method reproduces well-known solutions 
(e.g. \cite{Chakrabarti1985,DeVilliers03b}) even with black hole tilt. 4) The first self-consistent, self-gravitating, and 
tilted BHD solutions are presented that satisfy not only the constraint equations but the \textit{whole} Einstein 
system.
 
The starting point of Tsokaros {\it et~al.~} \cite{Tsokaros2018a}  is a line element in its general 3+1 form 
(Eq. (\ref{eq:dstpo})) together with a conformal metric decomposition as in Eq. (\ref{eq:confmetric}). Instead of
a maximal slicing and the Dirac gauge Eq. (\ref{eq:dirac}), the coordinates now used were Kerr-Schild with 
\be
K = K_{\ks}=\frac{2H\GA^3_{\ks}}{r}\left(1+H+\frac{2H^2 r}{m}\right),  \label{eq:trk}
\ee
and
\be
\zD_i \TUU{\GG}{i}{j} = \zD_i h^{ij}_{\ks} .
\label{eq:fi}
\ee
Here $\GA_{\ks},\, h^{ij}_{\ks}$ are the lapse and non flat part of the conformal metric in Kerr-Schild coordinates
$ds^2=(\GH_{\mu\nu}+2H\ell_\mu\ell_\nu)dx^\mu dx^\nu$.
Assuming Kerr-Schild boundary conditions for the potentials $\GC,\,\GA,\,\tGB_i,\, h_{ij}$ the authors found 
that the corresponding system \cite{Shibata04a,Bonazzola:2003dm} could not converge in the neighborhood outside 
the horizon. The failure was due to the fact that the equations for $h_{ij}$ were losing their elliptic character 
close to the horizon. The region of nonconvergence was larger for higher spins but still relatively small (less than twice the 
horizon radius). In order to evercome this difficulty, and even more to obtain horizon penetrating solutions the authors
introduced a new decomposition for the traceless extrinsic curvature as
\be
\TDD{A}{i}{j} = \tilde{A}^{\ks}_{ij} + \tGS\TWD{i}{j}  ,  \label{eq:ctt}
\ee
where $\tilde{A}^{\ks}_{ij}$ is the Kerr-Schild part, $\tW_i$ an unknown spatial vector, and $\tGS$ a scalar. Here 
$\tilde{\mathbb L}$ is the conformal Killing operator:
$\TWD{i}{j}=\TD{D}{i}\TD{W}{j}+\TD{D}{j}\TD{W}{i}-\frac{2}{3}\TDD{\GG}{i}{j}\TD{D}{k}\TU{W}{k}$.
Equation (\ref{eq:ctt}) implies that the momentum constraint is now solved twice once
for the shift vector $\tGB_i$ and once for $\tW_i$.
Solving for $\GB^i$ is necessary since it will be used in the computation of $\mathcal{L}_{\GA \Bn}K$.
On the other hand decomposition (\ref{eq:ctt}) with zero boundary conditions for $\tW_i$ on the
horizon results into convergence for the $h_{ij}$ potentials even inside the horizon.
The Kerr-Schild gauge conditions Eq. (\ref{eq:fi}) are satisfied through a transformation of the form 
Eq. (\ref{eq:gauge_transf}) which introduces 3 more potentials $\xi^i$. The augmented system of the 17 elliptic 
equations with zero boundary conditions for the gauge potentials and the vector $\TD{W}{i}$ converges smoothly in 
vaccum or in the presence of matter (like a massive disk), even for near maximally-spinning black holes, thus not only 
solving for the constraint equations but providing also a way to control the gravitational wave content of the 
initial data in a self-consistent way.

For the Euler equations the authors assumed a stationary and axisymmetric fluid flow which in the presence of a tilted
black hole (with respect to the angular momentum of the disk) is a valid approximation if the inner point of the disk is 
further away from the black hole horizon. Using the KEH scheme for black holes \cite{Tsokaros2007}, as implemented within the \cocal
code \cite{Uryu2012} the authors computed self-gravitating tilted BHD solutions with a disk mass larger 
than the black hole mass and with the black hole having almost extremal spin.


\section{Magnetized rotating neutron stars}
\label{sec:mrns}

While the typical surface magnetic field of pulsars is $\sim 10^{12}-10^{13}\;\rm G$ \cite{lyne_graham-smith_2012}
there exist neutron stars, the so-called magnetars, with extremely large magnetic fields $\gtrsim 10^{14}\;\rm G$
\cite{Duncan1992,ThompsonDuncan1995,ThompsonDuncan1996}. The magnetar model
has been invoked to explain soft $\GG$-ray repeaters  (object that emits large 
bursts of $\GG$-rays and X-rays at irregular intervals) and the related anomalous X-ray pulsars
\cite{Mazets1979,Paczynski92,Duncan1992,ThompsonDuncan1995,ThompsonDuncan1996,Kouveliotou1994,Kouveliotou1998,
Kouveliotou1999,Mereghetti2008}. Soft $\GG$-ray repeater bursts are caused by the cracking of the neutron star crust 
due to magnetic stresses, which leads to injection of Alfv\'en waves into the magnetosphere, particle acceleration, 
and formation of an optically thick pair plasma. The decay of the magnetic field heats the neutron-star interior, 
and gives rise to persistent thermal soft X-ray emission from the surface. In addition, the strong magnetic field causes
the spindown of the neutron star which may end up with rotation periods of $\sim 10\;\rm s$.
From the microscopic point of view when the cyclotron energy equals the electron rest mass, one reaches the
quantum critical magnetic field strength of $\sim 10^{13}\;\rm G$ beyond which the magnetic field
affects physical processes and is responsible for many exotic phenomena, such as vacuum birefringence, photon splitting, 
and the distortion of atoms (see \cite{Harding2006} for an extensive review).

Magnetic fields beyond magnetar strength that can reach values of $\sim 10^{17}\;\rm G$ can be developed 
in the merger of two neutron stars due to a number of different mechanisms: (i) The Kelvin-Helmholtz instability 
\cite{Chandrasekhar81,Bodo1994} which occurs in the shear layer that forms between the two neutron stars and can grow on a timescale 
of a couple of milliseconds \cite{Price06,Anderson2008,Kiuchi2015a}.
(ii) The magnetorotational instability \cite{Velikhov1959,Chandrasekhar1960,Balbus1991}
during the merger as well as in the postmerger compact object \cite{Shibata:2005mz,Duez2006a,Siegel2013,Kiuchi2015}. 
(iii) Magnetic winding which is due to differential rotation  \cite{Baumgarte00bb,Shapiro00,Kiuchi2015a}. 
Differential rotation generates toroidal Alfv\'en waves which 
convert rotational kinetic energy into magnetic field energy. 
In fact all of the mechanisms above are responsible for converting poloidal to toroidal
magnetic fields leading to a remnant with comparable poloidal and toroidal components. 
Strong magnetic fields affect the neutron star in at least two ways. First, 
they result in an anisotropy through a modification of the energy-momentum tensor.
Second, they affect the EOS due to Landau quantization of the constituent
particles, as pointed out in Bandyopadhyay, Chakraborty \& Pal \cite{Bandyopadhyay1997}. Therefore one
expects that the EOS, and thus a number of observational quantities such as the neutron star maximum
mass, to be affected too.
The magnetar scenario has also been recently employed to explain the blue radioactive ejecta in the BNS merger
event GW170817 \cite{Metzger2018}. The authors proposed that the source for these ejecta was a magnetized
neutrino-irradiated wind, which emerges from the hypermassive neutron star remnant over $\approx 0.1-1\;\rm s$ prior 
to its collapse to a black hole. The role of strong magnetic fields was instrumental in explaining the high ejecta mass 
and the observed velocities. 

Therefore the ab initio calculation of self-gravitating magnetars 
can greatly facilitate the study of these objects and understanding of their gravitational and electromagnetic signatures.
On the other hand equilibrium solutions does not mean that they are necessarily stable. The first general
relativistic MHD simulations with either purely toroidal magnetic fields \cite{Kiuchi2008b} or purely poloidal magnetic fields
\cite{Ciolfi2011,Lasky2011,Ciolfi2012,Lasky2012} confirmed the unstable nature of these solutions predicted decades ago
\cite{Tayler_1957,Tayler1973,Wright1973,Markey1973,Markey1974,Flowers1977}.  In
\cite{Ciolfi2011,Lasky2011,Ciolfi2012,Lasky2012} the initial
conditions were based on the self-consistent poloidal solutions of
\cite{Bocquet1995}, and the Cowling approximation was used, while in \cite{Kiuchi2008b} the initial toroidal conditions were
those of \cite{Kiuchi2008} and an axisymmetric general relativistic MHD simulation was employed.

In this section, we review theoretical works on the structure of MRNS in the framework of general relativity.  
While for BNSs and BHDs we focused only on numerical methods that model the whole or part of the Einstein-Euler system
(but still without truncating any equation),
for MRNS we will mention pertubative methods as well. Perturbative modelling of MRNSs is important because in
many astrophysically realistic scenarios the contribution of the magnetic field to the equilibrium of a compact 
star may be small enough to be treated as a perturbation.
In Table \ref{tab:MRNS}, we present a classification of mathematical 
modelings for relativistic magnetized compact stars based on three characteristics. 
In most of the cases stationarity and axisymmetry are assumed, except for a few models whose 
magnetic axis is tilted with respect to the rotation axis.

\renewcommand{\arraystretch}{1}
\begin{table}[h]
\begin{tabular}{lll}
\hline \\[-3mm]
Theory  & Types & Properties \\
\hline \\[-3mm]
\multirow{4}{*}{Flow fields} & no flows  & \\[-2mm]
& 
\multirow{4}{*}
{$\left. \!\!\!\!\! 
\begin{tabular}{l} 
toroidal (rotation) \\
poloidal (meridional circulation) \\
mixed  \\
\end{tabular} \!\!\!\!\right\} $} 
& 
\multirow{4}{*}
{$\left\{ \!\!\!\! 
\begin{tabular}{l} 
slow rotation (perturbation) \\ 
\ with $\Omega=$ const. or $\Omega=\Omega(r)$ \\ 
rapid rotation (numerical) \\ 
\ with $\Omega=$ const. or differential \\
\end{tabular} \right. $} \\
& & \\
& & \\  
& & \\ 
\\[-3mm]
\hline
\\[-3mm]
\multirow{4}{*}
{\!\!\!\!\!\! \begin{tabular}{l} 
EM field \\ 
\ components \\
\end{tabular}} 
& no fields &   \\ 
& 
\multirow{3}{*}
{$\left. \!\!\!\!\! 
\begin{tabular}{l} 
poloidal \\
toroidal \\
mixed    \\
\end{tabular} \!\!\!\!\right\} $}
&
\multirow{3}{*}
{$\left\{ \!\!\!\!\! 
\begin{tabular}{l} 
weak (perturbation)\\ 
strong (numerical) \\
\end{tabular} \right.$} \\
&  &  \\
&  &  \\
\\[-3mm]
\hline
\\[-3mm]
\multirow{3}{*}
{\!\!\!\!\!\! \begin{tabular}{l} 
EM field \\
\ configurations \\
\end{tabular}} 
& 
\multirow{3}{*}
{$\left. \!\!\!\!\! 
\begin{tabular}{l} 
confined, outside vaccum \\
extended, outside magnetovacuum \\
extended, outside magnetosphere \\
\end{tabular} \!\!\!\!\right\} $}
&  \\
&  &  arbitrary functions for the currents \\
&  &  \\
\\[-3mm]
\hline
\\[-3mm]
       & fixed spherical background & \\
Metric &
\multirow{2}{*}
{$\left. \!\!\!\!\! 
\begin{tabular}{l} 
truncated \\
total \\
\end{tabular} \!\!\!\!\right\} $}
& 
\multirow{2}{*}
{$\left\{ \!\!\!\!\! 
\begin{tabular}{l} 
small deformation (perturbation)\\ 
large deformation (numerical) \\
\end{tabular} \right.$} \\
&  &  \\
\\[-3mm]
\hline
\\[-3mm]
Stress energy tensor 
& no fields &   \\ 
& 
\multirow{3}{*}
{$\left. \!\!\!\!\! 
\begin{tabular}{l} 
poloidal \\
toroidal \\
mixed    \\
\end{tabular} \!\!\!\!\right\} $}
&
\multirow{3}{*}
{$\left\{ \!\!\!\!\! 
\begin{tabular}{l} 
weak (perturbation)\\ 
strong (numerical) \\
\end{tabular} \right.$} \\
&  &  \\
&  &  \\
\\[-3mm]
\hline
\\[-3mm]
                      & \multirow{3}{*}{$\left. \!\!\!\!\! \begin{tabular}{l} 
                                              perfect fluid (PF) \\ PF + EM \\ PF + EM + magnetization \end{tabular} \!\!\!\!\right\} $} 
                      & \multirow{3}{*}{ \!\!\!\!\! \begin{tabular}{l}  EOS (polytropic, realistic,  \\   magnetized, hypothetical) \end{tabular}  } \\
Stress energy tensor  &     &   \\         
                      &     &   \\
\hline
\end{tabular}
\caption{
A summary of mathematical modelings for magnetized compact stars. EM stands for electromagnetism. 
Brackets signify that any line in the second column can be combined with any line 
in the third column, since most models in the literature are computed using one
of the types with one of the properties.
}
\label{tab:MRNS}
\end{table}
\renewcommand{\arraystretch}{1}

\subsection{Magnetized rotating neutron stars with purely poloidal magnetic fields}
\label{sec:poloidal}

In their pioneering work Bocquet, {\it et~al.} \cite{Bocquet1995} have constructed
stationary and axisymmetric rotating neutron stars \cite{Bonazzola1993} 
(spacetime metric Eq. (\ref{eq:shibatads})) having circular flows, and electric currents that
induce a poloidal magnetic field. This work is 
not only the first self-consistent numerical construction (made with a highly precise
spectral code) of relativistic rotating stars associated with strong 
poloidal (electro)magnetic fields (where the exterior region of the star 
is a magnetovacuum/electrovacuum), but it is also achieved earlier than any 
perturbative study in the framework of general relativity.  

Carter has shown \cite{Carter73_Stationary_BH} that
conditions (\ref{eq:intcon}) imposed on an electromagnetic energy-momentum tensor
\be
\TEMuab\,=\, \frac1{4\pi}\left(F^{\alpha\gamma}F^\beta{}_\gamma-\frac14\gabu\Fcdd\Fcdu\right) ,
\label{eq:emset}
\ee
will imply that
\be
j^{[\GA} t^\GB \GP^{\GG]} = 0,
\label{eq:cpf1}
\ee
i.e. a circular (toroidal) current $j^\GA=(j^t,0,0,j^\GP)$ similar to the 4-velocity (\ref{eq:c4vel}). 
Conversely, if the fluid circularity condition (\ref{eq:cpf}), and the current circularity 
condition (\ref{eq:cpf1}) are satisfied on a certain connected domain, then 
the metric circularity condition (\ref{eq:intcon}), and the electromagnetic field circularity 
condition 
\beq
F_\albe t^\alpha \phi^\beta= 0, \quad F_{[\albe} t_\gamma \phi_{\delta ]} = 0
\eeq
are satisfied on the same domain (generalized Papapetrou theorem).
As a corollary, the electomagnetic potential circularity condition 
\beq
A_{[\GA} t_\GB \GP_{\GG]} = 0,
\eeq
is also satisfied.  
Here $A_\alpha$ is the electromagnetic 1-form that derives the Faraday tensor 
$F_{\GA\GB}=\pd_\GA A_\GB - \pd_\GB A_\GA$.

In this setup magnetic field lines lie on surfaces $A_\GP={\rm const}$, while the electric and magnetic field 
as seen by the normal observers will have only $r,\,\GU$ components
$E_\GA=F_{\GA\GB} n^\GB =(0,E_r,E_\GU,0)$, and $B_\GA=\frac{1}{2}\GE_{\GA\GK\mu\nu}n^\GK F^{\mu\nu}=(0,B_r,B_\GU,0) $ 

Using a stationary and axisymmetric metric, Eq. (\ref{eq:shibatads}), and the 3+1 formulation of the Einstein equations
one can calculate the gravitational potentials $\GA,\,\GC,\,\GB,\,q$ from 4 elliptic equations \cite{Bonazzola1993,Shibata2007}. 
Regarding the nonrotating (static) solutions we note here that when the magnetic field is zero 
we have $\GB=0$ i.e. the timelike Killing vector $t^\GA$ is orthogonal to the
spatial hypersurfaces. In the case of poloidal magnetic fields a static spacetime implies not only a
non-rotating fluid, but also a vanishing electric charge. A nonzero charge creates an electrostatic 
field outside the star and therefore a nonzero Poynting vector that leads to nonzero angular momentum
although the fluid does not rotate.

Bocquet {\it et~al.~} assume a stress-energy tensor composed of a perfect fluid and an electromagnetic field 
$\Tabu \,=\, \TPFuab \,+\, \TEMuab$. The source-free Maxwell equations 
\be
\nabla_{[\GG}F_{\GA\GB]}=0
\label{eq:sfME}
\ee
are identically satisfied, while the ones with nonzero sources
\be
\nabla_\GB F^{\GA\GB}=4\pi j^\GA
\label{eq:sME}
\ee
result into two elliptic equations for $A_t$ and $A_\GP$
\be
 \zD_i\zD^i A_t = P(\cdots; A_t, A_\GP, j^t, j^\GP),\quad\mbox{and}\quad 
 \zD_i\zD^i A_\GP = Q(\cdots; A_t, A_\GP, j^t, j^\GP) \ .
\label{eq:atap}
\ee
The dots in the right-hand side of the Eqs. (\ref{eq:atap}) signify the nonlinear
dependence on the gravitational metric components. On the other hand Eqs. (\ref{eq:atap})
depend linearly on $j^t$ and $j^\GP$.
In order to satisfy the magnetohydrostatic equilibrium, the projection of the conservation 
of the stress-energy tensor implies
\be
\pd_i \ln\frac{h\GA}{\Gamma_L} -  \frac{j^\GP-\Omega j^t}{\GE+ p} \pd_i A_\GP = 0\ ,
\label{eq:fistint}
\ee
where $\Gamma_L$ is the Lorentz factor connecting the normal and comoving observers.
Equation (\ref{eq:fistint}) is analogous to Eq. (\ref{eq:feom}) and (\ref{eq:mei}). Integrability demands 
\be
j^\GP-\Omega j^t = (\GE+ p) f(A_\GP)
\label{eq:jtjp}
\ee
where $f$ an arbitrary function which the authors refer to as the ``current function'' since
it relates to a current associated with the electromagnetic potential $A_\GP$. Different choices
for $f$ lead to different magnetic field distributions. In addition from Ohm's law and assuming
that matter has infinite conductivity, the electric field as measured by the fluid comoving
observer must be zero,
\be
F_{\GA\GB}u^\GB = 0  \, .
\label{eq:infcon}
\ee
This implies that inside the star 
$\pd_i A_t + \Omega\pd_i A_\GP = 0$,
and for the rigidly rotating case, $\Omega =$ constant, we have
\be
A_t + \Omega A_\GP = C .
\label{eq:Arigrot}
\ee
Here $C$ is a constant that determines the total electric charge of the star.
Equation (\ref{eq:Arigrot}) is called the perfect conductivity equation.

Given an EOS and a current function $f$, a solution is obtained if one specifies the 
central enthalpy, and a value for the total charge. The first integral of Eq. (\ref{eq:fistint}) that
describes the magnetohydrostatic equilibrium can be used to find the angular velocity and the corresponding
constant of integration.
Given the gravitational potentials, $\GA,\,\GC,\,\GB,\,q$, 
one can use the four equations (\ref{eq:atap}), (\ref{eq:jtjp}), and (\ref{eq:Arigrot}) to solve
for $A_t,A_\GP,j^t$, and $j^\GP$. The iteration is based on the fact that the $\zD_i\zD^i A_t$ equation
is linear in $j^t$ and together with Eq. (\ref{eq:jtjp}) they are used to compute the new values of the 
currents $j^\GP$, $j^t$. Then $A_t$, and $A_\GP$ are obtained from the  $\zD_i\zD^i A_\GP$ equation
and the perfect conductivity Eq. (\ref{eq:Arigrot}).
The equation of $A_\GP$ is the easiest since imposing $A_\GP=0$ as $r\rightarrow\infty$ one finds 
a smooth solution everywhere. On the other hand, because of the assumption that the exterior 
region of the neutron star is vacuum, $A_t$ is not differentiable across the surface of the 
star. In order to make $A_t$ continuous at the stellar surface, 
a rotating perfect conductor is endowed with a surface charge density (hence the component of 
the electric field normal to the surface is discontinuous). The solution for $A_t$ proceeds in two steps.
Outside the star a value $A_t^{(1)}$ is obtained by assuming $A_t=0$ as $r\rightarrow\infty$. This 
solution in general does not agree on the surface of the star with the interior solution 
$A_t^{(0)}=-\Omega A_\GP$ obtained previously from the perfect conductivity equation. Thus a harmonic
($\zD_i\zD^i A_t^{(2)}=0$) function 
\be
A_t^{(2)} = \sum_{\ell=0}^{L} a_\ell \frac{P_\ell(\cos\GU)}{r^{\ell+1}}
\label{eq:harmonicA}
\ee
is added to $A_t^{(1)}$ in order for the exterior field, $A_t^{(1)}+A_t^{(2)}$, to satisfy the appropriate boundary conditions
on the star surface. At the end, the exterior solution $A_t^{(1)}+A_t^{(2)}$ matches the interior one
$A_t^{(0)}$ at the surface of the neutron star, and therefore a continuous component of $A_t$ is obtained in the 
whole domain. This solution though has a certain electric charge (as measured in the asymptotics) that
does not coincide with the desired one. If in addition one wants to fix the charge of the solution
then a further adjustment of the arbitrary constant $C$ in Eq. (\ref{eq:Arigrot}) is needed. 

The authors have calculated non-rotating as well as rotating magnetized models for different EOSs and 
with a zero total charge (which is a free parameter in the formulation). Sequences of solutions are 
parametrized either by their rest mass or magnetic dipole moment $\mathcal{M}$, which is measured 
from the leading term of the asymptotic behavior of the magnetic field as measured by the normal observer.
For the static solutions the authors employed a constant current function $f(x)=\rm const$
and found models with a magnetic field at the pole as large as $B_{\rm pole}=1.5\times 10^{18}\ \rm G$,
and a ratio of magnetic to gas pressure at the center of $\sim 1.0$. 
The gravitational mass of the magnetized solutions was an increasing function of $\mathcal{M}$, reaching 
differences $\lesssim 29\%$ from the corresponding nonmagnetized ones (at very large central magnetic 
fields $\sim 10^{18}\ \rm G$). The authors found that this increse in the maximum mass was EOS dependent, 
with some EOSs leading to more modest enhancements ($\sim 13\%$) than others.
Given the fact that rotation increases the maximum gravitational mass by at most $\sim 20\%$ 
\cite{Cook94c,Lasota1996,Breu2016}, Bocquet {\it et~al.~} showed explicitly for the first time that 
depending on the EOS the magnetic field can be more efficient in increasing the maximum nonrotating 
gravitational mass than rotation itself.
 
For the magnetized rotating configurations the authors experimented with different current functions
$f(x)$ such as
\be
f_1(x) = \frac{N}{1+x},\qquad\mbox{or}\qquad f_2(x)=N\left(1-\frac{1}{1+(\Lambda x)^2}\right) ,
\label{eq:curr}
\ee
where $x=A_\GP/A_\GP^0$, and $A_\GP^0$, $N$, and $\Lambda$ constants. 
For example choice $f_1(x)$ led to a current distribution more concentrated towards the star's center, 
relative to the choice $f(x)=\rm const$.
The authors compared the MRNS having a magnetic 
dipole moment ${\mathcal M} = 1.5\times 10^{32} \mbox{A m}^2$, and the non-magnetized 
RNS of the same central enthalpy and angular velocity.
They found that a magnetic field increases the baryonic mass for a fixed central enthalpy and 
angular velocity. In other words magnetic forces act in a centrifugal manner that help the star support
more baryons. At the same time the total angular momentum of the system increases linearly with 
respect to the angular velocity, reaching values $\sim 14\%$ larger than the zero magnetic field ones. 
In addition the Keplerian angular velocity at the mass shedding limit, $\omekep$, 
also increases with $\mathcal{M}$.
On the other hand, the equatorial circumferential radius shows an increase for small angular velocities (as it does for
the static cases where Lorentz forces stretch the star out), while for larger angular velocities the 
radius decreases. 
Bocquet {\it et~al.~} argue that this behavior can be explained in the following way. In order for 
equilibrium to be maintained the gravitational force  must counterbalance both the centrifugal
force which is greater at the periphery of the star, and the Lorentz force which is greater at the
center of the star. For the same central density and large rotation rate, an energetically favorable 
configuration happens at a smaller radius contrary to the slowler rotating case.

The overall conclusion of this study was that the magnetic field influences the star structure mostly
through the Lorentz forces and not through the gravitational field generated by the electromagnetic stress-energy tensor,
something which is anticipated since even a huge magnetic field of $\sim 10^{18}\,\rm G$ has energy density
$\sim 0.25\GR_{\rm nuc}c^2$, much smaller than the matter density at the neutron star center. Notwithstanding that,
for such high values of the magnetic field, the deformation of the star can be dramatic. This is due to the anisotropic character 
of the magnetic pressure similar to the way anisotropic centrifugal forces deform a rotating star even though 
its kinetic energy is much smaller than its gravitational one. On the other hand the
deformation of the star due to the magnetic stresses is important if $B\gtrsim 10^{15}\,\rm G$.


A more in depth analysis of magnetized static poloidal solutions ($\GB=A_t=J^t=0$) by Cardall, {\it et~al.~} \cite{Cardall2001}
employing more EOSs, and using a constant current function, found that the maximum mass of these configurations is noticeably larger 
than the maximum mass attained by uniform rotation for all EOS examined, and even larger than that reported in
Bocquet, {\it et~al.}. For a fixed number of baryons, 
maximum mass configurations are characterized by an off-center density maximum. In their study they used
the KEH method in a compactified domain \cite{Komatsu89,Cook94b}, and constructed a large number of sequences
of constant rest mass and constant magnetic dipole moment $\mathcal{M}$ as in \cite{Bocquet1995}\footnote{Even though
there is no principle of conservation of magnetic moment these sequences are expected to be astrophysically relevant,
at least in a certain timeframe, since the timescale of magnetic field decay is expected to be large.}.


A method for generating exact static and axisymmetric interior solutions 
with a pure poloidal magnetic field has been developed by Yazadjiev \cite{Yazadjiev2011}. The author employs
an anisotropic stress energy tensor and finds solutions that are prolate in shape.

\subsection{Perturbative models for magnetized rotating neutron stars}

Perturbative techniques in the calculation of general relativistic MRNSs have been used first by
Konno, {\it et~al.} \cite{Konno1999} in order to calculate their deformation due to magnetic stresses.
They found that for nonrotating neutron stars the ellipticity ($(R_e-R_p)/R_m$, where $R_p$, $R_e$, and $R_m$ are 
the polar, equatorial and mean radius) for the dipole magnetic field becomes large as the compactness $M/R$ 
increases, for the same energy ratio magnetic energy to gravitational energy. In a subsequent 
work \cite{Konno2001}, Konno considered slow rotation of magnetically deformed stars on the symmetry
axis in order to define the moment of inertia and found that each principal moment is modified by a factor 
of 2 at most due to the general relativistic effects.

In an effort to compute equilibrium models that incorporate both poloidal and toroidal magnetic fields 
Ioka and Sasaki \cite{Ioka2003} presented a formalism for IMHD in stationary and axisymmetric spacetimes
that goes beyond circular flows. In this way they extended previous works on the 
Grad-Shafranov equation \cite{Lovelace1986,Mobarry1986,Nitta1991,Beskin1997} to noncircular spacetimes.
Their starting point was the work by
Bekenstein \& Oron \cite{Bekenstein1978,Bekenstein1979} who have shown the existence of 5 conserved quantities 
along a flow line. Given the ideal MHD condition $E_\GA=F_{\GA\GB}u^\GB=0$, and the assumptions of stationarity and
axisymmetry, one finds that the magnetic potential $\Psi:=A_\mu \GP^\mu = A_\GP$ and the electric potential
$\Phi:=A_\mu t^\mu = A_t$ are constant along each flow line, i.e. 
\be
u^\mu \nabla_\mu\Psi = u^\mu\nabla_\mu\Phi = 0 \, .
\label{eq:ebpot}
\ee
As a consequence, one can label each flow line by $\Psi$, which is called the flux function. Surfaces with $\Psi=\rm const$
are called flux surfaces and are generated by rotating the magnetic field lines or equivalently
the flow lines, about the axis of symmetry. It also implies that the electric and magnetic potentials are 
dependent, i.e. $\Phi=\Phi(\Psi)$. Introducing the function $\bar{\Omega}=-d\Phi/d\Psi$, and since
\be
E_A = (\bar{\Omega}-\Omega)u^t\pd_A\Psi + F_{AB} u^B = 0
\label{eq:EA}
\ee
one finds that $\bar{\Omega}$ coincides with the angular velocity only when toroidal fields are absent, i.e. $F_{12}=0$.
Using the continuity equation one can write the components of the Faraday tensor
\begin{align}
F_{tA} & = \bar{\Omega} F_{A\GP},  &  F_{t\GP} & = 0,  \label{eq:is1} \\
F_{12} & = C\sqrt{-g} \GR (u^\GP-\bar{\Omega} u^t),  & F_{1\GP} & = C\sqrt{-g}\GR u^2  = \pd_1 \Psi,   \label{eq:is2}  \\
F_{2\GP} & = C\sqrt{-g}\GR u^1 = \pd_2 \Psi,  &  &     \label{eq:is3} 
\end{align}
where $\bar{\Omega}=\bar{\Omega}(\Psi)$ and $C=C(\Psi)$ are conserved along each flow line. 
The magnetic field can be written in terms of the 4-velocity as
\be
B^\mu = -C\GR [(u_t+\bar{\Omega} u_\GP)u^\mu + t^\mu + \bar{\Omega} \GP^\mu] .
\label{eq:Bsa}
\ee
Similarly to the conserved quantities $C$ and $\bar{\Omega}$, it can be shown that
\begin{eqnarray}
E & = & -\left(h+\frac{b^2}{\GR}\right)u_t - C(u_t+\bar{\Omega} u_\GP)b_t \, ,   \label{eq:Econ}\\
L & = & \left(h+\frac{b^2}{\GR}\right)u_\GP + C(u_t+\bar{\Omega} u_\GP)b_\GP \, ,    \label{eq:Lcon}\\
D & = & -h(u_t + \bar{\Omega} u_\GP) \, ,   \label{eq:Dcon}
\end{eqnarray} 
where $h$ the specific enthalpy, are all conserved along the flow lines. Note here that not all of the quantities 
are independent and it is $D=E-\bar{\Omega} L$. Since they are essentially the first
integrals of motion, their specification characterizes the configuration of the electromagnetic field and the fluid flow.

In order to describe noncircular, stationary, and axisymmetric spacetime the authors use the (2+1)+1 formalism 
by Gourgoulhon \& Bonazzola \cite{Gourgoulhon1993a} for the 10 metric components $g_{\mu\nu}$. 
Under these assumptions the Euler equations reduce to the Grad-Shafranov equation for the flux function $\Psi$
\begin{eqnarray}
&& J^\GP -\bar{\Omega} J^t + \frac{1}{C\sqrt{-g}}[\pd_2(hu_1)-\pd_1(hu_2)]  + \GR T\frac{ds}{d\Psi}
\qquad\qquad\qquad\nonumber \\ 
&&\qquad\qquad\qquad - \GR u^0\left[\frac{dE}{d\Psi}-\Lambda\frac{d(C\bar{\Omega})}{d\Psi}\right]  + 
   \GR u^\GP\left[\frac{dL}{d\Psi}-\Lambda\frac{dC}{d\Psi}\right] = 0 \, ,
\label{eq:isGS}
\end{eqnarray}
where $\Lambda=(u_t B_\GP - u_\GP B_t)/4\pi$ and
\be
J^\GS= \frac{1}{4\pi\GA \GZ} \mathcal{D}_A (\GA \GZ F^{\GS A}),\qquad \mbox{for}\qquad \GS=\{t,\GP\} .
\label{eq:GSj}
\ee
In Eq. (\ref{eq:isGS})  $s$ is the specific entropy, while in Eq. (\ref{eq:GSj}), $\GZ$ is the ``lapse''
function that defines the unit spacelike 4-vector $m^\GA$ orthogonal to the $t=\rm const$ and $\GP=\rm const$ 
surfaces $\Sigma_{t\GP}$. It is  $m_\GA = \GZ \mathcal{D}_\GA \GP$, with $\mathcal{D}$ being the covariant derivative with respect to the 
induced 2-metric on $\Sigma_{t\GP}$.
The Grad-Shafranov equation is a second order nonlinear partial differential equation for $\Psi$ due to the first 3 terms
($J^\GP$, $J^t$, and the term in square brackets) which include first order derivatives of $\Psi$.
In addition to Eq. (\ref{eq:isGS}) one has the so-called wind equation from the normalization of the 4-velocity
that can be used to compute one of the thermodynamic variables. Given the conserved functions 
$E(\Psi)$, $L(\Psi)$, $\bar{\Omega}(\Psi)$, $C(\Psi)$, $s(\Psi)$, and the metric $g_{\mu\nu}$, Eq.
(\ref{eq:isGS}) together with the wind equation describe fully the magnetohydrostatic equilibrium. 

The authors proceed in a perturbative way to compute models of magnetized neutron stars \cite{Ioka04}
with mixed poloidal and toroidal 
internal magnetic fields where meridional circulation is present. They employed a polytropic EOS with $\Gamma=2$ 
and considered the Grad-Shafranov equation
in the weak magnetic field limit, with the flux function $\Psi$ being the perturbation parameter, similar in spirit
to the slow rotation limit of Hartle \& Thorne \cite{Hartle68}. For the metric they solved the perturbation
equations for $\Delta g_{\mu\nu}$ around a spherically symmetric metric, which turn out to be $O(\Psi^2)$. 
Models with toroidal magnetic fields were found
to distort prolately contrary to the oblate distortion in the pure poloidal case \cite{Bocquet1995}. 
For fixed baryonic mass and magnetic helicity \cite{Carter1979_conservation_laws,Bekenstein1987}
\be
\mathbb{H} = \int_{\Sigma_t}H^\GA dS_\GA ,\qquad \mbox{where}\qquad 
H^\GA=\frac{1}{2}\GE^{\GA\GB\GG\GD}A_\GB F_{\GG\GD},
\label{eq:hel}
\ee 
more spherical stars were found to have lower energy. In addition, the authors report on two new types of frame
dragging that differ from the familiar one in Kerr black holes. These effects that violated reflection symmetry with respect
to the equatorial plane, were due to the meridional flow and the toroidal magnetic field. 


One instability that has been argued to be operating in MRNSs is the Parker instability,
or the so-called magnetic buoyancy instability \cite{Parker1966}, according to which magnetic flux
tubes are subject to magnetic buoyancy and are forced to move toward the surface, destabilizing
the star. Many authors have argued that stable stratification is necessary for magnetized equilibria
to be stable \cite{Kiuchi2011,Reisenegger2008}. For sufficiently cold neutron stars, the proton-neutron 
composition gradient is a candidate for such stratification \cite{Reisenegger1992}.
Stable stratified neutron stars in general relativity have been computed by Yoshida, {\it et~al.} \cite{Yoshida2012} as well
as by Yoshida \cite{Yoshida2019}, although in their models buoyancy results from entropy gradients, and
not composition ones.
From the conservation of the stress energy tensor Eq. (\ref{eq:coset}), and the conservation of baryon mass, 
it is
\be
u^\GA \nabla_\GA s = 0 .
\label{eq:conentro}
\ee
If we restrict to stationary and axisymmetric systems Eq. (\ref{eq:conentro}) yields $u^A \nabla_A s = 0 $. 
In other words the specific entropy has to be constant along the streamlines on the stellar meridional plane, 
unless $u^A=0$, i.e. meridional flows do not exist. The authors argue that the specific entropy cannot be
constant along the streamlines in a meridional plane since in general these are closed curves for stationary 
and axisymmetric systems, thus it is inevitable for regions with $\nabla^\GA p \nabla_\GA s <0$ to exist.
Therefore for a stably stratified star one has
\be
 u^A = 0
\label{eq:merflow0} 
\ee
i.e. circular flows $u^\GA=u^t(t^\GA+\Omega\GP^\GA)$. In this case Maxwell equations, the perfect conductivity
equation $F_{\GA\GB}u^\GA=0$, and the projection $(g_{\GA\GG}+u_\GA u_\GG)\nabla_\GB \TPFuab=0$ yield
a set of equations similar to 
Eqs. (\ref{eq:is1}), (\ref{eq:is2}), (\ref{eq:is3}) and the Grad-Shafranov Eq. (\ref{eq:isGS}) with $\Omega=\bar{\Omega}$.
The arbitrary function of $\Psi$ that enter the equations are specified as in Ioka \& Sasaki \cite{Ioka2003}.
For the EOS the authors assume a parametric representation of the type
\be
p = k \GR^{1+1/n},\qquad\mbox{and}\qquad e = \frac{1}{\Gamma-1}\frac{p}{\GR},
\label{eq:ykseos}
\ee
where $k$, and $n$ are the polytropic constant and index respectively, and $\Gamma$ the adiabatic index,
which in general is not equal to $1+1/n$. For stars whose density profile satisfies $d\GR/dr<0$, the
condition for stable statification is 
\be
\Gamma > 1+ \frac{1}{n} \, .
\label{eq:yksgamma}
\ee 
The authors assumed $\Gamma=2.1$ (along with $\Gamma=2$ for comparison purposes), and using perturbative methods 
they calculated models that have both poloidal
and toroidal magnetic fields of comparable strength.
Building on this work, Yoshida \cite{Yoshida2012,Yoshida2019} was able to calculate new equilibria with 
magnetic fields whose toroidal components are much larger than the poloidal ones.

In the above works by Konno, {\it et~al.} \cite{Konno1999,Konno2001}, 
Ioka \& Sasaki \cite{Ioka2003,Ioka04}, and Yoshida, {\it et~al.} \cite{Yoshida2012,Yoshida2019}, 
the authors have taken into account not only the magnetic fields, but also the metric perturbations.
In this way they were able to compute the deformation of the neutron star due to the magnetic field 
and rotation. On the other hand, some of the studies below do not incorporate metric perturbations, 
but introduce a larger variety of magnetic field configurations.  
Another common feature between the perturbed models of Ioka \& Sasaki and Yoshida, {\it et~al.} is 
the assumption that both toroidal and poloidal magnetic field components are entirely 
confined inside the neutron stars, while outside it is vacuum without any electromagnetic fields.



Colaiuda, {\it et~al.} \cite{Colaiuda2008}
extended the work of Konno, {\it et~al.} \cite{Konno1999} to include toroidal magnetic fields with an amplitude
comparable to that of the poloidal fields. The authors pay special attention to the boundary conditions
and the matching of the interior solution of the Grad-Shafranov equation to the exterior solution of Maxwell's equation 
(magnetovacuum solution).
Because of their choice in the integrability condition, the toroidal field 
has a non-zero value at the stellar surface.  Since the toroidal field is not allowed 
under the assumption of magnetovacuum (no electric current) outside the neutron star, the authors implicitly 
introduced a surface electric current, and switched off the toroidal field outside.  They calculated various 
magnetic field configurations, including those distributed in the entire interior of the neutron star, as well 
as those localized in the crust. For the latter case, the surface deformation is much larger than in the former. 
They also calculated the stellar deformation due to the magnetic field and rotation, changing the mass, the EOS, 
and the magnetic field configuration.

Twisted torus configurations where the toroidal magnetic field component
is confined inside the neutron star, while the poloidal component extends to the exterior have been presented by
Ciolfi, {\it et~al.} \cite{Ciolfi2009} who built their equilibria based on the formalism
developed by Konno \cite{Konno1999}, Ioka \& Sasaki \cite{Ioka04} and Colaiuda, {\it et~al.} \cite{Colaiuda2008}. 
In their work the Grad-Shafranov equation is solved to first order in the magnetic field. 
In the solutions presented although the magnitude of the toroidal 
magnetic field is of the same order as the poloidal one, the contribution of the toroidal energy to the 
total magnetic energy is $\lesssim 10\%$, because the toroidal field is non-vanishing only in a small
region inside the neutron star.  
Unlike in Ioka \& Sasaki \cite{Ioka04}, the poloidal field smoothly extends outside 
of the star, and unlike in Colaiuda, {\it et~al.} \cite{Colaiuda2008} 
the toroidal field is smoothly confined inside the neutron star, and therefore consistent with 
the magnetovacuum exterior (no magnetosphere).
These equilibria have been built under the assumptions of a linear relation in the 
flux function between the poloidal and the toroidal components of the magnetic field, with their ratio being estimated 
by determining the configuration of minimal energy at fixed magnetic helicity. The contribution of higher
than $\ell=1$ multipoles was taken to be minimum outside the star. The latter assumption was removed in
Ciolfi, {\it et~al.} \cite{Ciolfi2010}, in addition with a more general parametrization of the relation 
between the toroidal and poloidal fields. The new configurations had a much smaller poloidal field near the 
symmetry axis, and a larger toroidal field near the stellar surface, so that the toroidal field energy never 
surpassing $\sim 13\%$ of the total magnetic energy inside the neutron star. A toroidal magnetic field that 
contains much less energy from the poloidal one can be problematic for a number of reasons. First, 
simulations indicate that poloidal-field-dominated geometries are unstable on Alfv\'en time-scales
\cite{Braithwaite2009,Ciolfi2011,Lasky2011,Lander2012,Ciolfi2012,Tsokaros2021}.
Second, MHD simulations of core-collapse supernovae show that the toroidal magnetic fields can be 
efficiently amplified due to the winding if the core rotates differentially. After core bounce, 
the toroidal fields generically dominate over the poloidal ones even if there is no toroidal
field initially (see \cite{Kotake_2006} for an extensive review). A similar mechanism is responsible for 
the creation of a large toroidal magnetic field in BNS mergers.
In order to address these limitations Ciolfi \& Rezzolla \cite{Ciolfi2013} adopted a new prescription for
the azimuthal currents that led to more generic twisted-torus configurations, where the
toroidal-to-total magnetic field energy ratio can be as high as $90\%$. 
The authors found that for a fixed exterior magnetic field strength, a higher relative content of
toroidal field energy implies a much higher total magnetic energy in the star, which can have a strong 
impact on the expected electromagnetic and gravitational wave emission properties of magnetars.

Finally, a series of works \cite{Rezzolla2001,Rezzolla2004,Rezzolla2001_err,Abdikamalov2009,Morozova2010,
Rezzolla2016a,Turimov2018,Turimov2021} focused mainly on the electromagnetic configuration outside of a neutron star.  
In the last work \cite{Turimov2021}, the authors reformulated a perturbative method for slowly rotating models of MRNS, 
and studied both interior and exterior magnetic configurations.  In particular, they demonstrated the exterior 
electromagnetic wave solutions of MRNS.  


\subsection{Magnetized rotating neutron stars with purely toroidal magnetic fields}
\label{sec:tormf}

As we discussed in the previous section toroidal magnetic fields appear naturally both in BNS mergers
as well as in core-collapse supernova. Therefore it is important to study
neutron stars with significant toroidal magnetic fields. Demanding a circular flow for a 
generic electromagnetic tensor leads to circular currents (Eq. (\ref{eq:cpf1})) that can sustain
only poloidal fields. Oron \cite{Oron2002} realized that if one restricts to an
IMHD stress-energy tensor $\TEMuab$, Eq. (\ref{eq:imhdset}),
then conditions (\ref{eq:intcon}) imply 
\be
B_t B^{[\GA} t^\GB \GP^{\GG]} = 0, \qquad\mbox{and}\qquad B_\GP B^{[\GA} t^\GB \GP^{\GG]} = 0 ,
\label{eq:oroncon}
\ee
i.e. either a circular magnetic field $B^\mu=(B^t,0,0,B^\GP)$, or a purely poloidal magnetic field
$B_t=B_\GP=0$ . Note that the magnetic field here is $B^\GA = -\frac{1}{2}\GE^{\GA\GB\GG\GD}u_\GB F_{\GG\GD}$, 
therefore $B^\GA u_\GA=0$. This implies that there is only one independent magnetic field component, since
\be
 B_t + \Omega B_\GP =  0.
\label{eq:Bu0}
\ee
In summary Oron \cite{Oron2002} has proved the following theorem: A spacetime containing a stationary
and axisymmetric purely toroidal flow of a perfect infinitely conducting fluid carrying a magnetic
field, will be circular, if and only if, the magnetic field is either purely poloidal, or purely toroidal.


Based on Oron's theorem, Kiuchi \& Yoshida \cite{Kiuchi2008} calculated 
the first fully general relativistic models of neutron stars with purely toroidal magnetic fields
using the KEH method \cite{Komatsu89,Cook94b}.  
Contrary to the purely poloidal case where there are 2 extra elliptic equations that need to be solved 
(see Eqs. (\ref{eq:atap})), here  there is only one independent magnetic component, which can be freely 
chosen due to the magnetohydrostatic equilibrium. The authors assume this independent component 
to be $F_{r\GU}$. The corresponding vector
potential is of the form $A_\mu=(0,A_r,A_\GU,0)$ and the relativistic Euler equation reduces to 
\be
\pd_{\scriptscriptstyle A} \ln\frac{h}{u^t} + \frac{1}{4\pi\GR h g_2} \sqrt{\frac{g_2}{g_1}}F_{12}
\pd_{\scriptscriptstyle A}\left(\sqrt{\frac{g_2}{g_1}}F_{12}\right) = 0 ,
\label{eq:kyEuler}
\ee
where $g_1=g_{rr}g_{\GU\GU}$, $g_2=g_{t\GP}^2-g_{tt}g_{\GP\GP}$.
Integrability of Eq. (\ref{eq:kyEuler}) demands 
\be
\sqrt{\frac{g_2}{g_1}}F_{12} = f(\GR h g_2), 
\ee
where $f$ an arbitrary function. The magnetic field with respect to the fluid observer will be
\be
B_\mu = u^t f(q) (-\Omega,0,0,1),\qquad\mbox{with}\qquad  q=\GR h g_2
\label{eq:kuBfield}
\ee
while the electromagnetic current
\be
j^\GA = \frac{1}{4\pi} \frac{1}{\sqrt{-g}}\pd_\GB (\sqrt{-g}F^{\GA\GB}) = 
\frac{1}{4\pi\sqrt{g_1 g_2}} (0, \pd_\GU f,  - \pd_r f, 0)  \, .
\label{eq:kyj}
\ee

The authors employed a simple polytropic EOS 
with $\Gamma=2$ and explored the whole parameter space of both nonrotating as well as rotating magnetized 
equilibria. For the free function $f$ they used 
\be
f(w) = b w^k ,
\label{eq:kyKfun}
\ee
where $b,\, k$ constants. Regularity of $B^\GA$ on the magnetic axis requires $k\geq 1$,
thus the authors explored two values; $k=1$ and $k=2$. For the choice $k=1$ the magnetic pressure
dominates over the matter pressure near the stellar surface, while for $k=2$ the opposite happens.
For such values of $k$ the magnetic field vanishes on the star surface therefore no boundary
conditions will be needed there.
Sequences of constant baryon mass and magnetic flux ($\Phi$) 
with respect to the meridional cross section $\Pi(\GT)$ 
\be
\Phi = \int_{\Pi(\tau)} F_{\mu\nu}dx^\mu\wedge dx^\nu = \int_0^R dr \int_0^\pi d\GU F_{r\GU}
\label{eq:EMflux}
\ee
are presented. These sequences of equilibria may model isolated neutron stars that are adiabatically losing 
angular momentum via gravitational radiation. In reality during such a process the arbitrary 
function $f$ will also change, therefore the evolutionary sequences computed 
can only hold in an appropriate timescale. 

Assuming a stress energy tensor as in Eq. (\ref{eq:imhdset}) the total energy momentum $p^\GA$ measured by the
observers with 4-velocity $u^\GA$ is $p^\GA=T^\GA_{\ \GB}u^\GB=-(\GE + \frac{1}{8\pi}B^\mu B_\mu)u^\GA$,
where $\GE$ the total energy density that includes the baryon mass contribution and the internal energy.
Due to this decomposition one can define a proper fluid energy and a magnetic energy as
\be
M_p := \int_\Sigma \GE u^\GA dS_\GA   \qquad\mbox{and}\qquad  
H := \frac{1}{8\pi} \int_\Sigma B_\mu B^\mu u^\GA dS_\GA ,
\ee
so that the gravitational potential energy will be $|W|=M_p + T + H - M$, where $T$ is the rotational energy, 
and $M$ the ADM mass of the system. 

As shown in Newtonian studies \cite{Trehan1972} the toroidal magnetic field tends to distort a neutron star prolately,
since the toroidal field lines act like ``rubber belts'' pulling in the matter around the
magnetic axis. This deformation exists even for stars rotating at the mass shedding limit 
as well as in static (nonrotating) magnetized equilibria. 
This is in contrast with the poloidal magnetic fields that result in oblate deformations.
If in addition the neutron star is rotating with its rotation axis in an angle with respect to the magnetic 
one, then such systems can be a source of gravitational waves where the
wobbling angle grows on a dissipation timescale until they become othogonal \cite{Jones1975,Cutler2002}. 
The authors measure the  prolateness of the neutron star by the parameter
\be
\bar{e} = \frac{I_{zz}-I_{xx}}{I_{zz}},
\label{eq:kyebar}
\ee
where $I_{xx},\, I_{zz}$ the principle moments of inertia around the x and z axes.
For oblate shapes $\bar{e}$ is positive (typical RNSs) while for prolate ones $\bar{e}$ is 
negative. Static solutions with a magnetic field as high as
$B_{\rm max}=1.168\times 10^{18}\rm G$, ratio of magnetic to gravitational potential energy
$H/|W|=0.2186$ and deformation parameter $\bar{e}=-1.012$ were presented.

For the rotating models the authors compute highly magnetized solutions with $H/|W|$ larger
than the ratio of kinetic to gravitational energy $T/|W|$ even at the mass shedding limit. The 
shape of the stellar surface of RNSs becomes oblate because of the centrifugal force 
while close to the core the matter distribution can still be prolate. Strong toroidal fields
result in mass shedding at lower values of angular velocity $\Omega$ or $T/|W|$ since the
central concentration of matter rises with the toroidal magnetic field. 


In the follow-up paper Kiuchi, {\it et~al.} \cite{Kiuchi2009a} computed toroidal
equilibria using different realistic EOSs in order to investigate possible
differences from the simple polytropic EOS used in
\cite{Kiuchi2008}. They considered equilibria only with $k=1$ (Eq. (\ref{eq:kyKfun}))
since general relativistic MHD simulations with $k\neq 1$ found them to be unstable
against axisymmetric perturbations \cite{Kiuchi2008b}.
One such difference was that along the maximum gravitational
mass sequences, the stars with realistic EOSs are not as prolate as the
corresponding ones with the polytropic EOS. 
The reason for this is that the magnetic belt effects subside in the vicinity of the equatorial 
plane $\lesssim 10$ km, where the adiabatic indices are generally higher than $2$
for all EOSs examined. This means that matter is stiffer there than the polytropic $\Gamma=2$ case \cite{Kiuchi2008},
leading to a smaller deformation. The dependence of the
mass-shedding angular velocity, along a sequence of constant rest mass and
magnetic flux is determined from the nonmagnetized case.  For some EOSs (like
Shen \cite{shen98}) the mass-shedding limit along the sequence is reached at smaller angular
velocity while for others (like FPS \cite{Pandharipande1989}) at larger.  Equilibrium configurations of
supramassive sequences are generally oblate in shape, although prolate shapes
exist in a narrow space of parameters that depends on the EOS\footnote{Regarding the 
definition of normal versus supramassive sequences \cite{Cook92b} there exists a difference with
respect to the nonmagnetized cases. A nonmagnetized normal sequence (has rest mass less than the maximum rest mass)
always starts from a spherical model and 
extends all the way to mass shedding \cite{Cook92b}. On the other hand it is possible a magnetized normal sequence 
not to include any nonrotating solution. }.
As with simple polytopic EOS magnetized equilibria with realistic EOSs reach mass shedding
at smaller angular velocities, since at the stellar surface the Lorenz force exerted on matter
has the same direction as the centrifugal force.
Similar to unmagnetized equilibria the spin-up effect \cite{Cook92b} is also present here.
Angular velocities $\Omega_{\rm up}$ above which the stars start to spin up as
they lose angular momentum, are found to depend sharply on the realistic EOSs.
In particular for the LS \cite{Lattimer91}, Shen \cite{shen98}, SLy \cite{Douchin01}, 
and FPS \cite{Pandharipande1989} EOSs examined the authors found that 
$\Omega_{\rm up}^{\rm SLy}>\Omega_{\rm up}^{\rm FPS}>\Omega_{\rm up}^{\rm LS}>\Omega_{\rm up}^{\rm Shen}$
even for sequences with strong magnetic fields.  In summary the authors suggest
that the EOSs of such magnetized equilibria can be constrained by observing the
angular velocities, the gravitational waves, and the signature of the spin-up.


Equilibria with strong toroidal magnetic fields have been constructed for
hybrid stars having a hadronic matter mantle and a quark core by Yasutake, {\it et~al.} \cite{Yasutake2010}.  
In particular the authors model the EOS
with a first-order transition by bridging the MIT bag model \cite{chodos1974} for the description
of quark matter with the Shen EOS \cite{shen98} using two matching densities $n_1$ and $n_2$. 
For $n<n_1$ the Shen EOS is assumed while for $n>n_2$ the MIT bag model.
For $n_1<n<n_2$ the authors compute a mixed phase in chemical
equilibrium under $\beta$-decay with vanishing neutrino chemical potentials.
For the free function  (Eq. (\ref{eq:kyKfun})) that appears in the magnetohydrostatic                                                   
equilibrium, the authors assumed $k=1$. Equilibrium sequences are
constructed by varying the central density, the axis ratio $r_p/r_e$, and the
magnetic field strength, $b$, in Eq. (\ref{eq:kyKfun}).
In general hybrid stars are more compact than neutron stars, due to the softness of the
EOS. Given the same baryon mass, the gravitational mass of hybrid stars is
smaller than of neutron stars, reflecting the smaller energy of the quark matter (or the
smaller binding energy). Also since the compression of the matter is more
enhanced for a small bag constant, hybrid stars with smaller bag constant
become more compact.

Yasutake, {\it et~al.} found that the maximum magnetic fields in hybrid stars are
$30\%$  larger  than in neutron stars. In particular the frozen-in  magnetic fields,  are
compressed by the presence of the quark phase leading to a pinching of the
field lines. Hybrid stars with smaller bag constant become more prolate
(smaller $\bar{e}$), and their maximum mass  becomes smaller 
than those with a larger bag constant.

The authors pay special attention to the possible evolutionary track of a rapidly rotating 
neutron star to a slowing rotating hybrid star due to the spin-down
via gravitational radiation and/or magnetic breaking \cite{Yasutake2004}. 
The formation of a quark core can happen through the conversion of an 
initially metastable hadronic matter through the increase of the central density
due to mass accretion, spin-down or cooling. In the process a large amount of
gravitational binding energy is released which can be a source 
of $\gamma$-ray bursts \cite{Bombaci2000}, or help explain various transient phenomena
such as glitches, magnetar flares, and super-bursts \cite{Alford2006}.
The authors found that the maximum energy release is $\lesssim 0.01 M_\odot$ which is
equivalent to $\lesssim 10^{52}\,\rm erg$, much smaller than those predicted in 
other studies \cite{gourgoulhon1999,Yasutake2004,Zdunik2006}.
In association with this energy release gravitational waves can be produced with peak
amplitudes as large as  $h\sim 10^{-18}-10^{-19}$ at frequencies of $\sim$kHz with
the source being at the Galactic center ($\sim 10\,\rm kpc$).
 

Equilibrium relativistic stars with toroidal magnetic fields have also been computed by Frieben \& Rezzolla
\cite{Frieben2012} using the \lorene~spectral code \cite{lorene}. Although the basic set up with respect to
the stress-energy tensor is the same as in
Oron \cite{Oron2002} or Kiuchi \& Yoshida \cite{Kiuchi2008}, the general relativistic formulation as well as the numerical 
implementation is different. In particular using an element of the form Eq. (\ref{eq:shibatads}),
the authors employ the 3+1 formulation under axisymmetry and stationarity
to solve the 4 elliptic equations as in Bonazzola, {\it et~al.} \cite{Bonazzola1993} or Shibata \cite{Shibata2007}. 
For the arbitrary function which appears in the integrability condition of the magnetohydrostatic euqilibrium, Eq.~(\ref{eq:kyEuler}) 
(see, Eqs.(\ref{eq:kuBfield}) and (\ref{eq:kyKfun}), the authors chose
\be
B_\GP = \GL_0 \GR h \GA e^{-2q} r^2 \sin^2\GU
\label{eq:frBphi}
\ee
which yields a magnetic potential 
\be
\tilde{M} = \frac{\GL^2_0}{4\pi} \GR h \GA^2 e^{-2q} r^2 \sin^2\GU  .
\label{eq:frMag}
\ee
Here $\pd_{\scriptscriptstyle A}\tilde{M}$ is the Lorentz force term in Eq. (\ref{eq:kyEuler}), and $\GL_0$
a constant parameter that controls the magnitude of the magnetic field and is called the magnetization parameter. 
In order to make a systematic study Frieben \& Rezzolla introduce the surface deformation or apparent oblateness 
\be
e_s = \frac{r_e}{r_p} -1 ,
\label{eq:frsdef}
\ee
and the quadrupole deformation
\be
e=-\frac{3}{2}\frac{\mathscr{I}_{zz}}{I} .
\label{eq:frqdef}
\ee
Here $r_e,\, r_p$ are the equatorial and polar coordinate radii respectively, $\mathscr{I}_{zz}$
is the quadrupole moment measured in some asymptotically Cartesian mass-centered system \cite{Thorne80b},
and $I=I_{zz}$ is the moment of inertia defined as $I=J/\Omega$, $J$ being the total (ADM) angular momentum
of the star. Positive values for $e$ or $e_s$ signify oblateness, and for a spherical unmagnetized star one has
$e_s=e=0$.
Depending on the rotation and magnetization levels the authors identify 3 regions
which they call PP, PO, and OO for prolate-prolate, prolate-oblate, and oblate-oblate respectively.
The PP region has both deformations negative $e<0$ and $e_s<0$ and corresponds to highly magnetized
toroidal solutions where centrifugal forces are subdominant to the magnetic ones. The OO region 
(typical RNS) corresponds
to the opposite scenario $e>0$ and $e_s>0$ where magnetic forces are the subdominant ones. Between these 
two regions there exists the PO region with $e<0$ and $e_s>0$ where the surface of the star is oblate due
to centrifugal forces, while the matter distribution is prolate due to electromagnetic forces. The PO region is bounded
by the $e=0$ (no quadrupole distortion) and $e_s=0$ (no surface deformation) neutral lines.

As in \cite{Kiuchi2008} the authors find that mass-shedding for magnetized
rotating equilibria happens at lower frequencies than the corresponding unmagnetized ones. The toroidal
magnetic field is acting as a source of additional pressure which not only deforms prolately the surface
and matter distribution, but causes an expansion of the star. This means that mass-shedding can set-in 
at lower angular velocities. At a given angular velocity the mass-shedding model coincides with the
one of maximum magnetization which develops the characteristic cusp on the equator.

One important new finding of \cite{Frieben2012} is that for nonrotating magnetized neutron stars no upper limit was
found to the magnetization parameter $\GL_0$, with stellar models becoming increasingly prolate and
extended as $\GL_0$ is increased independent of the EOS. The authors presented static solutions with 
circumferencial radius $\sim 102\rm km$ and $H/|W|\sim 0.5$. 
The general behavior of the nonrotating models of realistic EOSs was quite similar to the polytropic one.
For increasing magnetization a maximum value of the magnetic field appears, $\sim 10^{18}\rm G$, beyond 
which the magnetic field decreases. This nonmonotonic behavior of the mean magnetic field strength in 
terms of the magnetization parameter $\GL_0$ found in nonrotating models extends to the rotating ones as 
well. The authors present approximate relations for both the quadrupole and the surface deformation 
\be
\GE_s = b_{\scriptscriptstyle \Omega}\Omega^2 - b_{\scriptscriptstyle B} \langle B^2_{15} \rangle , \qquad\mbox{and}\qquad
\GE = c_{\scriptscriptstyle \Omega} \Omega^2 - c_{\scriptscriptstyle B} \langle B_{15}^2 \rangle ,
\label{eq:frdefo}
\ee
where $B_{15}=B/10^{15}\,\rm G$, and $b_{\scriptscriptstyle \Omega}$, $b_{\scriptscriptstyle B}$, 
$c_{\scriptscriptstyle \Omega}$, and $c_{\scriptscriptstyle B}$ are (positive) parameters that depend on the EOS.
The symbol $\langle\rangle$ denotes average values. 
These relations generalize Newtonian analogues \cite{Wentzel1960,Ostriker1969} and express
the fact that the surface and quadrupole deformation are approximately linear functions of $\Omega^2$
and $B^2$.
 

\subsection{Magnetized rotating neutron stars with various magnetic field configurations}

Bucciantini, {\it et~al.} have developed the XNS code for computing magnetized equilibriums, as a part of the X-ECHO 
code project for general relativistic MHD \cite{Bucciantini2011,Pili2014}.  
In \cite{Pili2014}, non-rotating and axisymmetric general relativistic magnetized equilibria have been constructed. 
Using the 3+1 formulation, the authors solved 2 components $\GA,\,\GC$ in the metric  Eq.~(\ref{eq:shibatads}) 
from the Hamiltonian constraint and the spatial trace of Einstein's equation.
A stress-energy tensor as in  Eq.~(\ref{eq:imhdset}) is assumed, 
and purely poloidal, purely toroidal as well as mixed magnetic field configurations are obtained.

The purely toroidal configurations are constructed along the
same lines as in \cite{Kiuchi2008}, and \cite{Frieben2012} with
\be
\GA B_\GP = b (\GR h \varpi^2)^k
\label{eq:piliKfun}
\ee
similar to Eq.~(\ref{eq:kyKfun}). The authors assumed for the magnetic exponent $k=1,\, 2$. Although the magnetic field 
distribution is similar for these values of $k$, the authors found that for higher values of $k$ the magnetic field 
reaches its maximum at larger radii. A magnetic field concentrated at larger radii will produce smaller effects, than
the same magnetic field, buried deeper inside, or alternatively, currents in the outer layers have minor effects
with respect to those residing in the deeper interior.
Despite their approximate scheme Pili, {\it et~al.} \cite{Pili2014} found perfect agreement with 
the results of Frieben \& Rezzolla
\cite{Frieben2012} and some differences with respect to those of Kiuchi \& Yoshida \cite{Kiuchi2008}. In particular
$B_{\rm max}$ is not a monotonic function of the magnetization constant $b$; increasing initially until a maximum
and then decreasing. This behavior is due to the expansion of the star for large values of $b$ with a corresponding
decrease of $B_{\rm max}$.

For poloidal configurations 
the authors employ the Grad-Shafranov equation \cite{Shafranov1958,Grad1960,Shafranov1966} in order to calculate $A_\GP$, 
\be
\nabla^2 \tilde{A}_\GP + \frac{\pd A_\GP \pd\ln(\GA\GC^{-2})}{r\sin\GU}
+ \GC^8 r\sin\GU\left(\GR h\frac{d\mathcal{M}}{dA_\GP} +
  \frac{\mathcal{I}}{\varpi^2}\frac{d\mathcal{I}}{dA_\GP}\right) = \frac{\tilde{A}_\GP}{r^2\sin^2\GU}
\label{eq:pbzAphi}
\ee
where $\tilde{A}_\GP=A_\GP/(r\sin\GU)$, $\varpi^2 = N^2\GC^4r^2\sin^2\GU$ and 
$\pd f\pd g = \pd_r f\pd_r g + \pd_\GU f \pd_\GU g/r^2$.
In Eq. (\ref{eq:pbzAphi}) there appear 2 free functions, the magnetization function 
$\mathcal{M}=\mathcal{M}(A_\GP)$, and $\mathcal{I}=\mathcal{I}(A_\GP)$, which are
both dependent on $A_\GP(r,\GU)$ only. 
In particular $\mathcal{M}$ appears in the magnetohydrostatic 
equilibrium Eq. (\ref{eq:kyEuler}) which can be written as
$\ln(h\GA)-\mathcal{M}=\rm const$. On the other hand  $\mathcal{I}$ is derived from the 
requirement that the $\GP$ component of the Lorentz force must vanish
$B^i\pd_i (\GA B_\GP)=0$ (axisymmetry), and hence  $\GA B_\GP = \mathcal{I}(A_\GP)$. 

As in \cite{Ciolfi2009} the authors used a second order polynomial for the magnetization function
\be
\mathcal{M}(A_\GP) = k_{\rm pol} \left(A_\GP+\frac{\xi}{2}A_\GP^2\right) ,
\label{eq:pbzMfun}
\ee
while for the function $\mathcal{I}(A_\GP)$, 
\be
\mathcal{I}(A_\GP) = \frac{a}{\GZ+1}\Theta[A_\GP-A_\GP^{\rm max}](A_\GP-A_\GP^{\rm max})^{\GZ+1} .
\label{eq:pbzIfun}
\ee
The authors managed to compute not only purely poloidal magnetized neutron stars, but also 
equilibria with mixed poloidal and toroidal components, the so-called twisted torus solutions 
using the same form of functions (\ref{eq:pbzMfun}) and (\ref{eq:pbzIfun}). 
Here, $k_{\rm pol}$ is the poloidal magnetization constant, $\xi$ is a nonlinear poloidal constant,
$\Theta$ is the Heaviside function, $a$ is the twisted torus magnetization parameter ($a=0$
for purely poloidal configurations), and $\GZ$
the twisted torus magnetization index. The choice of Eqs. (\ref{eq:pbzMfun}), (\ref{eq:pbzIfun})
guarantees that the currents are all confined within the star.

A comparison with the purely poloidal models of Bocquet, {\it et~al.} \cite{Bocquet1995} showed
excellent agreement despite the approximate scheme used. This means that for at least static 
solutions, even with extreme magnetic fields, the conformal flat approximation is very accurate.
Similar to rotation, poloidal magnetic fields lead to oblate deformations with a peak magnetic
field in the core of the neutron star. Contrary to the purely toroidal case here the magnetic 
field extends smoothly outside the neutron star. Also the maximum magnetic field appeared at the maximum
magnetization while its behavior for even larger magnetizations was unclear since such 
models could not be constructed. Therefore the nonmonotonic behavior present in toroidal
magnetic fields is not observed for the poloidal case. 

Beyond purely poloidal and purely toroidal magnetic field geometries the authors also constructed mixed field
the so-called twisted torus configurations. In order to do so they used the same magnetization function
as in the poloidal models Eq. (\ref{eq:pbzMfun}) including only linear terms for the toroidal currents,
$\xi=0$. The toroidal magnetic field is generated by the current Eq. (\ref{eq:pbzIfun}) and $\GZ=0$.
The structure of the resulted poloidal magnetic field is similar with the purely poloidal case,
threading the entire star, reaching its maximum value at the center, and vanishing only in a ring-like region 
on the equatorial plane. On the other hand the toroidal component has a different geometry than in the purely
toroidal case. It is confined in a torus tangent to the stellar surface at the equator and does not fill 
completely the interior of the star, while it reaches its maximum exactly in the ring-like region where the poloidal 
component vanishes. The toroidal component is subdominant to the poloidal one, which is mainly responsible for the
deformation of the star. All models presented had ratio of toroidal to total magnetic energy less than $0.07$.


In a sequel work, Bucciantini, {\it et~al.} \cite{Bucciantini2015} made a thorough investigation of the 
role of current distributions in general relativistic equilibria of magnetized neutron stars. In particular they assumed  fixed 
spherically symmetric distributions of metric and matter of a nonrotating neutron star in isotropic coordinates
(i.e. solve only for the conformal factor and lapse function) 
and solved only the Grad-Shafranov equation over the background. 
The magnetic field was of the form
\be
B^r = \frac{\pd_\GU A_\GP}{\sqrt{\GG}}, \qquad
B^\GU = -\frac{\pd_r A_\GP}{\sqrt{\GG}}, \qquad
B^\GP = \frac{\GC^2\mathcal{I}(A_\GP)}{\GA\sqrt{\GG}\sin\GU}, 
\label{eq:BuccBi}
\ee
where $\GG={\rm det}(\GG_{ij})$, $A_\GP$ the magnetic flux function, and $\mathcal{I}(A_\GP)$ the free current function.
The conduction current $J^i = \GE^{ijk}\pd_j(\GA B_k)/\GA$ depends on the two free functions $\mathcal{M}(A_\GP)$
and $\mathcal{I}(A_\GP)$. The determination of the flux function $A_\GP$ is done through the Grad-Shafranov Eq. (\ref{eq:pbzAphi})
which determines hydromagnetic equilibrium inside the star. It's solution can be extended outside the star as well
by neglecting the terms associated with the fluid rest-mass density. The authors extend the results of \cite{Pili2014}
by considering a magnetization functional form of
\be
\mathcal{M}(A_\GP) = k_{\rm pol} A_\GP\left[1+\frac{\xi}{\nu+1}\left(\frac{A_\GP}{A_\GP^{\rm max}}\right)^\nu \right]
\label{eq:BuccM}
\ee
and a current function of the form
\be
\mathcal{I}(A_\GP) = \frac{a}{\GZ+1}\Theta(A_\GP-A_\GP^{\rm surf})
\frac{(A_\GP-A_\GP^{\rm surf})^{\GZ+1}}{(A_\GP^{\rm surf})^\GZ}, 
\label{eq:BuccI1}
\ee
or
\be
\mathcal{I}(A_\GP) = \frac{a}{\GZ+1}\Theta(A_\GP-A_\GP^{\rm surf})
\frac{(A_\GP-A_\GP^{\rm surf})^{\GZ+1} (A_\GP^{\rm max}-A_\GP)^{\GZ+1}  }{(A_\GP^{\rm surf} A_\GP^{\rm max})^{\GZ+1/2}} .
\label{eq:BuccI2}
\ee
As in Eq. (\ref{eq:pbzMfun}) $k_{\rm pol}$ is the poloidal magnetization constant, while constant $\nu$ is the poloidal
magnetization index that generalizes the exponent $\nu=2$ in Eq. (\ref{eq:pbzMfun}). Similarly to Eq. (\ref{eq:pbzIfun}),
$a$ and $\GZ$ in Eqs. (\ref{eq:BuccI1}) and (\ref{eq:BuccI2}) are the toroidal magnetization constant and the toroidal 
magnetization index.
The magnetization function $\mathcal{M}$ vanishes outside the surface of the neutron star, while the toroidal magnetic field
is fully confined within the star. Equation (\ref{eq:BuccI1}) corresponds to a twisted torus configuration, where 
the azimuthal current has the same sign over its domain and the toroidal field reaches its maximum where the poloidal
field vanishes, while Eq. (\ref{eq:BuccI2}) corresponds to a twisted ring  configuration, where the current changes 
its sign, and the toroidal field vanishes in the same place where the poloidal field goes to zero.

For purely poloidal fields, i.e. $B^\GP=\mathcal{I}=0$, the main conclusions were:
(i) Subtractive currents ($\xi<0$), confine the magnetic field towards the axis, leaving large unmagnetized regions
inside the neutron star. The surface magnetic field is concentrated in a polar region of $\sim 20^\circ$ from the pole, while 
at lower latitudes it can be a factor of $\sim 10$ smaller than at the pole.
(ii) Additive currents ($\xi>0$) tend to concentrate the field in the outer layer of the neutron star. The field strength reaches 
its maximum closer to the surface, while its strength at the centre can be even more than a factor of 2 smaller.
The structure of the field at the equator can be qualitatively different from a dipole.

For mixed  toroidal and poloidal magnetic fields the authors found that despite using two
families of currents representative of a large class of configurations, in neither case they could
obtain magnetic field distributions where the energetics were dominated by the toroidal component. In particular
$\mathcal{H}_{\rm tor}/\mathcal{H}\lesssim 0.1$ in contrast with the results of Ciolfi \& Rezzolla \cite{Ciolfi2013}.
A possible origin of this difference, is related perhaps to the choice of boundary conditions, but further work is
needed to clarify this issue.
On the other hand, the ratio $\mathcal{H}_{\rm tor}/\mathcal{H}$ increases with the total mass of the neutron star. It appears that 
the rest-mass density stratification \cite{Glampedakis2012} regulates the relative importance of $\mathcal{I}$ and $\mathcal{M}$, 
and the net outcome in terms of energetics of the toroidal and poloidal components.

In \cite{Pili2015}, the same group was able to calculate a twisted magnetic field threading both the interior of the 
neutron star and the exterior magnetosphere. 
To do so, the Grad-Shafranov equation was solved over the background spherical solution as in 
\cite{Bucciantini2015} for both the interior and the exterior of the star 
with the use of a generalized current of the form \cite{Glampedakis2014}
\be
\mathcal{I}(A_\GP) = \frac{a}{\GZ+1}\Theta(A_\GP-A_\GP^{\rm ext})
\frac{(A_\GP-A_\GP^{\rm ext})^{\GZ+1}}{(A_\GP^{\rm max})^{\GZ+1/2}}.
\label{eq:BuccI1ext}
\ee
Here $A_\GP^{\rm ext}$ is the maximum value it reaches at a distance $r=\GL r_e$ from the star, where $r_e$
the equatorial radius. Parameter $\GL$ controls the size of the twisted magnetosphere outside the star and thus
the equilibria presented are generalizations of the twisted torus models of \cite{Pili2014}. 
In all the obtained configurations, the energy of the external toroidal magnetic field is, at most, $\sim 25\%$ 
of the total magnetic energy in the magnetosphere which is thus dominated by the poloidal field.


MRNS equilibria  associated with strong purely poloidal or purely toroidal magnetic fields in general relativity have been 
presented by the same Florence group in \cite{Pili2017}. 
For the gravity sector a 3+1 form of a metric is assumed and the IWM formulation is employed  
to solve for the lapse $\GA$, the conformal factor $\GC$, and the shift $\GB^\GP$. In the electromagnetic sector purely poloidal
magnetic fields are computed using the formalism of Bocquet, {\it et~al.~} \cite{Bocquet1995} where the Maxwell-Gauss
and the Maxwell-Amp\`ere equations are written as elliptic equations for the electromagnetic potential $\Phi$ and the
magnetic flux $\Psi=A_\GP$. These equations determine the electromagnetic field everywhere once the charge and current distributions
are known, independently of the fluid properties. Hydrostatic equilibrium depends on the magnetization function
for which the authors assume Eq. (\ref{eq:BuccM}) with $\xi=0$. On the other hand for purely toroidal magnetic fields
a choice similar to \cite{Kiuchi2008,Frieben2012}  is adopted.
A large number of sequences is presented and special attention is paid in the quantities like the surface ellipticity
$e_s$ and mean deformation $\bar{e}$, Eq. (\ref{eq:kyebar}). 

Purely toroidal equilibria show an increase in the gravitational and the baryonic mass, at a given central density, 
a result of the growth of the stellar radius caused by rotation and the magnetic field. 
Rapidly RNS appear as oblate ellipsoids. At higher magnetization the mass shedding limit occurs at
higher densities with respect to the nonmagnetized case. This happens because
the toroidal magnetic field significantly expands and rarefies the outer layers
of the star making them volatile to centrifugal effects.
At low magnetization, the surface shape is always oblate, as expected for an unmagnetized RNS. 
As the magnetic field increases, the oblateness diminishes and the shape becomes prolate. As the 
magnetic field begins to inflate the outer layer of the star, the local centrifugal support is 
enhanced, and the star becomes oblate again. At the mass shedding all models show
apparent oblateness with the equatorial radius being larger than the polar one. 
Bilinear relations that approximate the surface and mean deformation in terms of $B^2$ and $\Omega^2$
similar to Eqs. (\ref{eq:frdefo}), as well as in terms of $T/W$ and $H/W$ are derived. Although such empirical relations
are equivalent the authors find that the latter hold with the same accuracy for a $\sim 50\%$ larger 
range of magnetic field  strengths and rotation rates. Also their accuracy ($\lesssim 5\%$) holds up to 
the full non-linear regime.

The effect on the baryonic and gravitational mass of a purely poloidal magnetic field follows the same
behavior as with the toroidal one. Both masses increase with the magnetization and with the rotational 
frequency. The difference in the poloidal case is that the magnetic field acts in the same way as the
centrifugal force flattening the star in the direction of the equatorial plane. Therefore the configurations
are oblate and the surface ellipticity $e_s$ positive. The poloidal field does not inflate the outer
layers of the star even though the equatorial radius grows. The poloidal field enhances the stability 
against the Keplerian limit since the equatorial Lorentz force points outward in the inner region of the 
star causing its deformations, but points inward in the outer layers playing a confining role.
At high magnetization, the magnetic force can expel matter from the core so that the density reaches its 
maximum  in a ring located in the equatorial plane (rather than at the center) similar to a differentially
RNS. As pointed out by Cardall, {\it et~al.~} \cite{Cardall2001}, at even higher magnetization no stationary 
solution can be found because the magnetic field pushes off-center a sufficient amount of mass that results
in the gravitational force pointing outward near the center of the star.


\subsection{Magnetized matter and magnetic field dependent equation of state}

A step forward in understanding the interplay between the magnetic field and matter, was achieved by 
Chatterjee, {\it et~al.~}\cite{Chatterjee2014}. In their work the effect 
of the magnetic field on the EOS and the interaction of the electromagnetic field with matter were investigated in a self-consistent
manner. In particular building on the work of Bocquet, {\it et~al.~} \cite{Bocquet1995} the authors compute
magnetized equilibria with a pure poloidal magnetic field and a generalized energy-momentum tensor of the form
\be
T^{\GA\GB} = \TPFuab + \TEMuab + \TFMuab \qquad\mbox{with}\qquad
\TFMuab = \frac{1}{2}(F^\GA_{\ \GG}M^{\GG\GB} + F^\GB_{\ \GG}M^{\GG\GA}),
\label{eq:magset}
\ee
being the term that represents the interaction of the electromagnetic field with matter, and has been derived from the
interaction Lagrangian in a self-consistent way. $M^{\GA\GB}$ is the magnetization tensor (not to be confused
with the magnetization free function) which is defined as the derivative of the grand canonical potential 
with respect to the electromagnetic tensor. $\TPFuab$ is the perfect fluid stress-energy tensor, Eq. (\ref{eq:pfset}),
and $\TEMuab$ the IMHD stress-energy tensor, Eq. (\ref{eq:imhdset}).
For the magnetization tensor  the authors adopt the following form
\be
M_{\GA\GB} = \GE_{\mu\nu\GA\GB}m^\mu u^\nu \, ,\qquad\mbox{and}\qquad
m^\GA = w b^\GA ,
\label{eq:magnetizationtensor}
\ee
where $m^\GA$ is the magnetization 4-vector and $w$ a scalar quantity. Under such assumptions
$\TFMuab = w (b^\GA b^\GB-b^2(u^\GA u^\GB + g^{\GA\GB}))$.

The second important ingredient in the calculation of Chatterjee, {\it et~al.~} is that the EOS depends on the
magnetic field, i.e. 
\be
p=p(h,b),\qquad \GE=\GE(h,b),\qquad \GR(h,b),\qquad \mbox{and}\qquad w(h,b). 
\label{eq:MEOS}
\ee
The evaluation of the thermodynamic
variables in the presence of magnetic field is described in \cite{Noronha2007,Sinha2013}, while the authors
employ the quark model in the MCFL (Magnetic Colour-Flavour-Locked) phase to describe the neutron star interior \cite{Noronha2007}.
Elliptic equations for $A_t$ and $A_\GP$ are written similarly to Eqs. (\ref{eq:atap}) in Bocquet {\it et~al.}
although extra terms that depend on the magnetization $w$ are now present. On the other hand the equation of
magnetohydrostatic equilibrium is now the same as in Eq. (\ref{eq:fistint}) due to the specific form of the 
magnetization tensor in accordance with Blandford \& Hernquist \cite{Blandford1982}.
In general the authors found that the effect of inclusion of the magnetic field dependence on the EOS does not 
change significantly the stellar structure. Quantities like the polar magnetic field, the gravitational mass and 
the compactness for static and uniformly rotating magnetars  are only slightly modified even for the strongest 
magnetic fields considered, well above the values that are considered realistic from present magnetar observations.



In order to explore further the effects of a magnetic-field-dependent EOS and
magnetization, Franzon, {\it et~al.} \cite{Franzon2015} compute
equilibria with pure poloidal magnetic fields using the \lorene\ code
\cite{Bonazzola1993} as in \cite{Chatterjee2014,Bocquet1995} but with an EOS
that describes magnetized hybrid stars containing nucleons, hyperons, and
quarks, and takes into account the anomalous magnetic moment for all hadrons.
This EOS  \cite{Hempel:2013,Papazoglou:1998,Dexheimer:2008} is an extended
hadronic and quark SU(3) non-linear realization of the sigma model that
describes magnetized hybrid stars containing nucleons, hyperons, and quarks.
Despite the fact that they can reach a magnetization approximately 10 times
higher than in \cite{Chatterjee2014}, the neutron star structure, like its
mass-radius relationship, is not modified drastically.  On the other hand, the
magnetic field causes the central density in these objects to be reduced,
inducing major changes in the populated degrees of freedom and, potentially,
converting a hybrid star into a hadronic star.

In \cite{Franzon2016a} the same group  investigated the effects of strong
magnetic fields on a hot and rapidly rotating proto-neutron star.  Different
from typical cold neutron stars, proto-neutron stars can have temperatures up to $50\,\rm
MeV$, and are lepton-rich as well as  optically thick to neutrinos, which are
temporarily trapped within the star.  The magnetic field can affect the amount
of trapped neutrinos and prevent or favor exotic phases with hyperons or
quarks. For the EOS the authors use the hadronic chiral SU(3) model
\cite{Hempel:2013,Papazoglou:1998,Dexheimer:2008} explicitly including trapped
neutrinos and fixed entropy per baryon. The cold and hot EOSs are then
calculated at finite temperature and over a range of entropies and neutrino
fractions, while equilibria with pure poloidal magnetic fields are computed using
the methods of \cite{Bocquet1995,Franzon2015}.  Their results suggested that
spherical hot stars with trapped neutrinos are less massive than the same
stars in $\GB$-equilibrium or their cold counterparts.  The primary effect of
the magnetic field decay is to increase the amount of neutrinos and the
strangeness at the stellar core. Assuming that the magnetic field decays over
time, the temperature in the equatorial plane increases in the inner core while
it decreases in the outer core. This fact is related to the Lorenz force, which
reverses its direction in the equatorial plane.  For rotating proto-neutron
stars the electron neutrino distribution does not differ much from their
nonrotating counterpart since the centrifugal forces act mainly on the outer
layers of the star. However, the amount of hyperons is reduced inside these
objects, which may affect the cooling of these stars. As expected, the
reduction in the central densities is even more pronounced, and magnetic fields
suppress exotic phases in rotating proto-neutron stars even further, as in the
case of cold neutron stars.

Effects of the magnetic field on the crust structure of neutron stars was investigated 
by  Franzon, {\it et~al.~} \cite{Franzon2017}. The authors define 
the crust thickness as the difference between the stellar surface
radius and the radius at the base of the crust where the crust-core transition
takes place, for which they assume  a baryon number density of $0.076\,\rm fm^{-3}$.
They found that on average the crust thickness as a function of the poloidal magnetic field
decreases first, before it starts to increase. This is in contrast with rotationally deformed axially symmetric
neutron stars, where the crust gets always thicker as a function of rotation.
The authors argue that the reason behind this behavior lies in the dual role of the
electromagnetic field in a general relativistic scenario. On one hand the energy of the electromagnetic field 
contributes to the curvature of spacetime and on the other it generates additional forces 
that modify the equilibrium of the star. In particular the two competing effects are the Lorentz force 
that tends to make the crust thinner, and the gravitational contribution of the
magnetic field that tends to make the crust thicker. For moderately high magnetic
fields, the former wins, and the crust gets thinner on average, whereas for
extreme values of magnetic fields, the latter is dominant, making the crust thicker overall.
This change in crust geometry may be relevant to the overall cooling of neutron stars,
as well as their deformability during the late inspiral in a BNS merger.


\subsection{A general formulation for magnetized rotating neutron stars and numerical solutions}

The fully general models for a stationary and axisymmetric magnetized equilibriums 
are those associated with mixed toroidal and poloidal magnetic fields as well as 
mixed circular and meridional matter flows.  The stress-energy tensor of such 
models does not satisfy conditions (\ref{eq:intcon}), and hence the spacetime metric 
cannot be described in the form (\ref{eq:circular}) in case such mixed electromagnetic 
fields and/or matter flows dominate. Full exact formulations for such models are 
derived in \cite{Gourgoulhon:2011gz,Uryu2019},  and numerical solutions for such 
strongly magnetized rotating relativistic equilibriums are presented by Ury\=u, {\it et~al.}
\cite{Uryu2014,Uryu2019}.  

The formulation consists of three parts, that for the gravitational fields, 
for the electromagnetic fields and for the magneto-hydrostationary equilibrium.  
For the gravitational fields, a formulation developed for computing initial 
data of BNSs using a fully general form of the metric, Eq. (\ref{eq:dstpo}), is applied 
(see section \ref{sec:ncfid}) \cite{Shibata04a,Bonazzola:2003dm,Uryu2006,Uryu:2009ye}.

An analogous idea is used to write the 3+1 decomposed Maxwell's equations as a system of 
elliptic partial differential equations for the electromagnetic potential 1-form $A_\alpha$
\cite{Uryu2019}.

For the formulation to compute the MHD equilibrium 
one may assume an IMHD condition $\Fabd u^\beta=0$ as mentioned in the purely poloidal or toroidal cases.  
A set of first integrals, and several integrability conditions, 
of the IMHD equations can be derived for the case with mixed poloidal and
toroidal fields under the assumptions of stationarity and axisymmetry. 
Those conditions amount to express several quantities in terms of 
a master potential $\Upsilon$ \cite{Gourgoulhon:2011gz,Uryu2019} as 
\beqn
A_t = A_t(\Upsilon), 
\qquad 
A_\phi &=& A_\phi(\Upsilon), 
\qquad
\sqrt{-g}\Psi = [\sqrt{-g}\Psi](\Upsilon)
\\[2mm]
-[\sqrt{-g}\Psi]'hu_\phi
&+& \frac1{4\pi}A'_\phi B\sqrt{-g}
\,=\, [\sqrt{-g}\Lambda_\phi](\Upsilon),
\\[2mm]
A'_\phi hu_t &-& \,A'_t hu_\phi\,=\,\Lambda(\Upsilon),
\label{eq:MHD-Euler_tphi}
\eeqn
where $\sqrt{-g}\Psi$ is a weighted stream function of the meridional flow, 
$h$ the relativistic enthalpy, and $B = -F^{xz}$.  
$A_t(\Upsilon)$, $A_\phi(\Upsilon)$, $[\sqrt{-g}\Psi](\Upsilon)$, and 
$\Lambda(\Upsilon)$ are arbitrary functions of $\Upsilon$, and primed 
functions such as $A'_t(\Upsilon)$ are the derivatives with respect to 
$\Upsilon$ of those functions.  
In the computations presented in \cite{Uryu2019,Uryu2014}, the master potential 
$\Upsilon$ is chosen to be $\Upsilon = A_\phi$ for simplicity.  In this case, 
the relativistic enthalpy and the components of 4-velocity are calculated as
\beqn
u^A &=& \frac1{\rho\sqrt{-g}}[\sqrt{-g}\Psi]'\epsilon^{AB}\pa_B A_\phi \, , 
\\[1mm]
u^t \,&=& \frac1{[-\gabd(\alpha n^\alpha+\beta^\alpha+v^\alpha)
                     (\alpha n^\beta+\beta^\beta+v^\beta)]^{1/2}}  \, ,
\\[1mm]
u^\phi &=& \frac{[\sqrt{-g}\Psi]'B_\phi}{\rho\sqrt{-g}} - A'_t u^t  \, ,
\\[1mm]
h\ \,&=& \frac{\Lambda}{u_t - A'_t u_\phi} \, .
\eeqn
The equation for $u^t$ is derived from the normalization condition 
of the 4-velocity, $u_\alpha u^\alpha=-1$, that for $u^\phi$ from 
the meridional components of the IMHD condition $\Fabd u^\beta=0$, 
and that for $h$ from Eq.(\ref{eq:MHD-Euler_tphi}).  Here the 4-velocity 
is decomposed as $u^\alpha = u^t(t^\alpha+v^\alpha)$.  
 
Under the IMHD condition, the electric current does not have a dynamical 
degree of freedom, and therefore it can be written in terms of the above 
integrability conditions as in the cases of purely poloidal or toroidal magnetic 
fields.  Substituting those expressions of the components of the current $j^\alpha$ 
to Maxwell's equations, one can derive the transfield equation for the master 
potential $\Upsilon$, which fully determines a mixed poloidal-toroidal magnetic field
configuration \cite{Gourgoulhon:2011gz}.  Different from this formulation, 
all components of Maxwell's equations are solved in \cite{Uryu2019,Uryu2014} 
as mentioned above.  The expressions of the components of the current in terms of 
the integrability conditions are written, for the case with $\Upsilon = A_\phi$, 
\beqn
j^A \sqrt{-g} &=& 
\left([\sqrt{-g}\Psi]'' hu_\phi + [\sqrt{-g}\Lambda_\phi]' \right)\delta^{AB}B_B
- [\sqrt{-g}\Psi]'\delta^{AB}\omega_B, 
\\[2mm]
j^\phi \sqrt{-g} &+& A'_t j^t \sqrt{-g} \,=\, 
\left([\sqrt{-g}\Psi]'' hu_\phi + [\sqrt{-g}\Lambda_\phi]' \right)B_\phi
- [\sqrt{-g}\Psi]'\omega_\phi 
\nonumber \\[1mm]
&& \qquad \qquad\ \ 
- \left(A''_t h u_\phi + \Lambda'\right)\rho u^t\sqrt{-g}
- s'T\rho\sqrt{-g}, 
\eeqn
where $s = s(A_\phi)$ the entropy per unit baryon mass, which is taken 
to be constant. These current components 
are substituted into the source terms of the 3+1 decomposed Maxwell's equations.  

In the computations of \cite{Uryu2019,Uryu2014}, a one-parameter EOS $p = p(\rho)$ 
is used, and the arbitrary functions of integrability conditions are assumed as follows: 
\beqn
A_t(A_\phi) &=& -\Omega_c A_\phi + C_e, 
\\[1mm]
\Lambda(A_\phi) &=& -\Lambda_0\Xi(A_\phi) - \Lambda_1 A_\phi -\cal{E}, 
\\[1mm]
[\sqrt{-g}\Lambda_\phi](A_\phi) &=& \Lambda_{\phi0}\Xi(A_\phi), 
\\[1mm]
[\sqrt{-g}\Psi](A_\phi) &=& \mbox{constant}.  
\eeqn
The derivative of $\Xi(A_\phi)$ is the smoothed step (``sigmoid'')  function defined by 
\beq
\Xi'(A_\phi) \,:=\, \frac12\left[\tanh\left(\frac1{b}
\frac{A_\phi-A_{\phi,\rm S}^{\rm max}}{A_\phi^{\rm max}-A_{\phi,\rm S}^{\rm max}}
-c\right)+1\right], 
\eeq
where $A_\phi \in [A_{\phi,\rm S}^{\rm max}, A_\phi^{\rm max}]$, and $b$, $c$ are 
parameters such that $0 < b < 1$, $0 < c < 1$.  
The prescribed constants $\Lambda_0$, $\Lambda_1$ and $\Lambda_{\phi0}$ 
control the magnitude and configuration of the electromagnetic fields. On the other hand,
$\Omega_c$ and $\cal{E}$ are constants to be determined from the rotating 
equilibrium, while constant $C_e$ from charge neutrality.  

Several numerical solutions of strongly magnetized rotating equilibria 
associated with mixed poloidal and toroidal magnetic fields are 
demonstrated in \cite{Uryu2019,Uryu2014}. In the case with the strongest 
toroidal field, the authors show that the magnetic pressure and energy 
density dominate over those of the fluid, and that the matter is expelled 
from the toroidal region. Therefore, it is expected that compact stars with an extremely 
strong toroidal magnetic field may exhibit an internal toroidal electromagnetic vacuum tunnel.  


\section{Conclusions}
\label{sec:concl}

A number of recent breakthroughs, from the BNS event GW170817 \cite{GW170817prl} to the 
Event Horizon Telescope observations of the core of the galaxy M87 \cite{Akiyama2019_L1}, 
have shown that BNSs, BHDs, and MRNSs play a central role in understanding the physics of compact objects and, more generally, 
the physics of matter under extreme conditions.  
In order to simulate accurately such systems, one needs self-consistent models as initial data. These models can be
thought as ``snapshots'' of the system during an evolutionary process. The assumptions that lead to such a snapshot cannot
be underestimated.
In this review we summarized studies for the numerical construction of self-gravitating (quasi)equilibria for the 3 
aforementioned compact objects. Our focus was to present an overview of the basic equations that govern these 
(quasi)equilibria along with the crucial assumptions that led to them, as well as the corresponding numerical results.
Despite the different nature of the problems, common strategies are identified and different methods are underlined.

There are a lot of works that couldn't be covered in this article. Those include related studies in the framework of 
Newtonian gravity, including 
\cite{Eriguchi1983,Hachisu1984a,Eriguchi1985,Hachisu86a,Uryu98c,Uryu98b,Uryu98c} for BNSs, 
\cite{Otani2009} for self-gravitating BHDs, and 
\cite{Tomimura2005, Yoshida2006, Yoshida2006b, Yoshida2011, Fujisawa2012, Fujisawa2013a, Fujisawa2013,
Lander:2009, Lander2012a, Glampedakis2012, Lander2012, Lander2013, Palapanidis2015, Lander2021} for MRNSs.  
These studies often treat more advanced and astrophysically realistic problems, and many of their ideas are transferred to 
the relativistic problems introduced in this review.  
Finally, we did not cover studies that go beyond general relativistic gravity.  For example, 
BNS models in scalar-tensor theories \cite{Taniguchi2015}, or 
\cite{Soldateschi2020, Soldateschi2021, Soldateschi2021a} for MRNSs.  
As observations of gravitational waves and  electromagnetic fields from compact objects are expected to become increasingly 
more accurate in the future, their modelling in alternative theories of gravity will gain momentum similarly.
Hopefully a future version of this article will close this gap.

%
%

%
\section*{Conflict of interest}

 The authors declare that they have no conflict of interest.

\bibliographystyle{spphys}       

\bibliography{atreferences}

\begin{thebibliography}{100}
\providecommand{\url}[1]{{#1}}
\providecommand{\urlprefix}{URL }
\expandafter\ifx\csname urlstyle\endcsname\relax
  \providecommand{\doi}[1]{DOI \discretionary{}{}{}#1}\else
  \providecommand{\doi}{DOI \discretionary{}{}{}\begingroup
  \urlstyle{rm}\Url}\fi

\bibitem{Hewish1968}
A.~{Hewish}, S.J. {Bell}, J.D.H. {Pilkington}, P.F. {Scott}, R.A. {Collins},
  Nature \textbf{217}, 709 (1968).
\newblock \doi{10.1038/217709a0}

\bibitem{lyne_graham-smith_2012}
A.~Lyne, F.~Graham-Smith, \emph{Pulsar Astronomy}, 4th edn.
\newblock Cambridge Astrophysics (Cambridge University Press, 2012).
\newblock \doi{10.1017/CBO9780511844584}

\bibitem{HulseTaylor74}
R.~Hulse, J.~Taylor, Astrophys. J. \textbf{195}, L51 (1975)

\bibitem{Zhu2020}
X.J. Zhu, G.~Ashton, Astrophys. J. Lett. \textbf{902}(1), L12 (2020).
\newblock \doi{10.3847/2041-8213/abb6ea}

\bibitem{MillerColeman2016}
M.~Coleman~Miller, Gen. Rel. Grav. \textbf{48}(7), 95 (2016).
\newblock \doi{10.1007/s10714-016-2088-4}

\bibitem{Abbott2019}
B.~Abbott, et~al., Phys. Rev. D \textbf{100}(10), 104036 (2019).
\newblock \doi{10.1103/PhysRevD.100.104036}

\bibitem{Yagi2013b}
K.~{Yagi}, N.~{Yunes}, Phys. Rev. D \textbf{88}(2), 023009 (2013).
\newblock \doi{10.1103/PhysRevD.88.023009}

\bibitem{lorimer:lr}
D.R. Lorimer, Living Rev. Relativ. \textbf{4}(5) (2001).
\newblock \urlprefix\url{http://www.livingreviews.org/lrr-2001-5}

\bibitem{Lattimer2016}
J.M. {Lattimer}, M.~{Prakash}, Physics Reports \textbf{621}, 127 (2016).
\newblock \doi{10.1016/j.physrep.2015.12.005}

\bibitem{Ozel2016}
F.~{{\"O}zel}, P.~{Freire}, Annual Review of Astronomy and Astrophysics
  \textbf{54}, 401 (2016).
\newblock \doi{10.1146/annurev-astro-081915-023322}

\bibitem{Eichler89}
D.~{Eichler}, M.~{Livio}, T.~{Piran}, D.N. {Schramm}, Nature \textbf{340}, 126
  (1989).
\newblock \doi{10.1038/340126a0}

\bibitem{Narayan92}
R.~{Narayan}, B.~{Paczynski}, T.~{Piran}, Astrophys. J. Lett. \textbf{395}, L83
  (1992).
\newblock \doi{10.1086/186493}

\bibitem{Fong2013}
W.~Fong, E.~Berger, B.D. Metzger, R.~Margutti, R.~Chornock, G.~Migliori, R.J.
  Foley, B.A. Zauderer, R.~Lunnan, T.~Laskar, S.J. Desch, K.J. Meech,
  S.~Sonnett, C.~Dickey, A.~Hedlund, P.~Harding, The Astrophysical Journal
  \textbf{780}(2), 118 (2013).
\newblock \doi{10.1088/0004-637x/780/2/118}.
\newblock \urlprefix\url{https://doi.org/10.1088%2F0004-637x%2F780%2F2%2F118}

\bibitem{Lattimer74}
J.M. {Lattimer}, D.N. {Schramm}, Astrophys. J. Lett. \textbf{192}, L145 (1974).
\newblock \doi{10.1086/181612}

\bibitem{Lattimer76}
J.M. Lattimer, D.N. Schramm, Astrophys. J. \textbf{210}, 549 (1976)

\bibitem{Symbalisty82}
E.M.D. Symbalisty, D.N. Schramm, Astrophys. Lett. \textbf{22}, 143 (1982)

\bibitem{Burbidge1957}
E.M. {Burbidge}, G.R. {Burbidge}, W.A. {Fowler}, F.~{Hoyle}, Rev. Mod. Phys.
  \textbf{29}, 547 (1957).
\newblock \doi{10.1103/RevModPhys.29.547}

\bibitem{Cameron1957}
A.G.W. Cameron, Publications of the Astronomical Society of the Pacific
  \textbf{69}, 201 (1957).
\newblock \doi{10.1086/127051}.
\newblock \urlprefix\url{https://doi.org/10.1086%2F127051}

\bibitem{GW170817prl}
B.P. Abbott, et~al., Phys. Rev. Lett. \textbf{119}(16), 161101 (2017).
\newblock \doi{10.1103/PhysRevLett.119.161101}

\bibitem{Goldstein2017}
A.~{Goldstein}, P.~{Veres}, E.~{Burns}, M.S. {Briggs}, R.~{Hamburg},
  D.~{Kocevski}, C.A. {Wilson-Hodge}, R.D. {Preece}, S.~{Poolakkil}, O.J.
  {Roberts}, C.M. {Hui}, V.~{Connaughton}, J.~{Racusin}, A.~{von Kienlin},
  T.~{Dal Canton}, N.~{Christensen}, T.~{Littenberg}, K.~{Siellez},
  L.~{Blackburn}, J.~{Broida}, E.~{Bissaldi}, W.H. {Cleveland}, M.H. {Gibby},
  M.M. {Giles}, R.M. {Kippen}, S.~{McBreen}, J.~{McEnery}, C.A. {Meegan}, W.S.
  {Paciesas}, M.~{Stanbro}, Astrophys. J. Letters \textbf{848}, L14 (2017).
\newblock \doi{10.3847/2041-8213/aa8f41}

\bibitem{Savchenko2017}
V.~{Savchenko}, C.~{Ferrigno}, E.~{Kuulkers}, A.~{Bazzano}, E.~{Bozzo},
  S.~{Brandt}, J.~{Chenevez}, T.J.L. {Courvoisier}, R.~{Diehl}, A.~{Domingo},
  L.~{Hanlon}, E.~{Jourdain}, A.~{von Kienlin}, P.~{Laurent}, F.~{Lebrun},
  A.~{Lutovinov}, A.~{Martin-Carrillo}, S.~{Mereghetti}, L.~{Natalucci},
  J.~{Rodi}, J.P. {Roques}, R.~{Sunyaev}, P.~{Ubertini}, Astrophys. J. Letters
  \textbf{848}, L15 (2017).
\newblock \doi{10.3847/2041-8213/aa8f94}

\bibitem{Abbott2017d}
B.P. Abbott, et~al., Astrophys. J. Lett. \textbf{848}(2), L13 (2017).
\newblock \doi{10.3847/2041-8213/aa920c}

\bibitem{Coulter2017}
D.A. {Coulter}, R.J. {Foley}, C.D. {Kilpatrick}, M.R. {Drout}, A.L. {Piro},
  B.J. {Shappee}, M.R. {Siebert}, J.D. {Simon}, N.~{Ulloa}, D.~{Kasen}, B.F.
  {Madore}, A.~{Murguia-Berthier}, Y.C. {Pan}, J.X. {Prochaska},
  E.~{Ramirez-Ruiz}, A.~{Rest}, C.~{Rojas-Bravo}, Science \textbf{358}, 1556
  (2017).
\newblock \doi{10.1126/science.aap9811}

\bibitem{Soares-Santos2017}
M.~{Soares-Santos}, D.E. {Holz}, J.~{Annis}, R.~{Chornock}, K.~{Herner},
  E.~{Berger}, D.~{Brout}, H.Y. {Chen}, R.~{Kessler}, M.~{Sako}, S.~{Allam},
  D.L. {Tucker}, R.E. {Butler}, A.~{Palmese}, Z.~{Doctor}, H.T. {Diehl},
  J.~{Frieman}, B.~{Yanny}, H.~{Lin}, D.~{Scolnic}, P.~{Cowperthwaite},
  E.~{Neilsen}, J.~{Marriner}, {Dark Energy Survey}, {Dark Energy Camera GW-EM
  Collaboration}, Astrophys. J. Letters \textbf{848}, L16 (2017).
\newblock \doi{10.3847/2041-8213/aa9059}

\bibitem{Arcavi2017}
I.~{Arcavi}, G.~{Hosseinzadeh}, D.A. {Howell}, C.~{McCully}, D.~{Poznanski},
  D.~{Kasen}, J.~{Barnes}, M.~{Zaltzman}, S.~{Vasylyev}, D.~{Maoz},
  S.~{Valenti}, Nature \textbf{551}, 64 (2017).
\newblock \doi{10.1038/nature24291}

\bibitem{Diaz2017}
M.C. D{\'{\i}}az, L.M. Macri, D.G. Lambas, C.M. de~Oliveira, J.L.N.
  Castell{\'{o}}n, T.~Ribeiro, B.~S{\'{a}}nchez, W.~Schoenell, L.R. Abramo,
  S.~Akras, J.S. Alcaniz, R.~Artola, M.~Beroiz, S.~Bonoli, J.~Cabral,
  R.~Camuccio, M.~Castillo, V.~Chavushyan, P.~Coelho, C.~Colazo, M.V.
  Costa-Duarte, H.C. Larenas, D.L. DePoy, M.D. Romero, D.~Dultzin,
  D.~Fern{\'{a}}ndez, J.~Garc{\'{\i}}a, C.~Girardini, D.R. Gon{\c{c}}alves,
  T.S. Gon{\c{c}}alves, S.~Gurovich, Y.~Jim{\'{e}}nez-Teja, A.~Kanaan,
  M.~Lares, R.L. de~Oliveira, O.~L{\'{o}}pez-Cruz, J.L. Marshall, R.~Melia,
  A.~Molino, N.~Padilla, T.~Pe{\~{n}}uela, V.M. Placco, C.~Qui{\~{n}}ones, A.R.
  Rivera, V.~Renzi, L.~Riguccini, E.~R{\'{\i}}os-L{\'{o}}pez, H.~Rodriguez,
  L.~Sampedro, M.~Schneiter, L.~Sodr{\'{e}}, M.~Starck, S.~Torres-Flores,
  M.~Tornatore, A.~Zadro{\.{z}}ny, The Astrophysical Journal \textbf{848}(2),
  L29 (2017).
\newblock \doi{10.3847/2041-8213/aa9060}.
\newblock \urlprefix\url{https://doi.org/10.3847%2F2041-8213%2Faa9060}

\bibitem{Abbott2017c}
B.P. {Abbott}, et~al., Astrophys. J. Lett. \textbf{850}, L39 (2017).
\newblock \doi{10.3847/2041-8213/aa9478}

\bibitem{Metzger:2010}
B.D. {Metzger}, G.~{Mart{\'{\i}}nez-Pinedo}, S.~{Darbha}, E.~{Quataert},
  A.~{Arcones}, D.~{Kasen}, R.~{Thomas}, P.~{Nugent}, I.V. {Panov}, N.T.
  {Zinner}, Mon. Not. R. Astron. Soc. \textbf{406}, 2650 (2010).
\newblock \doi{10.1111/j.1365-2966.2010.16864.x}

\bibitem{Li:1998}
L.X. Li, B.~Paczynski, Astrophys. J. \textbf{507}, L59 (1998).
\newblock \doi{10.1086/311680}

\bibitem{Metzger2019}
B.D. Metzger, Living Rev. Rel. \textbf{23}(1), 1 (2020).
\newblock \doi{10.1007/s41114-019-0024-0}

\bibitem{ShibataTaniguchilrr-2011-6}
M.~Shibata, K.~Taniguchi, Living Rev. Relativity \textbf{14}(6) (2011).
\newblock \urlprefix\url{http://www.livingreviews.org/lrr-2011-6}

\bibitem{Berger2013b}
E.~{Berger}, Annual Review of Astron. and Astrophys. \textbf{52}, 43 (2014).
\newblock \doi{10.1146/annurev-astro-081913-035926}

\bibitem{Baiotti2016}
L.~Baiotti, L.~Rezzolla, Rept. Prog. Phys. \textbf{80}(9), 096901 (2017).
\newblock \doi{10.1088/1361-6633/aa67bb}

\bibitem{Paschalidis2016}
V.~{Paschalidis}, Classical and Quantum Gravity \textbf{34}(8), 084002 (2017).
\newblock \doi{10.1088/1361-6382/aa61ce}

\bibitem{Shibata2019a}
M.~Shibata, K.~Hotokezaka, Ann. Rev. Nucl. Part. Sci. \textbf{69}, 41 (2019).
\newblock \doi{10.1146/annurev-nucl-101918-023625}

\bibitem{Meszaros2019}
P.~M\'esz\'aros, Mem. Soc. Ast. It. \textbf{90}(1-2), 57 (2019)

\bibitem{Friedman2020}
J.L. Friedman, N.~Stergioulas, Int. J. Mod. Phys. D \textbf{29}(11), 2041015
  (2020).
\newblock \doi{10.1142/S0218271820410151}

\bibitem{Foucart2020}
F.~Foucart, Front. Astron. Space Sci. \textbf{7}, 46 (2020).
\newblock \doi{10.3389/fspas.2020.00046}

\bibitem{Ciolfi2020}
R.~Ciolfi, Gen. Rel. Grav. \textbf{52}(6), 59 (2020).
\newblock \doi{10.1007/s10714-020-02714-x}

\bibitem{Radice2020}
D.~Radice, S.~Bernuzzi, A.~Perego, Ann. Rev. Nucl. Part. Sci. \textbf{70}, 95
  (2020).
\newblock \doi{10.1146/annurev-nucl-013120-114541}

\bibitem{Abramowicz2011}
M.A. Abramowicz, P.C. Fragile, Living Rev. Relativity \textbf{16}(1) (2013).
\newblock \urlprefix\url{http://www.livingreviews.org/lrr-2013-1}

\bibitem{Blanchet06}
L.~Blanchet, Living Rev. Relativ. \textbf{9}, 4 (2006)

\bibitem{Pretorius:2005gq}
F.~Pretorius, Phys. Rev. Lett. \textbf{95}, 121101 (2005)

\bibitem{Campanelli06}
M.~{Campanelli}, C.O. {Lousto}, P.~{Marronetti}, Y.~{Zlochower}, Phys. Rev.
  Lett. \textbf{96}(11), 111101 (2006).
\newblock \doi{10.1103/PhysRevLett.96.111101}

\bibitem{Baker:2005vv}
J.G. Baker, J.~Centrella, D.I. Choi, M.~Koppitz, J.~van Meter, Phys. Rev. Lett.
  \textbf{96}, 111102 (2006)

\bibitem{Wilson2003}
J.R. Wilson, G.J. Mathews, \emph{Relativistic numerical hydrodynamics}
  (Cambridge University Press, Cambridge, UK, 2003)

\bibitem{Alcubierre:2008}
M.~Alcubierre, \emph{Introduction to $3+1$ {N}umerical {R}elativity} (Oxford
  University Press, Oxford, UK, 2008).
\newblock \doi{10.1093/acprof:oso/9780199205677.001.0001}

\bibitem{Bona2009}
C.~Bona, C.~Palenzuela-Luque, C.~Bona-Casas, \emph{Elements of Numerical
  Relativity and Relativistic Hydrodynamics: From Einstein's Equations to
  Astrophysical Simulations}.
\newblock Lecture Notes in Physics (Springer, Berlin Heidelberg, 2009).
\newblock \urlprefix\url{http://books.google.co.uk/books?id=KgPGHaCUaAYC}

\bibitem{Baumgarte2010}
T.W. {Baumgarte}, S.L. {Shapiro}, \emph{{Numerical Relativity: Solving
  Einstein's Equations on the Computer}} (Cambridge University Press,
  Cambridge, UK, 2010).
\newblock \doi{10.1017/cbo9781139193344}

\bibitem{gourgoulhon20123+1}
{\'E}.~Gourgoulhon, \emph{3+1 Formalism in General Relativity: Bases of
  Numerical Relativity}.
\newblock Lecture Notes in Physics (Springer Berlin Heidelberg, 2012).
\newblock \urlprefix\url{https://books.google.com/books?id=HKcMBwAAQBAJ}

\bibitem{Friedman2012}
J.L. {Friedman}, N.~{Stergioulas}, \emph{{Rotating Relativistic Stars}}
  (Cambridge University Press, 2013).
\newblock \urlprefix\url{https://doi.org/10.1017/CBO9780511977596}

\bibitem{Rezzolla_book:2013}
L.~{Rezzolla}, O.~{Zanotti}, \emph{Relativistic Hydrodynamics} (Oxford
  University Press, Oxford, UK, 2013).
\newblock \doi{10.1093/acprof:oso/9780198528906.001.0001}

\bibitem{Lehner2014}
L.~Lehner, F.~Pretorius, Ann. Rev. Astron. Astrophys. \textbf{52}, 661 (2014).
\newblock \doi{10.1146/annurev-astro-081913-040031}

\bibitem{Tichy2016rev}
W.~Tichy, Rept. Prog. Phys. \textbf{80}(2), 026901 (2017).
\newblock \doi{10.1088/1361-6633/80/2/026901}

\bibitem{Paschalidis2017b}
V.~{Paschalidis}, N.~{Stergioulas}, Living Reviews in Relativity \textbf{20}, 7
  (2017).
\newblock \doi{10.1007/s41114-017-0008-x}

\bibitem{Duez2019}
M.D. {Duez}, Y.~{Zlochower}, Reports on Progress in Physics \textbf{82}(1),
  016902 (2019).
\newblock \doi{10.1088/1361-6633/aadb16}

\bibitem{Shibata99d}
M.~{Shibata}, K.~{Ury{\=u}}, Phys. Rev. D \textbf{61}(6), 064001 (2000).
\newblock \doi{10.1103/PhysRevD.61.064001}

\bibitem{Baiotti08}
L.~{Baiotti}, B.~{Giacomazzo}, L.~{Rezzolla}, Phys. Rev. D \textbf{78}(8),
  084033 (2008).
\newblock \doi{10.1103/PhysRevD.78.084033}

\bibitem{Anderson2007}
M.~{Anderson}, E.W. {Hirschmann}, L.~{Lehner}, S.L. {Liebling}, P.M. {Motl},
  D.~{Neilsen}, C.~{Palenzuela}, J.E. {Tohline}, Phys. Rev. D \textbf{77}(2),
  024006 (2008).
\newblock \doi{10.1103/PhysRevD.77.024006}

\bibitem{Liu:2008xy}
Y.T. Liu, S.L. Shapiro, Z.B. Etienne, K.~Taniguchi, Phys. Rev. D \textbf{78},
  024012 (2008).
\newblock \doi{10.1103/PhysRevD.78.024012}

\bibitem{Bernuzzi2011}
S.~{Bernuzzi}, M.~{Thierfelder}, B.~{Br{\"u}gmann}, Phys. Rev. D
  \textbf{85}(10), 104030 (2012).
\newblock \doi{10.1103/PhysRevD.85.104030}

\bibitem{Paschalidis2014}
V.~{Paschalidis}, M.~{Ruiz}, S.L. {Shapiro}, Astrophys. J. Lett. \textbf{806},
  L14 (2015).
\newblock \doi{10.1088/2041-8205/806/1/L14}

\bibitem{Ruiz2016}
M.~{Ruiz}, R.N. {Lang}, V.~{Paschalidis}, S.L. {Shapiro}, Astrophys. J. Lett.
  \textbf{824}, L6 (2016).
\newblock \doi{10.3847/2041-8205/824/1/L6}

\bibitem{Frank2002}
J.~{Frank}, A.~{King}, D.J. {Raine}, \emph{{Accretion Power in Astrophysics:
  Third Edition}} (Cambridge University Press, 2002)

\bibitem{Kato2008}
S.~{Kato}, J.~{Fukue}, S.~{Mineshige}, \emph{{Black-Hole Accretion Disks ---
  Towards a New Paradigm ---}} (Kyoto University Press, Kyoto, 2008)

\bibitem{Blanchet2013}
L.~Blanchet, Living Rev. Rel. \textbf{17}, 2 (2014).
\newblock \doi{10.12942/lrr-2014-2}

\bibitem{Isenberg08}
J.A. {Isenberg}, International Journal of Modern Physics D \textbf{17}, 265
  (2008).
\newblock \doi{10.1142/S0218271808011997}

\bibitem{Wilson89}
J.R. {Wilson}, G.J. {Mathews}, \emph{{Relativistic hydrodynamics.}} (Cambridge
  University Press, 1989), pp. 306--314

\bibitem{Wilson95}
J.R. Wilson, G.J. Mathews, Phys. Rev. Lett. \textbf{75}, 4161 (1995)

\bibitem{Wilson96}
J.R. {Wilson}, G.J. {Mathews}, P.~{Marronetti}, Phys. Rev. D \textbf{54}, 1317
  (1996).
\newblock \doi{10.1103/PhysRevD.54.1317}

\bibitem{Lichnerowicz44}
A.~Lichnerowicz, J. Math. Pures et Appl. \textbf{23}, 37 (1944)

\bibitem{Nakamura1994}
T.~Nakamura, in \emph{{8th Nishinomiya-Yukawa Memorial Symposium: Relativistic
  Cosmology}} (1994)

\bibitem{Bauswein2010}
A.~{Bauswein}, R.~{Oechslin}, H.T. {Janka}, Phys. Rev. D \textbf{81}(2), 024012
  (2010).
\newblock \doi{10.1103/PhysRevD.81.024012}

\bibitem{Bauswein2014}
A.~{Bauswein}, N.~{Stergioulas}, H.T. {Janka}, Phys. Rev. D \textbf{90}(2),
  023002 (2014).
\newblock \doi{10.1103/PhysRevD.90.023002}

\bibitem{Bonazzola:2003dm}
S.~Bonazzola, E.~Gourgoulhon, P.~Grandclement, J.~Novak, Phys. Rev. D
  \textbf{70}, 104007 (2004)

\bibitem{Shibata04a}
M.~Shibata, K.~Uryu, J.L. Friedman, Phys. Rev. \textbf{D70}, 044044 (2004).
\newblock \doi{10.1103/PhysRevD.70.044044, 10.1103/PhysRevD.70.129901}.
\newblock [Erratum: Phys. Rev.D70,129901(2004)]

\bibitem{Arnowitt1960}
R.~{Arnowitt}, S.~{Deser}, C.W. {Misner}, Physical Review \textbf{118}(4), 1100
  (1960).
\newblock \doi{10.1103/PhysRev.118.1100}

\bibitem{Arnowitt62unfindable}
R.~Arnowitt, S.~Deser, C.W. Misner, in \emph{Gravitation: An introduction to
  current research}, ed. by L.~Witten (John Wiley, New York, 1962), pp.
  227--265

\bibitem{Friedman01a}
J.L. Friedman, K.~Uryu, M.~Shibata, Phys. Rev. \textbf{D65}, 064035 (2002).
\newblock \doi{10.1103/PhysRevD.70.129904, 10.1103/PhysRevD.65.064035}.
\newblock [Erratum: Phys. Rev.D70,129904(2004)]

\bibitem{Baumgarte97}
T.W. Baumgarte, G.B. Cook, M.A. Scheel, S.L. Shapiro, S.A. Teukolsky, Phys.
  Rev. Lett. \textbf{79}, 1182 (1997)

\bibitem{Baumgarte98b}
T.W. {Baumgarte}, G.B. {Cook}, M.A. {Scheel}, S.L. {Shapiro}, S.A. {Teukolsky},
  Phys. Rev. D \textbf{57}, 7299 (1998).
\newblock \doi{10.1103/PhysRevD.57.7299}

\bibitem{Thorne1967}
K.S. Thorne, pp 259-441 of High Energy Astrophysics. Vol. III. DeWitt, C.
  Schatzman, E. Veron, P. (eds.). New York, Gordon and Breach, Science
  Publishers, 1967.  (1968)

\bibitem{Kochanek92}
C.S. {Kochanek}, Astrophys. J. \textbf{398}, 234 (1992).
\newblock \doi{10.1086/171851}

\bibitem{Bildsten92}
L.~{Bildsten}, C.~{Cutler}, Astrophys. J. \textbf{400}, 175 (1992).
\newblock \doi{10.1086/171983}

\bibitem{Mathews97}
G.~Mathews, J.~Wilson, Astrophys. J. \textbf{482}, 929 (1997)

\bibitem{Lai96}
D.~Lai, Phys. Rev. Lett. \textbf{76}, 4878 (1996)

\bibitem{Wiseman97}
A.G. Wiseman, Phys. Rev. Lett. \textbf{79}, 1189 (1997)

\bibitem{Brady97}
P.R. Brady, S.A. Hughes, Phys. Rev. Lett. \textbf{79}, 1186 (1997)

\bibitem{Shibata98b}
M.~Shibata, T.W. Baumgarte, S.L. Shapiro, Phys. Rev. D \textbf{58}, 023002
  (1998)

\bibitem{Baumgarte98c}
T.W. Baumgarte, G.B. Cook, M.A. Scheel, S.L. Shapiro, S.A. Teukolsky, Phys.
  Rev. D \textbf{57}(10), 6181 (1998)

\bibitem{Flanagan:1998c}
E.E. Flanagan, Phys. Rev. Lett. \textbf{82}, 1354 (1999).
\newblock \doi{10.1103/PhysRevLett.82.1354}

\bibitem{Mathews00}
G.J. Mathews, J.R. Wilson, Phys. Rev. D \textbf{61}, 127304 (2000)

\bibitem{Read:2009a}
J.S. {Read}, B.D. {Lackey}, B.J. {Owen}, J.L. {Friedman}, Phys. Rev. D
  \textbf{79}(12), 124032 (2009).
\newblock \doi{10.1103/PhysRevD.79.124032}

\bibitem{Mathews98}
G.J. Mathews, P.~Marronetti, J.R. Wilson, Phys. Rev. D \textbf{58}, 043003
  (1998)

\bibitem{Marronetti98}
P.~Marronetti, G.J. Mathews, J.R. Wilson, Phys. Rev. D \textbf{58}, 107503
  (1998)

\bibitem{Miller:2003vc}
M.~Miller, P.~Gressman, W.M. Suen, Phys. Rev. D \textbf{69}, 064026 (2004)

\bibitem{Bonazzola97}
S.~{Bonazzola}, E.~{Gourgoulhon}, J.A. {Marck}, Phys. Rev. D \textbf{56}, 7740
  (1997).
\newblock \doi{10.1103/PhysRevD.56.7740}

\bibitem{Asada1998}
H.~{Asada}, Phys. Rev. D \textbf{57}, 7292 (1998).
\newblock \doi{10.1103/PhysRevD.57.7292}

\bibitem{Shibata98}
M.~{Shibata}, Phys. Rev. D \textbf{58}(2), 024012 (1998).
\newblock \doi{10.1103/PhysRevD.58.024012}

\bibitem{Teukolsky98}
S.A. {Teukolsky}, Astrophys. J. \textbf{504}, 442 (1998).
\newblock \doi{10.1086/306082}

\bibitem{Gourgoulhon01}
E.~Gourgoulhon, P.~Grandcl{\'e}ment, K.~Taniguchi, J.A. Marck, S.~Bonazzola,
  Phys. Rev. D \textbf{63}, 064029 (2001)

\bibitem{Bonazzola98b}
S.~Bonazzola, E.~Gourgoulhon, J.A. Marck, Phys. Rev. Lett. \textbf{82}, 892
  (1999)

\bibitem{Marronetti99}
P.~Marronetti, G.J. Mathews, J.R. Wilson, Phys. Rev. D \textbf{60}, 087301
  (1999)

\bibitem{Uryu00}
K.~{Ury{\=u}}, Y.~{Eriguchi}, Phys. Rev. D \textbf{61}(12), 124023 (2000).
\newblock \doi{10.1103/PhysRevD.61.124023}

\bibitem{Gourgoulhon-etal-2000:2ns-initial-data}
E.~{Gourgoulhon}, P.~{Grandcl{\'e}ment}, K.~{Taniguchi}, J.A. {Marck},
  S.~{Bonazzola}, Phys. Rev. D \textbf{63}(6), 064029 (2001).
\newblock \doi{10.1103/PhysRevD.63.064029}

\bibitem{Uryu00a}
K.~{Ury{\=u}}, M.~{Shibata}, Y.~{Eriguchi}, Phys. Rev. D \textbf{62}(10),
  104015 (2000).
\newblock \doi{10.1103/PhysRevD.62.104015}

\bibitem{Usui2000}
F.~{Usui}, K.~{Ury{\={u}}}, Y.~{Eriguchi}, Phys. Rev. D \textbf{61}(2), 024039
  (2000).
\newblock \doi{10.1103/PhysRevD.61.024039}

\bibitem{Taniguchi02b}
K.~{Taniguchi}, E.~{Gourgoulhon}, Phys. Rev. D \textbf{66}(10), 104019 (2002).
\newblock \doi{10.1103/PhysRevD.66.104019}

\bibitem{Faber:2002zn}
J.A. Faber, P.~Grandclement, F.A. Rasio, K.~Taniguchi, Phys. Rev. Lett.
  \textbf{89}, 231102 (2002).
\newblock \doi{10.1103/PhysRevLett.89.231102}

\bibitem{Taniguchi03}
K.~{Taniguchi}, E.~{Gourgoulhon}, Phys. Rev. D \textbf{68}(12), 124025 (2003).
\newblock \doi{10.1103/PhysRevD.68.124025}

\bibitem{Usui2002}
F.~Usui, Y.~Eriguchi, Phys. Rev. D \textbf{65}, 064030 (2002).
\newblock \doi{10.1103/PhysRevD.65.064030}.
\newblock \urlprefix\url{https://link.aps.org/doi/10.1103/PhysRevD.65.064030}

\bibitem{Bejger04}
M.~Bejger, D.~Gondek-Rosinska, E.~Gourgoulhon, P.~Haensel, K.~Taniguchi,
  J.~Zdunik, Astron. Astrophys. \textbf{431}, 297 (2005).
\newblock \doi{10.1051/0004-6361:20041441}

\bibitem{lorene_web}
{LORENE} {Langage Objet pour la RElativit\'e Num\'eriquE}.
\newblock {\tt http://www.lorene.obspm.fr}

\bibitem{Taniguchi2010}
K.~{Taniguchi}, M.~{Shibata}, Astrophys. J., Supp. \textbf{188}, 187 (2010).
\newblock \doi{10.1088/0067-0049/188/1/187}

\bibitem{Hessels2006}
J.W. {Hessels}, S.M. {Ransom}, I.H. {Stairs}, P.C. {Freire}, V.M. {Kaspi},
  F.~{Camillo}, Science \textbf{311}(5769), 1901 (2006).
\newblock \doi{10.1126/science.1123430}

\bibitem{Tauris2017}
T.M. {Tauris}, M.~{Kramer}, P.C.C. {Freire}, N.~{Wex}, H.T. {Janka},
  N.~{Langer}, P.~{Podsiadlowski}, E.~{Bozzo}, S.~{Chaty}, M.U. {Kruckow},
  E.P.J. {van den Heuvel}, J.~{Antoniadis}, R.P. {Breton}, D.J. {Champion},
  Astrophys. J. \textbf{846}, 170 (2017).
\newblock \doi{10.3847/1538-4357/aa7e89}

\bibitem{ZhuX2018}
X.~{Zhu}, E.~{Thrane}, Y.~{Os{\l}owski}, Stefan~and{Levin}, P.D. {Lasky}, Phys.
  Rev. D \textbf{98}, 043002 (2018).
\newblock \doi{10.1103/PhysRevD.98.043002}

\bibitem{Stovall2018}
K.~{Stovall}, P.C.C. {Freire}, S.~{Chatterjee}, P.B. {Demorest}, D.R.
  {Lorimer}, M.A. {McLaughlin}, N.~{Pol}, J.~{van Leeuwen}, R.S. {Wharton},
  B.~{Allen}, M.~{Boyce}, A.~{Brazier}, K.~{Caballero}, F.~{Camilo},
  R.~{Camuccio}, J.M. {Cordes}, F.~{Crawford}, J.S. {Deneva}, R.D. {Ferdman},
  J.W.T. {Hessels}, F.A. {Jenet}, V.M. {Kaspi}, B.~{Knispel}, P.~{Lazarus},
  R.~{Lynch}, E.~{Parent}, C.~{Patel}, Z.~{Pleunis}, S.M. {Ransom},
  P.~{Scholz}, A.~{Seymour}, X.~{Siemens}, I.H. {Stairs}, J.~{Swiggum}, W.W.
  {Zhu}, Astrophys. J. Letters \textbf{854}(2), L22 (2018).
\newblock \doi{10.3847/2041-8213/aaad06}

\bibitem{Cameron2018}
A.D. {Cameron}, D.J. {Champion}, M.~{Kramer}, M.~{Bailes}, E.D. {Barr}, C.G.
  {Bassa}, S.~{Bhandari}, N.D.R. {Bhat}, M.~{Burgay}, S.~{Burke-Spolaor}, R.P.
  {Eatough}, C.M.L. {Flynn}, P.C.C. {Freire}, A.~{Jameson}, S.~{Johnston},
  R.~{Karuppusamy}, M.J. {Keith}, L.~{Levin}, D.R. {Lorimer}, A.G. {Lyne}, M.A.
  {McLaughlin}, C.~{Ng}, E.~{Petroff}, A.~{Possenti}, A.~{Ridolfi}, B.W.
  {Stappers}, W.~{van Straten}, T.M. {Tauris}, C.~{Tiburzi}, N.~{Wex}, Mon.
  Not. R. Astron. Soc. \textbf{475}(1), L57 (2018).
\newblock \doi{10.1093/mnrasl/sly003}

\bibitem{Kramer2006}
M.~{Kramer}, I.H. {Stairs}, R.N. {Manchester}, M.A. {McLaughlin}, A.G. {Lyne},
  R.D. {Ferdman}, M.~{Burgay}, D.R. {Lorimer}, A.~{Possenti}, N.~{D'Amico},
  J.M. {Sarkissian}, G.B. {Hobbs}, J.E. {Reynolds}, P.C.C. {Freire},
  F.~{Camilo}, Science \textbf{314}, 97 (2006).
\newblock \doi{10.1126/science.1132305}

\bibitem{Marronetti03}
P.~{Marronetti}, S.L. {Shapiro}, Phys. Rev. D \textbf{68}(10), 104024 (2003).
\newblock \doi{10.1103/PhysRevD.68.104024}

\bibitem{Tichy11}
W.~{Tichy}, Phys. Rev. D \textbf{84}(2), 024041 (2011).
\newblock \doi{10.1103/PhysRevD.84.024041}

\bibitem{Carter1979_conservation_laws}
B.~{Carter}, in \emph{Active Galactic Nuclei}, ed. by C.~{Hazard}, S.~{Mitton}
  (Cambridge University Press, Cambridge, England, 1979), pp. 273--300

\bibitem{Baumgarte:2009}
T.W. {Baumgarte}, S.L. {Shapiro}, Phys. Rev. D \textbf{80}(6), 064009 (2009).
\newblock \doi{10.1103/PhysRevD.80.064009}

\bibitem{Baumgarte:2009e}
T.W. {Baumgarte}, S.L. {Shapiro}, Phys. Rev. D \textbf{80}, 089901 (2009).
\newblock \doi{10.1103/PhysRevD.80.089901}

\bibitem{Tsokaros2018}
A.~Tsokaros, K.~Uryu, M.~Ruiz, S.L. Shapiro, Phys. Rev. \textbf{D98}(12),
  124019 (2018).
\newblock \doi{10.1103/PhysRevD.98.124019}

\bibitem{Tichy12}
W.~{Tichy}, Phys. Rev. D \textbf{86}(6), 064024 (2012).
\newblock \doi{10.1103/PhysRevD.86.064024}

\bibitem{Tichy:2009}
W.~Tichy, Class. Quant. Grav. \textbf{26}, 175018 (2009).
\newblock \doi{10.1088/0264-9381/26/17/175018}

\bibitem{Dietrich:2015b}
T.~{Dietrich}, N.~{Moldenhauer}, N.K. {Johnson-McDaniel}, S.~{Bernuzzi}, C.M.
  {Markakis}, B.~{Br{\"u}gmann}, W.~{Tichy}, Phys. Rev. D \textbf{92}(12),
  124007 (2015).
\newblock \doi{10.1103/PhysRevD.92.124007}

\bibitem{Tichy2019}
W.~Tichy, A.~Rashti, T.~Dietrich, R.~Dudi, B.~Br\"ugmann, Phys. Rev. D
  \textbf{100}(12), 124046 (2019).
\newblock \doi{10.1103/PhysRevD.100.124046}

\bibitem{Tsatsin2013}
P.~Tsatsin, P.~Marronetti, Phys. Rev. D \textbf{88}, 064060 (2013).
\newblock \doi{10.1103/PhysRevD.88.064060}.
\newblock \urlprefix\url{http://link.aps.org/doi/10.1103/PhysRevD.88.064060}

\bibitem{East2012d}
W.E. {East}, F.M. {Ramazano{\v g}lu}, F.~{Pretorius}, Phys. Rev. D
  \textbf{86}(10), 104053 (2012).
\newblock \doi{10.1103/PhysRevD.86.104053}

\bibitem{Kastaun2013}
W.~{Kastaun}, F.~{Galeazzi}, D.~{Alic}, L.~{Rezzolla}, J.A. {Font}, Phys. Rev.
  D \textbf{88}(2), 021501 (2013).
\newblock \doi{10.1103/PhysRevD.88.021501}

\bibitem{Tsokaros2015}
A.~{Tsokaros}, K.~{Ury{\={u}}}, L.~{Rezzolla}, Phys. Rev. D \textbf{91}(10),
  104030 (2015).
\newblock \doi{10.1103/PhysRevD.91.104030}

\bibitem{Huang08}
X.~Huang, C.~Markakis, N.~Sugiyama, K.~Ury{\={u}}, Phys. Rev. D \textbf{D78},
  124023 (2008).
\newblock \doi{10.1103/PhysRevD.78.124023}

\bibitem{Uryu2016a}
K.~Ury{\={u}}, A.~Tsokaros, F.~Galeazzi, H.~Hotta, M.~Sugimura, K.~Taniguchi,
  S.~Yoshida, Phys. Rev. D \textbf{D93}(4), 044056 (2016).
\newblock \doi{10.1103/PhysRevD.93.044056}

\bibitem{Uryu2016b}
K.~{Ury{\=u}}, A.~{Tsokaros}, L.~{Baiotti}, F.~{Galeazzi}, N.~{Sugiyama},
  K.~{Taniguchi}, S.~{Yoshida}, Phys. Rev. D \textbf{94}(10), 101302 (2016).
\newblock \doi{10.1103/PhysRevD.94.101302}

\bibitem{Uryu2017}
K.~{Uryu}, A.~{Tsokaros}, L.~{Baiotti}, F.~{Galeazzi}, K.~{Taniguchi},
  S.~{Yoshida}, ArXiv e-prints  (2017)

\bibitem{Zhou2017xhf}
E.~Zhou, A.~Tsokaros, L.~Rezzolla, R.~Xu, K.~Uryū, Phys. Rev. \textbf{D97}(2),
  023013 (2018).
\newblock \doi{10.1103/PhysRevD.97.023013}

\bibitem{Zhou2019hyy}
E.~Zhou, A.~Tsokaros, K.~Uryu, R.~Xu, M.~Shibata, Phys. Rev. \textbf{D100}(4),
  043015 (2019).
\newblock \doi{10.1103/PhysRevD.100.043015}

\bibitem{Uryu2012}
K.~{Ury{\={u}}}, A.~{Tsokaros}, Phys. Rev. D \textbf{85}(6), 064014 (2012).
\newblock \doi{10.1103/PhysRevD.85.064014}

\bibitem{Tsokaros2012}
A.~Tsokaros, K.~Ury{\={u}}, Journal of Engineering Mathematics \textbf{82}(1),
  133 (2012).
\newblock \doi{10.1007/s10665-012-9585-6}.
\newblock \urlprefix\url{http://dx.doi.org/10.1007/s10665-012-9585-6}

\bibitem{Uryu:2012b}
K.~Ury{\={u}}, A.~Tsokaros, P.~Grandclement, Phys. Rev. D \textbf{D86}, 104001
  (2012).
\newblock \doi{10.1103/PhysRevD.86.104001}

\bibitem{Uryu2014}
K.~Uryu, E.~Gourgoulhon, C.~Markakis, K.~Fujisawa, A.~Tsokaros, Y.~Eriguchi,
  Phys. Rev. \textbf{D90}(10), 101501 (2014).
\newblock \doi{10.1103/PhysRevD.90.101501}

\bibitem{Uryu2019}
K.~Uryu, S.~Yoshida, E.~Gourgoulhon, C.~Markakis, K.~Fujisawa, A.~Tsokaros,
  K.~Taniguchi, Y.~Eriguchi, Phys. Rev. \textbf{D100}(12), 123019 (2019).
\newblock \doi{10.1103/PhysRevD.100.123019}

\bibitem{Tsokaros2018a}
A.~Tsokaros, K.~Uryu, S.L. Shapiro, Phys. Rev. \textbf{D99}(4), 041501 (2019).
\newblock \doi{10.1103/PhysRevD.99.041501}

\bibitem{Tsokaros2007}
A.A. Tsokaros, K.~Ury{\={u}}, Phys. Rev. D \textbf{75}, 044026 (2007).
\newblock \doi{10.1103/PhysRevD.75.044026}.
\newblock \urlprefix\url{http://link.aps.org/doi/10.1103/PhysRevD.75.044026}

\bibitem{Tsokaros2016}
A.~{Tsokaros}, B.C. {Mundim}, F.~{Galeazzi}, L.~{Rezzolla}, K.~{Ury{\=u}},
  arXiv:1605.07205  (2016)

\bibitem{Tsokaros:2019lnx}
A.~Tsokaros, M.~Ruiz, S.L. Shapiro, L.~Sun, K.~Ury\={u}, Phys. Rev. Lett.
  \textbf{124}(7), 071101 (2020).
\newblock \doi{10.1103/PhysRevLett.124.071101}

\bibitem{Tacik15}
N.~{Tacik}, F.~{Foucart}, H.P. {Pfeiffer}, R.~{Haas}, S.~{Ossokine},
  J.~{Kaplan}, C.~{Muhlberger}, M.D. {Duez}, L.E. {Kidder}, M.A. {Scheel},
  B.~{Szil{\'a}gyi}, Phys. Rev. D \textbf{92}(12), 124012 (2015).
\newblock \doi{10.1103/PhysRevD.92.124012}

\bibitem{Tacik16}
N.~Tacik, F.~Foucart, H.P. Pfeiffer, R.~Haas, S.~Ossokine, J.~Kaplan,
  C.~Muhlberger, M.D. Duez, L.E. Kidder, M.A. Scheel, B.~Szil\'agyi, Phys. Rev.
  D \textbf{94}, 049903 (2016).
\newblock \doi{10.1103/PhysRevD.94.049903}.
\newblock \urlprefix\url{https://link.aps.org/doi/10.1103/PhysRevD.94.049903}

\bibitem{Pfeiffer:2002wt}
H.P. Pfeiffer, L.E. Kidder, M.A. Scheel, S.A. Teukolsky, Comput. Phys. Commun.
  \textbf{152}, 253 (2003)

\bibitem{Foucart2008}
F.~{Foucart}, L.E. {Kidder}, H.P. {Pfeiffer}, S.A. {Teukolsky}, Phys. Rev. D
  \textbf{77}(12), 124051 (2008).
\newblock \doi{10.1103/PhysRevD.77.124051}

\bibitem{Cook:2004kt}
G.B. Cook, H.P. Pfeiffer, Phys. Rev. D \textbf{70} (2004)

\bibitem{Pfeiffer:2004qz}
H.P. Pfeiffer, L.E. Kidder, M.A. Scheel, D.~Shoemaker, Phys. Rev. D
  \textbf{71}, 024020 (2005)

\bibitem{Caudill:2006hw}
M.~Caudill, G.B. Cook, J.D. Grigsby, H.P. Pfeiffer, Phys. Rev. D \textbf{74},
  064011 (2006)

\bibitem{Boyle:2006ne}
M.~Boyle, L.~Lindblom, H.~Pfeiffer, M.~Scheel, L.E. Kidder, Phys.Rev.
  \textbf{D75}, 024006 (2007).
\newblock \doi{10.1103/PhysRevD.75.024006}

\bibitem{Lovelace2008c}
G.~{Lovelace}, R.~{Owen}, H.P. {Pfeiffer}, T.~{Chu}, Phys. Rev. D
  \textbf{78}(8), 084017 (2008).
\newblock \doi{10.1103/PhysRevD.78.084017}

\bibitem{Brown1993}
J.D. Brown, J.W. York, Phys. Rev. D \textbf{47}, 1407 (1993).
\newblock \doi{10.1103/PhysRevD.47.1407}.
\newblock \urlprefix\url{https://link.aps.org/doi/10.1103/PhysRevD.47.1407}

\bibitem{Ashtekar01a}
A.~Ashtekar, C.~Beetle, J.~Lewandowski, Phys. Rev. D \textbf{64}, 044016 (2001)

\bibitem{Ashtekar03a}
A.~Ashtekar, B.~Krishnan, Phys. Rev. D \textbf{68}, 104030 (2003)

\bibitem{Pfeiffer2007}
H.P. Pfeiffer, D.A. Brown, L.E. Kidder, L.~Lindblom, G.~Lovelace, M.A. Scheel,
  Class. Quant. Grav. \textbf{24}, S59 (2007).
\newblock \doi{10.1088/0264-9381/24/12/S06}

\bibitem{Tsao2020}
B.J. Tsao, R.~Haas, A.~Tsokaros, Class. Quant. Grav. \textbf{38}(13), 135008
  (2021).
\newblock \doi{10.1088/1361-6382/abfc29}

\bibitem{Towers2018}
J.D. Towers, Journal of Computational Physics \textbf{361}, 424  (2018).
\newblock \doi{https://doi.org/10.1016/j.jcp.2018.01.038}.
\newblock
  \urlprefix\url{http://www.sciencedirect.com/science/article/pii/S0021999118300482}

\bibitem{Papenfort2021}
L.J. Papenfort, S.D. Tootle, P.~Grandcl\'ement, E.R. Most, L.~Rezzolla, Phys.
  Rev. D \textbf{104}(2), 024057 (2021).
\newblock \doi{10.1103/PhysRevD.104.024057}

\bibitem{kadath}
\urlprefix\url{https://kadath.obspm.fr}

\bibitem{Grandclement09}
P.~Grandclement, J. Comput. Phys. \textbf{229}, 3334 (2010).
\newblock \doi{10.1016/j.jcp.2010.01.005}

\bibitem{Favata2014}
M.~{Favata}, Phys. Rev. Lett. \textbf{112}(10), 101101 (2014).
\newblock \doi{10.1103/PhysRevLett.112.101101}

\bibitem{Yagi2014c}
K.~Yagi, N.~Yunes, Phys. Rev. D \textbf{89}, 021303 (2014).
\newblock \doi{10.1103/PhysRevD.89.021303}.
\newblock \urlprefix\url{https://link.aps.org/doi/10.1103/PhysRevD.89.021303}

\bibitem{Wade2014}
L.~{Wade}, J.D.E. {Creighton}, E.~{Ochsner}, B.D. {Lackey}, B.F. {Farr}, T.B.
  {Littenberg}, V.~{Raymond}, Phys. Rev. D \textbf{89}(10), 103012 (2014).
\newblock \doi{10.1103/PhysRevD.89.103012}

\bibitem{Kyutoku2014}
K.~{Kyutoku}, M.~{Shibata}, K.~{Taniguchi}, Phys. Rev. D \textbf{90}(6), 064006
  (2014).
\newblock \doi{10.1103/PhysRevD.90.064006}

\bibitem{Boyle:2007ft}
M.~Boyle, D.A. Barrow, L.E. Kidder, A.H. {Mrou\'e}, H.P. Pfeiffer, M.A. Scheel,
  G.B. Cook, S.A. Teukolsky, Phys. Rev. D \textbf{76}, 124038 (2007).
\newblock \doi{10.1103/PhysRevD.76.124038}

\bibitem{Buonanno2011}
A.~{Buonanno}, L.E. {Kidder}, A.H. {Mrou{\'e}}, H.P. {Pfeiffer},
  A.~{Taracchini}, Phys. Rev. D \textbf{83}(10), 104034 (2011).
\newblock \doi{10.1103/PhysRevD.83.104034}

\bibitem{Radice2013b}
D.~{Radice}, L.~{Rezzolla}, F.~{Galeazzi}, Mon. Not. R. Astron. Soc. L.
  \textbf{437}, L46 (2014).
\newblock \doi{10.1093/mnrasl/slt137}

\bibitem{Radice2013c}
D.~{Radice}, L.~{Rezzolla}, F.~{Galeazzi}, Class. Quantum Grav. \textbf{31}(7),
  075012 (2014).
\newblock \doi{10.1088/0264-9381/31/7/075012}

\bibitem{Peters:1964}
P.C. Peters, Phys. Rev. \textbf{136}, B1224 (1964)

\bibitem{Oleary2009}
R.M. {O'Leary}, B.~{Kocsis}, A.~{Loeb}, Mon. Not. R. Astron. Soc. \textbf{395},
  2127 (2009).
\newblock \doi{10.1111/j.1365-2966.2009.14653.x}

\bibitem{Lee2010}
W.H. {Lee}, E.~{Ramirez-Ruiz}, G.~{van de Ven}, Astrophys. J. \textbf{720}, 953
  (2010).
\newblock \doi{10.1088/0004-637X/720/1/953}

\bibitem{Tsang2013}
D.~Tsang, Astrophys. J. \textbf{777}, 103 (2013).
\newblock \doi{10.1088/0004-637X/777/2/103}

\bibitem{Antonini2012}
F.~{Antonini}, H.B. {Perets}, Astrophys. J. \textbf{757}, 27 (2012).
\newblock \doi{10.1088/0004-637X/757/1/27}

\bibitem{Naoz2012}
S.~Naoz, B.~Kocsis, A.~Loeb, N.~Yunes, Astrophys. J. \textbf{773}, 187 (2013).
\newblock \doi{10.1088/0004-637X/773/2/187}

\bibitem{Seto2013}
N.~Seto, Phys. Rev. Lett. \textbf{111}, 061106 (2013).
\newblock \doi{10.1103/PhysRevLett.111.061106}.
\newblock
  \urlprefix\url{https://link.aps.org/doi/10.1103/PhysRevLett.111.061106}

\bibitem{Antognini2014}
J.M. Antognini, B.J. Shappee, T.A. Thompson, P.~Amaro-Seoane, Monthly Notices
  of the Royal Astronomical Society \textbf{439}(1), 1079 (2014).
\newblock \doi{10.1093/mnras/stu039}.
\newblock \urlprefix\url{https://doi.org/10.1093/mnras/stu039}

\bibitem{Gold2012}
R.~{Gold}, S.~{Bernuzzi}, M.~{Thierfelder}, B.~{Br{\"u}gmann}, F.~{Pretorius},
  Phys. Rev. D \textbf{86}(12), 121501 (2012).
\newblock \doi{10.1103/PhysRevD.86.121501}

\bibitem{East2012c}
W.E. {East}, F.~{Pretorius}, Astrophys. J. \textbf{760}, L4 (2012).
\newblock \doi{10.1088/2041-8205/760/1/L4}

\bibitem{Moldenhauer2014}
N.~{Moldenhauer}, C.M. {Markakis}, N.K. {Johnson-McDaniel}, W.~{Tichy},
  B.~{Br{\"u}gmann}, Phys. Rev. D \textbf{90}(8), 084043 (2014).
\newblock \doi{10.1103/PhysRevD.90.084043}

\bibitem{Shibata95}
M.~{Shibata}, T.~{Nakamura}, Phys. Rev. D \textbf{52}, 5428 (1995).
\newblock \doi{10.1103/PhysRevD.52.5428}

\bibitem{Baumgarte99}
T.W. {Baumgarte}, S.L. {Shapiro}, Phys. Rev. D \textbf{59}(2), 024007 (1999).
\newblock \doi{10.1103/PhysRevD.59.024007}

\bibitem{Dirac59a}
P.A.M. Dirac, Phys. Rev. \textbf{114}, 924 (1959)

\bibitem{Beig1978}
R.~{Beig}, Physics Letters A \textbf{69}(3), 153 (1978).
\newblock \doi{10.1016/0375-9601(78)90198-6}

\bibitem{Ashtekar79a}
A.~Ashtekar, A.~Magnon-Ashtekar, J. Math. Phys. \textbf{20}(5), 793 (1979)

\bibitem{Uryu2006}
K.~{Ury{\=u}}, F.~{Limousin}, J.L. {Friedman}, E.~{Gourgoulhon}, M.~{Shibata},
  Phys. Rev. Lett. \textbf{97}(17), 171101 (2006).
\newblock \doi{10.1103/PhysRevLett.97.171101}

\bibitem{Uryu:2009ye}
K.~Ury{\=u}, F.~Limousin, J.L. Friedman, E.~Gourgoulhon, M.~Shibata, Phys. Rev.
  D \textbf{80}, 124004 (2009).
\newblock \doi{10.1103/PhysRevD.80.124004}

\bibitem{Yoshida06a}
S.~Yoshida, B.C. Bromley, J.S. Read, K.~Uryu, J.L. Friedman, Class. Quant.
  Grav. \textbf{23}, S599 (2006).
\newblock \doi{10.1088/0264-9381/23/16/S16}

\bibitem{Woosley1993}
S.E. {Woosley}, Astrophys. J. \textbf{405}, 273 (1993).
\newblock \doi{10.1086/172359}

\bibitem{MacFadyen1999}
A.I. {MacFadyen}, S.E. {Woosley}, Astrop. J. \textbf{524}, 262 (1999).
\newblock \doi{10.1086/307790}

\bibitem{LyndenBell1969}
D.~{Lynden-Bell}, Nature \textbf{223}(5207), 690 (1969).
\newblock \doi{10.1038/223690a0}

\bibitem{Shakura1973}
N.I. {Shakura}, R.A. {Sunyaev}, Astron. Astrophys. \textbf{24}, 337 (1973)

\bibitem{Paczynski1978}
B.~{Paczynski}, Acta Astronomica \textbf{28}, 91 (1978)

\bibitem{Shibata06d}
M.~{Shibata}, K.~{Uryu}, Phys. Rev. D \textbf{74}, {121503 (R)} (2006).
\newblock \doi{10.1103/PhysRevD.74.121503}

\bibitem{Etienne2007b}
Z.B. {Etienne}, J.A. {Faber}, Y.T. {Liu}, S.L. {Shapiro}, K.~{Taniguchi}, T.W.
  {Baumgarte}, Phys. Rev. D \textbf{77}(8), 084002 (2008).
\newblock \doi{10.1103/PhysRevD.77.084002}

\bibitem{Rezzolla:2010}
L.~{Rezzolla}, L.~{Baiotti}, B.~{Giacomazzo}, D.~{Link}, J.A. {Font}, Class.
  Quantum Grav. \textbf{27}(11), 114105 (2010).
\newblock \doi{10.1088/0264-9381/27/11/114105}

\bibitem{Lovelace2013}
G.~{Lovelace}, M.D. {Duez}, F.~{Foucart}, L.E. {Kidder}, H.P. {Pfeiffer}, M.A.
  {Scheel}, B.~{Szil{\'a}gyi}, Class. Quantum Grav. \textbf{30}(13), 135004
  (2013).
\newblock \doi{10.1088/0264-9381/30/13/135004}

\bibitem{Amaro2017}
P.~{Amaro-Seoane}, H.~{Audley}, S.~{Babak}, J.~{Baker}, E.~{Barausse},
  P.~{Bender}, E.~{Berti}, P.~{Binetruy}, M.~{Born}, D.~{Bortoluzzi},
  J.~{Camp}, C.~{Caprini}, V.~{Cardoso}, M.~{Colpi}, J.~{Conklin},
  N.~{Cornish}, C.~{Cutler}, K.~{Danzmann}, R.~{Dolesi}, L.~{Ferraioli},
  V.~{Ferroni}, E.~{Fitzsimons}, J.~{Gair}, L.~{Gesa Bote}, D.~{Giardini},
  F.~{Gibert}, C.~{Grimani}, H.~{Halloin}, G.~{Heinzel}, T.~{Hertog},
  M.~{Hewitson}, K.~{Holley-Bockelmann}, D.~{Hollington}, M.~{Hueller},
  H.~{Inchauspe}, P.~{Jetzer}, N.~{Karnesis}, C.~{Killow}, A.~{Klein},
  B.~{Klipstein}, N.~{Korsakova}, S.L. {Larson}, J.~{Livas}, I.~{Lloro},
  N.~{Man}, D.~{Mance}, J.~{Martino}, I.~{Mateos}, K.~{McKenzie}, S.T.
  {McWilliams}, C.~{Miller}, G.~{Mueller}, G.~{Nardini}, G.~{Nelemans},
  M.~{Nofrarias}, A.~{Petiteau}, P.~{Pivato}, E.~{Plagnol}, E.~{Porter},
  J.~{Reiche}, D.~{Robertson}, N.~{Robertson}, E.~{Rossi}, G.~{Russano},
  B.~{Schutz}, A.~{Sesana}, D.~{Shoemaker}, J.~{Slutsky}, C.F. {Sopuerta},
  T.~{Sumner}, N.~{Tamanini}, I.~{Thorpe}, M.~{Troebs}, M.~{Vallisneri},
  A.~{Vecchio}, D.~{Vetrugno}, S.~{Vitale}, M.~{Volonteri}, G.~{Wanner},
  H.~{Ward}, P.~{Wass}, W.~{Weber}, J.~{Ziemer}, P.~{Zweifel}, arXiv e-prints
  arXiv:1702.00786 (2017)

\bibitem{Papaloizou84}
J.C.B. {Papaloizou}, J.E. {Pringle}, Mon. Not. R. Astron. Soc. \textbf{208},
  721 (1984)

\bibitem{Papaloizou85}
J.C.B. {Papaloizou}, J.E. {Pringle}, Mon. Not. R. Astron. Soc. \textbf{213},
  799 (1985)

\bibitem{Blaes1985}
O.M. {Blaes}, Mon. Not. R. Astron. Soc. \textbf{216}, 553 (1985).
\newblock \doi{10.1093/mnras/216.3.553}

\bibitem{Goldreich1986}
P.~{Goldreich}, J.~{Goodman}, R.~{Narayan}, Mon. Not. R. Astron. Soc.
  \textbf{221}, 339 (1986).
\newblock \doi{10.1093/mnras/221.2.339}

\bibitem{Narayan1987}
R.~{Narayan}, P.~{Goldreich}, J.~{Goodman}, Mon. Not. R. Astron. Soc.
  \textbf{228}, 1 (1987).
\newblock \doi{10.1093/mnras/228.1.1}

\bibitem{Abramowicz1980}
M.A. {Abramowicz}, M.~{Calvani}, L.~{Nobili}, Astrophys. J. \textbf{242}, 772
  (1980).
\newblock \doi{10.1086/158512}

\bibitem{Korobkin2011}
O.~{Korobkin}, E.B. {Abdikamalov}, E.~{Schnetter}, N.~{Stergioulas}, B.~{Zink},
  Phys. Rev. D \textbf{83}(4), 043007 (2011).
\newblock \doi{10.1103/PhysRevD.83.043007}

\bibitem{Korobkin2013}
O.~{Korobkin}, E.~{Abdikamalov}, N.~{Stergioulas}, E.~{Schnetter}, B.~{Zink},
  S.~{Rosswog}, C.D. {Ott}, Mon. Not. R. Astron. Soc. \textbf{431}, 349 (2013).
\newblock \doi{10.1093/mnras/stt166}

\bibitem{Mewes2016}
V.~{Mewes}, F.~{Galeazzi}, J.A. {Font}, P.J. {Montero}, N.~{Stergioulas}, Mon.
  Not. R. Astron. Soc. \textbf{461}, 2480 (2016).
\newblock \doi{10.1093/mnras/stw1490}

\bibitem{Kiuchi2011b}
K.~{Kiuchi}, M.~{Shibata}, P.J. {Montero}, J.A. {Font}, Phys. Rev. Lett.
  \textbf{106}(25), 251102 (2011).
\newblock \doi{10.1103/PhysRevLett.106.251102}

\bibitem{Wessel2020}
E.~Wessel, V.~Paschalidis, A.~Tsokaros, M.~Ruiz, S.L. Shapiro, Phys. Rev. D
  \textbf{103}(4), 043013 (2021).
\newblock \doi{10.1103/PhysRevD.103.043013}

\bibitem{Gergely:07}
L.{\'A}. {Gergely}, P.L. {Biermann}, Astrophys. J. \textbf{697}, 1621 (2009).
\newblock \doi{10.1088/0004-637X/697/2/1621}

\bibitem{Gammie:2003qi}
C.F. Gammie, S.L. Shapiro, J.C. McKinney, Astrophys. J. \textbf{602}, 312
  (2004)

\bibitem{Petrich89}
L.I. {Petrich}, S.L. {Shapiro}, R.F. {Stark}, S.A. {Teukolsky}, Astrophys. J.
  \textbf{336}, 313 (1989).
\newblock \doi{10.1086/167013}

\bibitem{Barausse2007}
E.~Barausse, L.~Rezzolla, Phys. Rev. D \textbf{77}, 104027 (2008).
\newblock \doi{10.1103/PhysRevD.77.104027}

\bibitem{Barausse2007a}
E.~Barausse, Mon. Not. Roy. Astron. Soc. \textbf{382}, 826 (2007).
\newblock \doi{10.1111/j.1365-2966.2007.12408.x}

\bibitem{Yunes2011}
N.~Yunes, B.~Kocsis, A.~Loeb, Z.~Haiman, Phys. Rev. Lett. \textbf{107}, 171103
  (2011).
\newblock \doi{10.1103/PhysRevLett.107.171103}

\bibitem{Abbott2020}
R.~Abbott, et~al., Phys. Rev. Lett. \textbf{125}(10), 101102 (2020).
\newblock \doi{10.1103/PhysRevLett.125.101102}

\bibitem{Shibata2021}
M.~Shibata, K.~Kiuchi, S.~Fujibayashi, Y.~Sekiguchi, Phys. Rev. D
  \textbf{103}(6), 063037 (2021).
\newblock \doi{10.1103/PhysRevD.103.063037}

\bibitem{Carter1970}
B.~{Carter}, Communications in Mathematical Physics \textbf{17}(3), 233 (1970).
\newblock \doi{10.1007/BF01647092}

\bibitem{Papapetrou66}
A.~Papapetrou, Ann.Inst.H. Poincar{\'e} A \textbf{4}, 83 (1966)

\bibitem{Kundt66}
W.~{Kundt}, M.~{Tr{\"u}mper}, Zeitschrift fur Physik \textbf{192}(4), 419
  (1966).
\newblock \doi{10.1007/BF01325677}

\bibitem{Wald84}
R.M. Wald, \emph{General relativity} (The University of Chicago Press, Chicago,
  1984)

\bibitem{Bardeen71}
J.M. {Bardeen}, R.V. {Wagoner}, Astrophys. J. \textbf{167}, 359 (1971).
\newblock \doi{10.1086/151039}

\bibitem{Shibata2007}
M.~Shibata, Phys. Rev. D \textbf{76}, 064035 (2007).
\newblock \doi{10.1103/PhysRevD.76.064035}

\bibitem{Fishbone76}
L.G. {Fishbone}, V.~{Moncrief}, Astrophys. J. \textbf{207}, 962 (1976)

\bibitem{Abramowicz78}
M.~{Abramowicz}, M.~{Jaroszynski}, M.~{Sikora}, Astron. Astrophys. \textbf{63},
  221 (1978)

\bibitem{Kozlowski1978}
M.~{Kozlowski}, M.~{Jaroszynski}, M.A. {Abramowicz}, Astron. and Astrophys.
  \textbf{63}, 209 (1978)

\bibitem{Bardeen73b}
B.C. J.M.~Bardeen, S.~Hawking, Commun. Math. Phys. \textbf{31}, 161 (1973)

\bibitem{Bardeen73_Stars_Disks_Black_Holes}
J.M. Bardeen, in \emph{Black Holes} (Gordon and Breach, New York, 1973), pp.
  245--289

\bibitem{Carter73_BH_Equilibrium_States}
B.~{Carter}, in \emph{Black Holes} (Gordon and Breach, New York, 1973), pp.
  59--214

\bibitem{Smarr73a}
L.L. Smarr, Phys. Rev. Lett. \textbf{30}, 71 (1973)

\bibitem{Lanza1992a}
A.~{Lanza}, Astrophys. J. \textbf{389}, 141 (1992).
\newblock \doi{10.1086/171193}

\bibitem{Abramowicz1984}
M.A. {Abramowicz}, A.~{Curir}, A.~{Schwarzenberg-Czerny}, R.E. {Wilson}, Mon.
  Not. R. Astron. Soc. \textbf{208}, 279 (1984).
\newblock \doi{10.1093/mnras/208.2.279}

\bibitem{Chakrabarti1988}
S.K. {Chakrabarti}, Journal of Astrophysics and Astronomy \textbf{9}, 49
  (1988).
\newblock \doi{10.1007/BF02715056}

\bibitem{Will1974}
C.M. {Will}, Astrophys. J. \textbf{191}, 521 (1974).
\newblock \doi{10.1086/152992}

\bibitem{Will1975}
C.M. {Will}, Astrophys. J. \textbf{196}, 41 (1975).
\newblock \doi{10.1086/153392}

\bibitem{NishidaEriguchi94:ToroidAroundBH}
S.~{Nishida}, Y.~{Eriguchi}, Astrophys. J. \textbf{427}, 429 (1994).
\newblock \doi{10.1086/174153}

\bibitem{Komatsu89}
H.~{Komatsu}, Y.~{Eriguchi}, I.~{Hachisu}, Mon. Not. R. Astron. Soc.
  \textbf{237}, 355 (1989).
\newblock \doi{10.1093/mnras/237.2.355}

\bibitem{Komatsu89b}
H.~{Komatsu}, Y.~{Eriguchi}, I.~{Hachisu}, Mon. Not. R. Astron. Soc.
  \textbf{239}, 153 (1989)

\bibitem{Ansorg05}
M.~Ansorg, D.~Petroff, Phys. Rev. D \textbf{72}, 024019 (2005).
\newblock \doi{10.1103/PhysRevD.72.024019}

\bibitem{Ansorg:2003br}
M.~Ansorg, A.~Kleinw{\"a}chter, R.~Meinel, Astron. Astrophys. \textbf{405}, 711
  (2003)

\bibitem{Ansorg2006}
M.~Ansorg, D.~Petroff, Class. Quant. Grav. \textbf{23}, L81 (2006).
\newblock \doi{10.1088/0264-9381/23/24/L01}

\bibitem{Stergioulas2011}
N.~{Stergioulas}, International Journal of Modern Physics D \textbf{20}, 1251
  (2011).
\newblock \doi{10.1142/S021827181101944X}

\bibitem{Stergioulas2011c}
N.~{Stergioulas}, in \emph{Journal of Physics Conference Series}, \emph{Journal
  of Physics Conference Series}, vol. 283 (2011), \emph{Journal of Physics
  Conference Series}, vol. 283, p. 012036.
\newblock \doi{10.1088/1742-6596/283/1/012036}

\bibitem{Cook94b}
G.B. {Cook}, S.L. {Shapiro}, S.A. {Teukolsky}, Astrophys. J. \textbf{422}, 227
  (1994).
\newblock \doi{10.1086/173721}

\bibitem{Krivan:1998td}
W.~Krivan, R.H. Price, Phys. Rev. D \textbf{58}, 104003 (1998)

\bibitem{Butterworth1975}
E.M. {Butterworth}, J.R. {Ipser}, Astrophys. J. Lett. \textbf{200}, L103
  (1975).
\newblock \doi{10.1086/181907}

\bibitem{Hawking73a}
S.W. Hawking, G.F.R. Ellis, \emph{The large scale structure of spacetime}
  (Cambridge University Press, Cambridge, England, 1973)

\bibitem{Christodoulou70}
D.~Christodoulou, Phys. Rev. Lett. \textbf{25}(22), 1596 (1970)

\bibitem{Karkowski2017}
J.~Karkowski, W.~Kulczycki, P.~Mach, E.~Malec, A.~Odrzywo\l~ek, M.~Piróg,
  Phys. Rev. D \textbf{97}(10), 104034 (2018).
\newblock \doi{10.1103/PhysRevD.97.104034}

\bibitem{Karkowski2018}
J.~Karkowski, W.~Kulczycki, P.~Mach, E.~Malec, A.~Odrzywo\l{}ek, M.~Pir\'og,
  Phys. Rev. D \textbf{97}, 104017 (2018).
\newblock \doi{10.1103/PhysRevD.97.104017}.
\newblock \urlprefix\url{https://link.aps.org/doi/10.1103/PhysRevD.97.104017}

\bibitem{Mach2019}
P.~Mach, S.~Gimeno-Soler, J.A. Font, A.~Odrzywo\l~ek, M.~Piróg, Phys. Rev. D
  \textbf{99}(10), 104063 (2019).
\newblock \doi{10.1103/PhysRevD.99.104063}

\bibitem{Kulczycki2019}
W.~Kulczycki, P.~Mach, E.~Malec, Phys. Rev. D \textbf{99}, 024004 (2019).
\newblock \doi{10.1103/PhysRevD.99.024004}.
\newblock \urlprefix\url{https://link.aps.org/doi/10.1103/PhysRevD.99.024004}

\bibitem{Dyba2019}
W.~Dyba, W.~Kulczycki, P.~Mach, Phys. Rev. D \textbf{101}(4), 044036 (2020).
\newblock \doi{10.1103/PhysRevD.101.044036}

\bibitem{Kulczycki2021}
W.~Kulczycki, P.~Mach, E.~Malec, Phys. Rev. D \textbf{104}(2), 024005 (2021).
\newblock \doi{10.1103/PhysRevD.104.024005}

\bibitem{vonZeipel1924}
H.~{von Zeipel}, Mon. Not. Roy. Soc. \textbf{84}, 665 (1924)

\bibitem{Tassoul-1978:theory-of-rotating-stars}
J.L. Tassoul, \emph{Theory of Rotating Stars} (Princeton University Press,
  1978)

\bibitem{Abramowicz1974}
M.A. {Abramowicz}, Acta Astronomica \textbf{24}, 45 (1974)

\bibitem{Seguin1975}
F.H. {Seguin}, Astrophys. J. \textbf{197}, 745 (1975).
\newblock \doi{10.1086/153563}

\bibitem{Komissarov2006a}
S.S. {Komissarov}, Mon. Not. R. Astron. Soc. \textbf{368}, 993 (2006).
\newblock \doi{10.1111/j.1365-2966.2006.10183.x}

\bibitem{York71}
J.W. York, Phys. Rev. Lett. \textbf{26}, 1656 (1971)

\bibitem{York72}
J.W. York, Phys. Rev. Lett. \textbf{28}, 1082 (1972)

\bibitem{Pfeiffer:2002iy}
H.P. Pfeiffer, J.W. York, Phys. Rev. D \textbf{67}, 044022 (2003)

\bibitem{Brandt97b}
S.~Brandt, B.~Br{\"u}gmann, Phys. Rev. Lett. \textbf{78}(19), 3606 (1997)

\bibitem{York99}
J.W. York, Phys. Rev. Lett. \textbf{82}, 1350 (1999)

\bibitem{Chakrabarti1985}
S.K. {Chakrabarti}, Astrophys. J. \textbf{288}, 1 (1985).
\newblock \doi{10.1086/162755}

\bibitem{DeVilliers03b}
J.P.D. Villiers, J.F. Hawley, J.H. Krolik, The Astrophysical Journal
  \textbf{599}(2), 1238 (2003).
\newblock \doi{10.1086/379509}.
\newblock \urlprefix\url{https://doi.org/10.1086%2F379509}

\bibitem{Duncan1992}
R.C. {Duncan}, C.~{Thompson}, Astrophys. J. Lett. \textbf{392}, L9 (1992).
\newblock \doi{10.1086/186413}

\bibitem{ThompsonDuncan1995}
C.~{Thompson}, R.C. {Duncan}, Mon. Not. R. Astron. Soc. \textbf{275}, 255
  (1995)

\bibitem{ThompsonDuncan1996}
C.~{Thompson}, R.C. {Duncan}, Astrophys. J. \textbf{473}, 322 (1996).
\newblock \doi{10.1086/178147}

\bibitem{Mazets1979}
E.P. {Mazets}, S.V. {Golentskii}, V.N. {Ilinskii}, R.L. {Aptekar}, I.A.
  {Guryan}, Nature \textbf{282}(5739), 587 (1979).
\newblock \doi{10.1038/282587a0}

\bibitem{Paczynski92}
B.~Paczynski, Acta. Astron. \textbf{42}, 145 (1992)

\bibitem{Kouveliotou1994}
C.~{Kouveliotou}, G.J. {Fishman}, C.A. {Meegan}, W.S. {Paciesas}, J.~{van
  Paradijs}, J.P. {Norris}, R.D. {Preece}, M.S. {Briggs}, J.M. {Horack}, G.N.
  {Pendleton}, D.A. {Green}, Nature \textbf{368}(6467), 125 (1994).
\newblock \doi{10.1038/368125a0}

\bibitem{Kouveliotou1998}
C.~{Kouveliotou}, S.~{Dieters}, T.~{Strohmayer}, J.~{van Paradijs}, G.J.
  {Fishman}, C.A. {Meegan}, K.~{Hurley}, J.~{Kommers}, I.~{Smith}, D.~{Frail},
  T.~{Murakami}, Nature \textbf{393}(6682), 235 (1998).
\newblock \doi{10.1038/30410}

\bibitem{Kouveliotou1999}
C.~{Kouveliotou}, T.~{Strohmayer}, K.~{Hurley}, J.~{van Paradijs}, M.H.
  {Finger}, S.~{Dieters}, P.~{Woods}, C.~{Thompson}, R.C. {Duncan}, Astrophys.
  J. Letters \textbf{510}(2), L115 (1999).
\newblock \doi{10.1086/311813}

\bibitem{Mereghetti2008}
S.~{Mereghetti}, Astron. Astrophys. Review \textbf{15}, 225 (2008).
\newblock \doi{10.1007/s00159-008-0011-z}

\bibitem{Harding2006}
A.K. {Harding}, D.~{Lai}, Reports on Progress in Physics \textbf{69}, 2631
  (2006).
\newblock \doi{10.1088/0034-4885/69/9/R03}

\bibitem{Chandrasekhar81}
S.~Chandrasekhar, \emph{Hydrodynamic and hydromagnetic stability} (Dover
  Edition, New York, USA, 1981)

\bibitem{Bodo1994}
G.~{Bodo}, S.~{Massaglia}, A.~{Ferrari}, E.~{Trussoni}, Astron. Astrophys.
  \textbf{283}, 655 (1994)

\bibitem{Price06}
D.J. {Price}, S.~{Rosswog}, Science \textbf{312}, 719 (2006).
\newblock \doi{10.1126/science.1125201}

\bibitem{Anderson2008}
M.~{Anderson}, E.W. {Hirschmann}, L.~{Lehner}, S.L. {Liebling}, P.M. {Motl},
  D.~{Neilsen}, C.~{Palenzuela}, J.E. {Tohline}, Phys. Rev. Lett.
  \textbf{100}(19), 191101 (2008).
\newblock \doi{10.1103/PhysRevLett.100.191101}

\bibitem{Kiuchi2015a}
K.~{Kiuchi}, P.~{Cerd{\'a}-Dur{\'a}n}, K.~{Kyutoku}, Y.~{Sekiguchi},
  M.~{Shibata}, Phys. Rev. D \textbf{92}(12), 124034 (2015).
\newblock \doi{10.1103/PhysRevD.92.124034}

\bibitem{Velikhov1959}
E.P. {Velikhov}, Sov. Phys. JETP \textbf{9}, 995 (1959)

\bibitem{Chandrasekhar1960}
S.~{Chandrasekhar}, Proc. Natl. Acad. Sci. \textbf{46}, 253 (1960).
\newblock \doi{10.1073/pnas.46.2.253}

\bibitem{Balbus1991}
S.A. {Balbus}, J.F. {Hawley}, Astrophys. J. \textbf{376}, 214 (1991).
\newblock \doi{10.1086/170270}

\bibitem{Shibata:2005mz}
M.~Shibata, M.D. Duez, Y.T. Liu, S.L. Shapiro, B.C. Stephens, Phys. Rev. Lett.
  \textbf{96}, 031102 (2006).
\newblock \doi{10.1103/PhysRevLett.96.031102}

\bibitem{Duez2006a}
M.D. {Duez}, Y.T. {Liu}, S.L. {Shapiro}, M.~{Shibata}, B.C. {Stephens}, Phys.
  Rev. Lett. \textbf{96}(3), 031101 (2006).
\newblock \doi{10.1103/PhysRevLett.96.031101}

\bibitem{Siegel2013}
D.M. {Siegel}, R.~{Ciolfi}, A.I. {Harte}, L.~{Rezzolla}, Phys. Rev. D R
  \textbf{87}(12), 121302 (2013).
\newblock \doi{10.1103/PhysRevD.87.121302}

\bibitem{Kiuchi2015}
K.~{Kiuchi}, Y.~{Sekiguchi}, K.~{Kyutoku}, M.~{Shibata}, K.~{Taniguchi},
  T.~{Wada}, Phys. Rev. D \textbf{92}(6), 064034 (2015).
\newblock \doi{10.1103/PhysRevD.92.064034}

\bibitem{Baumgarte00bb}
T.W. {Baumgarte}, S.L. {Shapiro}, M.~{Shibata}, Astrophys. J. Lett.
  \textbf{528}, L29 (2000).
\newblock \doi{10.1086/312425}

\bibitem{Shapiro00}
S.L. {Shapiro}, Astrophys. J. \textbf{544}, 397 (2000)

\bibitem{Bandyopadhyay1997}
D.~Bandyopadhyay, S.~Chakrabarty, S.~Pal, Phys. Rev. Lett. \textbf{79}, 2176
  (1997).
\newblock \doi{10.1103/PhysRevLett.79.2176}.
\newblock \urlprefix\url{https://link.aps.org/doi/10.1103/PhysRevLett.79.2176}

\bibitem{Metzger2018}
B.D. {Metzger}, T.A. {Thompson}, E.~{Quataert}, Astrophys. J. \textbf{856}, 101
  (2018).
\newblock \doi{10.3847/1538-4357/aab095}

\bibitem{Kiuchi2008b}
K.~{Kiuchi}, M.~{Shibata}, S.~{Yoshida}, Phys. Rev. D \textbf{78}(2), 024029
  (2008).
\newblock \doi{10.1103/PhysRevD.78.024029}

\bibitem{Ciolfi2011}
R.~{Ciolfi}, S.K. {Lander}, G.M. {Manca}, L.~{Rezzolla}, Astrophys. J.
  \textbf{736}, L6 (2011).
\newblock \doi{10.1088/2041-8205/736/1/L6}

\bibitem{Lasky2011}
P.D. {Lasky}, B.~{Zink}, K.D. {Kokkotas}, K.~{Glampedakis}, Astrophys. J.
  \textbf{735}, L20 (2011).
\newblock \doi{10.1088/2041-8205/735/1/L20}

\bibitem{Ciolfi2012}
R.~{Ciolfi}, L.~{Rezzolla}, Astrophys. J. \textbf{760}, 1 (2012).
\newblock \doi{10.1088/0004-637X/760/1/1}

\bibitem{Lasky2012}
P.D. {Lasky}, B.~{Zink}, K.D. {Kokkotas}, ArXiv e-prints
  \textbf{arXiv:1203.3590} (2012)

\bibitem{Tayler_1957}
R.J. Tayler, Proceedings of the Physical Society. Section B \textbf{70}(1), 31
  (1957).
\newblock \doi{10.1088/0370-1301/70/1/306}.
\newblock \urlprefix\url{https://doi.org/10.1088/0370-1301/70/1/306}

\bibitem{Tayler1973}
R.J. {Tayler}, Mon. Not. R. Astron. Soc. \textbf{161}, 365 (1973)

\bibitem{Wright1973}
G.A.E. {Wright}, Mon. Not. R. Astron. Soc. \textbf{162}, 339 (1973)

\bibitem{Markey1973}
P.~{Markey}, R.J. {Tayler}, Mon. Not. R. Astron. Soc. \textbf{163}, 77 (1973)

\bibitem{Markey1974}
P.~Markey, R.J. Tayler, Monthly Notices of the Royal Astronomical Society
  \textbf{168}(3), 505 (1974).
\newblock \doi{10.1093/mnras/168.3.505}.
\newblock \urlprefix\url{https://doi.org/10.1093/mnras/168.3.505}

\bibitem{Flowers1977}
E.~{Flowers}, M.A. {Ruderman}, Astrophys. J. \textbf{215}, 302 (1977).
\newblock \doi{10.1086/155359}

\bibitem{Bocquet1995}
M.~{Bocquet}, S.~{Bonazzola}, E.~{Gourgoulhon}, J.~{Novak}, Astron. and
  Astrophys. \textbf{301}, 757 (1995)

\bibitem{Kiuchi2008}
K.~{Kiuchi}, S.~{Yoshida}, Phys. Rev. D \textbf{78}(4), 044045 (2008).
\newblock \doi{10.1103/PhysRevD.78.044045}

\bibitem{Bonazzola1993}
S.~{Bonazzola}, E.~{Gourgoulhon}, M.~{Salgado}, J.A. {Marck}, Astron. and
  Astrophys. \textbf{278}, 421 (1993)

\bibitem{Carter73_Stationary_BH}
B.~{Carter}, in \emph{Black Holes} (Gordon and Breach, New York, 1973), pp.
  125--214

\bibitem{Cook94c}
G.B. {Cook}, S.L. {Shapiro}, S.A. {Teukolsky}, Astrophys. J. \textbf{424}, 823
  (1994)

\bibitem{Lasota1996}
J.P. {Lasota}, P.~{Haensel}, M.A. {Abramowicz}, Astrophys. J. \textbf{456}, 300
  (1996).
\newblock \doi{10.1086/176650}

\bibitem{Breu2016}
C.~{Breu}, L.~{Rezzolla}, Mon. Not. R. Astron. Soc. \textbf{459}, 646 (2016).
\newblock \doi{10.1093/mnras/stw575}

\bibitem{Cardall2001}
C.Y. {Cardall}, M.~{Prakash}, J.M. {Lattimer}, Astrophys. J. \textbf{554}, 322
  (2001).
\newblock \doi{10.1086/321370}

\bibitem{Yazadjiev2011}
S.~Yazadjiev, Phys. Rev. D \textbf{85}, 044030 (2012).
\newblock \doi{10.1103/PhysRevD.85.044030}

\bibitem{Konno1999}
K.~{Konno}, T.~{Obata}, Y.~{Kojima}, Astron. Astrophys. \textbf{352}, 211
  (1999)

\bibitem{Konno2001}
K.~{Konno}, Astron. Astrophys. \textbf{372}, 594 (2001).
\newblock \doi{10.1051/0004-6361:20010556}

\bibitem{Ioka2003}
K.~{Ioka}, M.~{Sasaki}, Phys. Rev. D \textbf{67}(12), 124026 (2003).
\newblock \doi{10.1103/PhysRevD.67.124026}

\bibitem{Lovelace1986}
R.V.E. {Lovelace}, C.~{Mehanian}, C.M. {Mobarry}, M.E. {Sulkanen}, The
  Astrophysical Journal Supplement Series \textbf{62}, 1 (1986).
\newblock \doi{10.1086/191132}

\bibitem{Mobarry1986}
C.M. {Mobarry}, R.V.E. {Lovelace}, Astrophys. J. \textbf{309}, 455 (1986).
\newblock \doi{10.1086/164617}

\bibitem{Nitta1991}
S.Y. {Nitta}, M.~{Takahashi}, A.~{Tomimatsu}, Phys. Rev. D \textbf{44}(8), 2295
  (1991).
\newblock \doi{10.1103/PhysRevD.44.2295}

\bibitem{Beskin1997}
V.S. {Beskin}, Soviet Physics Uspekhi \textbf{40}, 659 (1997)

\bibitem{Bekenstein1978}
J.D. {Bekenstein}, E.~{Oron}, Phys. Rev. D \textbf{18}, 1809 (1978).
\newblock \doi{10.1103/PhysRevD.18.1809}

\bibitem{Bekenstein1979}
J.D. {Bekenstein}, E.~{Oron}, Phys. Rev. D \textbf{19}(10), 2827 (1979).
\newblock \doi{10.1103/PhysRevD.19.2827}

\bibitem{Gourgoulhon1993a}
E.~{Gourgoulhon}, S.~{Bonazzola}, Phys. Rev. D \textbf{48}(6), 2635 (1993).
\newblock \doi{10.1103/PhysRevD.48.2635}

\bibitem{Ioka04}
K.~{Ioka}, M.~{Sasaki}, Astrophys. J. \textbf{600}, 296 (2004).
\newblock \doi{10.1086/379650}

\bibitem{Hartle68}
J.B. {Hartle}, K.S. {Thorne}, Astrophys. J. \textbf{153}, 807 (1968).
\newblock \doi{10.1086/149707}

\bibitem{Bekenstein1987}
J.D. {Bekenstein}, Astrophys. J. \textbf{319}, 207 (1987).
\newblock \doi{10.1086/165447}

\bibitem{Parker1966}
E.N. {Parker}, Astrophys. J. \textbf{145}, 811 (1966).
\newblock \doi{10.1086/148828}

\bibitem{Kiuchi2011}
K.~{Kiuchi}, S.~{Yoshida}, M.~{Shibata}, Astron. Astrophys. \textbf{532}, A30
  (2011).
\newblock \doi{10.1051/0004-6361/201016242}

\bibitem{Reisenegger2008}
A.~Reisenegger, Astron. Astrophys. \textbf{499}, 557 (2009).
\newblock \doi{10.1051/0004-6361/200810895}

\bibitem{Reisenegger1992}
A.~{Reisenegger}, P.~{Goldreich}, Astrophysical J. \textbf{395}, 240 (1992).
\newblock \doi{10.1086/171645}

\bibitem{Yoshida2012}
S.~{Yoshida}, K.~{Kiuchi}, M.~{Shibata}, ArXiv e-prints
  \textbf{arXiv:1207.1942} (2012)

\bibitem{Yoshida2019}
S.~Yoshida, Phys. Rev. D \textbf{99}(8), 084034 (2019).
\newblock \doi{10.1103/PhysRevD.99.084034}

\bibitem{Colaiuda2008}
A.~{Colaiuda}, V.~{Ferrari}, L.~{Gualtieri}, J.A. {Pons}, Mon. Not. R. Astron.
  Soc. \textbf{385}, 2080 (2008).
\newblock \doi{10.1111/j.1365-2966.2008.12966.x}

\bibitem{Ciolfi2009}
R.~{Ciolfi}, V.~{Ferrari}, L.~{Gualtieri}, J.A. {Pons}, Mon. Not. R. Astron.
  Soc. \textbf{397}, 913 (2009).
\newblock \doi{10.1111/j.1365-2966.2009.14990.x}

\bibitem{Ciolfi2010}
R.~{Ciolfi}, V.~{Ferrari}, L.~{Gualtieri}, Mon. Not. R. Astron. Soc.
  \textbf{406}, 2540 (2010).
\newblock \doi{10.1111/j.1365-2966.2010.16847.x}

\bibitem{Braithwaite2009}
J.~{Braithwaite}, Mon. Not. R. Astron. Soc. \textbf{397}, 763 (2009).
\newblock \doi{10.1111/j.1365-2966.2008.14034.x}

\bibitem{Lander2012}
S.K. {Lander}, D.I. {Jones}, Mon. Not. R. Astron. Soc. \textbf{424}, 482
  (2012).
\newblock \doi{10.1111/j.1365-2966.2012.21213.x}

\bibitem{Tsokaros2021}
A.~Tsokaros, M.~Ruiz, S.L. Shapiro, K.~Ury\={u}, arXiv:2111.00013  (2021)

\bibitem{Kotake_2006}
K.~Kotake, K.~Sato, K.~Takahashi, Reports on Progress in Physics
  \textbf{69}(4), 971 (2006).
\newblock \doi{10.1088/0034-4885/69/4/r03}.
\newblock \urlprefix\url{https://doi.org/10.1088%2F0034-4885%2F69%2F4%2Fr03}

\bibitem{Ciolfi2013}
R.~{Ciolfi}, L.~{Rezzolla}, Mon. Not. R. Astron. Soc. \textbf{435}, L43 (2013).
\newblock \doi{10.1093/mnrasl/slt092}

\bibitem{Rezzolla2001}
L.~{Rezzolla}, B.J. {Ahmedov}, J.C. {Miller}, Mon. Not. R. Astron. Soc.
  \textbf{322}, 723 (2001).
\newblock \doi{10.1046/j.1365-8711.2001.04161.x}

\bibitem{Rezzolla2004}
L.~{Rezzolla}, B.J. {Ahmedov}, Mon. Not. R. Astron. Soc. \textbf{352}, 1161
  (2004).
\newblock \doi{10.1111/j.1365-2966.2004.08006.x}

\bibitem{Rezzolla2001_err}
L.~{Rezzolla}, B.J. {Ahmedov}, J.C. {Miller}, Mon. Not. R. Astron. Soc.
  \textbf{338}, 816 (2003).
\newblock \doi{10.1046/j.1365-8711.2003.06261.x}

\bibitem{Abdikamalov2009}
E.B. {Abdikamalov}, B.J. {Ahmedov}, J.C. {Miller}, Mon. Not. R. Astron. Soc.
  \textbf{395}, 443 (2009).
\newblock \doi{10.1111/j.1365-2966.2009.14540.x}

\bibitem{Morozova2010}
V.S. {Morozova}, B.J. {Ahmedov}, O.~{Zanotti}, Mon. Not. R. Astron. Soc.
  \textbf{408}, 490 (2010).
\newblock \doi{10.1111/j.1365-2966.2010.17131.x}

\bibitem{Rezzolla2016a}
L.~{Rezzolla}, B.J. {Ahmedov}, Mon. Not. R. Astron. Soc. \textbf{459}(4), 4144
  (2016).
\newblock \doi{10.1093/mnras/stw864}

\bibitem{Turimov2018}
B.~{Turimov}, B.~{Ahmedov}, A.~{Abdujabbarov}, C.~{Bambi}, Phys. Rev. D
  \textbf{97}(12), 124005 (2018).
\newblock \doi{10.1103/PhysRevD.97.124005}

\bibitem{Turimov2021}
B.~Turimov, Z.c.v. Stuchl\'{\i}k, J.~Rayimbayev, A.~Abdujabbarov, Phys. Rev. D
  \textbf{103}, 124039 (2021).
\newblock \doi{10.1103/PhysRevD.103.124039}.
\newblock \urlprefix\url{https://link.aps.org/doi/10.1103/PhysRevD.103.124039}

\bibitem{Oron2002}
A.~{Oron}, Phys. Rev. D \textbf{66}(2), 023006 (2002).
\newblock \doi{10.1103/PhysRevD.66.023006}

\bibitem{Trehan1972}
S.K. {Trehan}, M.S. {Uberoi}, Astrophys. J. \textbf{175}, 161 (1972).
\newblock \doi{10.1086/151546}

\bibitem{Jones1975}
P.B. {Jones}, Astrophysics and Space Science \textbf{33}, 215 (1975).
\newblock \doi{10.1007/BF00646019}

\bibitem{Cutler2002}
C.~{Cutler}, Phys. Rev. D \textbf{66}(8), 084025 (2002).
\newblock \doi{10.1103/PhysRevD.66.084025}

\bibitem{Kiuchi2009a}
K.~{Kiuchi}, K.~{Kotake}, S.~{Yoshida}, Astrophys. J. \textbf{698}(1), 541
  (2009).
\newblock \doi{10.1088/0004-637X/698/1/541}

\bibitem{shen98}
H.~{Shen}, H.~{Toki}, K.~{Oyamatsu}, K.~{Sumiyoshi}, Nuclear Physics A
  \textbf{637}, 435 (1998).
\newblock \doi{10.1016/S0375-9474(98)00236-X}

\bibitem{Pandharipande1989}
V.R. {Pandharipande}, D.G. {Ravenhall}, in \emph{Nuclear Matter and Heavy Ion
  Collisions}, \emph{NATO Advanced Study Institute (ASI) Series B}, vol. 205
  (1989), \emph{NATO Advanced Study Institute (ASI) Series B}, vol. 205, p. 103

\bibitem{Cook92b}
G.B. {Cook}, S.L. {Shapiro}, S.A. {Teukolsky}, Astrophys. J. \textbf{398}, 203
  (1992)

\bibitem{Lattimer91}
J.M. Lattimer, F.D. Swesty, Nucl. Phys. A \textbf{535}, 331 (1991)

\bibitem{Douchin01}
F.~{Douchin}, P.~{Haensel}, Astron. Astrophys. \textbf{380}, 151 (2001).
\newblock \doi{10.1051/0004-6361:20011402}

\bibitem{Yasutake2010}
N.~{Yasutake}, K.~{Kiuchi}, K.~{Kotake}, Mon. Not. R. Astron. Soc.
  \textbf{401}(3), 2101 (2010).
\newblock \doi{10.1111/j.1365-2966.2009.15813.x}

\bibitem{chodos1974}
A.~{Chodos}, R.L. {Jaffe}, K.~{Johnson}, C.B. {Thorn}, V.F. {Weisskopf}, Phys.
  Rev. D \textbf{9}, 3471 (1974).
\newblock \doi{10.1103/PhysRevD.9.3471}

\bibitem{Yasutake2004}
N.~Yasutake, M.a. Hashimoto, Y.~Eriguchi, Prog. Theor. Phys. \textbf{113}, 953
  (2005).
\newblock \doi{10.1143/PTP.113.953}

\bibitem{Bombaci2000}
I.~Bombaci, B.~Datta, Astrophys. J. Lett. \textbf{530}, L69 (2000).
\newblock \doi{10.1086/312497}

\bibitem{Alford2006}
M.~Alford, D.~Blaschke, A.~Drago, T.~Klahn, G.~Pagliara, J.~Schaffner-Bielich,
  Nature \textbf{445}, E7 (2007).
\newblock \doi{10.1038/nature05582}

\bibitem{gourgoulhon1999}
E.~{Gourgoulhon}, P.~{Haensel}, R.~{Livine}, E.~{Paluch}, S.~{Bonazzola}, J.A.
  {Marck}, Astron. Astrophys. \textbf{349}, 851 (1999)

\bibitem{Zdunik2006}
J.L. Zdunik, M.~Bejger, P.~Haensel, E.~Gourgoulhon, Astron. Astrophys.
  \textbf{465}, 533 (2007).
\newblock \doi{10.1051/0004-6361:20066515}

\bibitem{Frieben2012}
J.~{Frieben}, L.~{Rezzolla}, Mon. Not. R. Astron. Soc. \textbf{427}, 3406
  (2012).
\newblock \doi{10.1111/j.1365-2966.2012.22027.x}

\bibitem{lorene}
\urlprefix\url{http://www.lorene.obspm.fr}.
\newblock Langage Objet pour la RElativit\'{e} Num\'{e}rique,
  \url{www.lorene.obspm.fr}

\bibitem{Thorne80b}
K.~Thorne, Rev. Mod. Phys. \textbf{52}(2), 299 (1980)

\bibitem{Wentzel1960}
D.G. {Wentzel}, The Astrophysical Journal Supplement Series \textbf{5}, 187
  (1960).
\newblock \doi{10.1086/190055}

\bibitem{Ostriker1969}
J.P. {Ostriker}, J.E. {Gunn}, Astrophys. J. \textbf{157}, 1395 (1969).
\newblock \doi{10.1086/150160}

\bibitem{Bucciantini2011}
N.~{Bucciantini}, L.~{Del Zanna}, Astron. Astrophys. \textbf{528}, A101 (2011).
\newblock \doi{10.1051/0004-6361/201015945}

\bibitem{Pili2014}
A.~Pili, N.~Bucciantini, L.~Del~Zanna, Mon. Not. Roy. Astron. Soc.
  \textbf{439}, 3541 (2014).
\newblock \doi{10.1093/mnras/stu215}

\bibitem{Shafranov1958}
V.D. {Shafranov}, Soviet Journal of Experimental and Theoretical Physics
  \textbf{6}, 545 (1958)

\bibitem{Grad1960}
H.~Grad, Rev. Mod. Phys. \textbf{32}, 830 (1960).
\newblock \doi{10.1103/RevModPhys.32.830}.
\newblock \urlprefix\url{https://link.aps.org/doi/10.1103/RevModPhys.32.830}

\bibitem{Shafranov1966}
V.D. {Shafranov}, Reviews of Plasma Physics \textbf{2}, 103 (1966)

\bibitem{Bucciantini2015}
N.~{Bucciantini}, A.G. {Pili}, L.~{Del Zanna}, Mon. Not. R. Astron. Soc.
  \textbf{447}(4), 3278 (2015).
\newblock \doi{10.1093/mnras/stu2689}

\bibitem{Glampedakis2012}
K.~{Glampedakis}, N.~{Andersson}, S.K. {Lander}, Mon. Not. R. Astron. Soc.
  \textbf{420}, 1263 (2012).
\newblock \doi{10.1111/j.1365-2966.2011.20112.x}

\bibitem{Pili2015}
A.G. {Pili}, N.~{Bucciantini}, L.~{Del Zanna}, Mon. Not. Roy. Astron. Soc.
  \textbf{447}(3), 2821 (2015).
\newblock \doi{10.1093/mnras/stu2628}

\bibitem{Glampedakis2014}
K.~Glampedakis, S.K. Lander, N.~Andersson, Mon. Not. Roy. Astron. Soc.
  \textbf{437}(1), 2 (2014).
\newblock \doi{10.1093/mnras/stt1814}

\bibitem{Pili2017}
A.~Pili, N.~Bucciantini, L.~Del~Zanna, Mon. Not. Roy. Astron. Soc.
  \textbf{470}(2), 2469 (2017).
\newblock \doi{10.1093/mnras/stx1176}

\bibitem{Chatterjee2014}
D.~Chatterjee, T.~Elghozi, J.~Novak, M.~Oertel, Mon. Not. Roy. Astron. Soc.
  \textbf{447}, 3785 (2015).
\newblock \doi{10.1093/mnras/stu2706}

\bibitem{Noronha2007}
J.L. Noronha, I.A. Shovkovy, Phys. Rev. D \textbf{76}, 105030 (2007).
\newblock \doi{10.1103/PhysRevD.76.105030}.
\newblock \urlprefix\url{https://link.aps.org/doi/10.1103/PhysRevD.76.105030}

\bibitem{Sinha2013}
M.~Sinha, X.G. Huang, A.~Sedrakian, Phys. Rev. D \textbf{88}, 025008 (2013).
\newblock \doi{10.1103/PhysRevD.88.025008}.
\newblock \urlprefix\url{https://link.aps.org/doi/10.1103/PhysRevD.88.025008}

\bibitem{Blandford1982}
R.D. {Blandford}, L.~{Hernquist}, Journal of Physics C Solid State Physics
  \textbf{15}(30), 6233 (1982).
\newblock \doi{10.1088/0022-3719/15/30/017}

\bibitem{Franzon2015}
B.~Franzon, V.~Dexheimer, S.~Schramm, Mon. Not. Roy. Astron. Soc.
  \textbf{456}(3), 2937 (2016).
\newblock \doi{10.1093/mnras/stv2606}

\bibitem{Hempel:2013}
M.~Hempel, V.~Dexheimer, S.~Schramm, I.~Iosilevskiy, Phys. Rev.
  \textbf{C88}(1), 014906 (2013).
\newblock \doi{10.1103/PhysRevC.88.014906}

\bibitem{Papazoglou:1998}
P.~Papazoglou, D.~Zschiesche, S.~Schramm, J.~Schaffner-Bielich, H.~Stoecker,
  W.~Greiner, Phys. Rev. \textbf{C59}, 411 (1999).
\newblock \doi{10.1103/PhysRevC.59.411}

\bibitem{Dexheimer:2008}
V.~Dexheimer, S.~Schramm, Astrophys. J. \textbf{683}, 943 (2008).
\newblock \doi{10.1086/589735}

\bibitem{Franzon2016a}
B.~Franzon, V.~Dexheimer, S.~Schramm, Phys. Rev. D \textbf{94}(4), 044018
  (2016).
\newblock \doi{10.1103/PhysRevD.94.044018}

\bibitem{Franzon2017}
B.~Franzon, R.~Negreiros, S.~Schramm, Phys. Rev. D \textbf{96}(12), 123005
  (2017).
\newblock \doi{10.1103/PhysRevD.96.123005}

\bibitem{Gourgoulhon:2011gz}
E.~Gourgoulhon, C.~Markakis, K.~Uryu, Y.~Eriguchi, Phys. Rev. \textbf{D83},
  104007 (2011).
\newblock \doi{10.1103/PhysRevD.83.104007}

\bibitem{Akiyama2019_L1}
{Event Horizon Telescope Collaboration}, K.~{Akiyama}, A.~{Alberdi}, W.~{Alef},
  K.~{Asada}, R.~{Azulay}, A.K. {Baczko}, D.~{Ball}, M.~{Balokovi{\'c}},
  J.~{Barrett}, D.~{Bintley}, L.~{Blackburn}, W.~{Boland}, K.L. {Bouman}, G.C.
  {Bower}, M.~{Bremer}, C.D. {Brinkerink}, R.~{Brissenden}, S.~{Britzen}, A.E.
  {Broderick}, D.~{Broguiere}, T.~{Bronzwaer}, D.Y. {Byun}, J.E. {Carlstrom},
  A.~{Chael}, C.k. {Chan}, S.~{Chatterjee}, K.~{Chatterjee}, M.T. {Chen},
  Y.~{Chen}, I.~{Cho}, P.~{Christian}, J.E. {Conway}, J.M. {Cordes}, G.B.
  {Crew}, Y.~{Cui}, J.~{Davelaar}, M.~{De Laurentis}, R.~{Deane}, J.~{Dempsey},
  G.~{Desvignes}, J.~{Dexter}, S.S. {Doeleman}, R.P. {Eatough}, H.~{Falcke},
  V.L. {Fish}, E.~{Fomalont}, R.~{Fraga-Encinas}, W.T. {Freeman}, P.~{Friberg},
  C.M. {Fromm}, J.L. {G{\'o}mez}, P.~{Galison}, C.F. {Gammie},
  R.~{Garc{\'\i}a}, O.~{Gentaz}, B.~{Georgiev}, C.~{Goddi}, R.~{Gold}, M.~{Gu},
  M.~{Gurwell}, K.~{Hada}, M.H. {Hecht}, R.~{Hesper}, L.C. {Ho}, P.~{Ho},
  M.~{Honma}, C.W.L. {Huang}, L.~{Huang}, D.H. {Hughes}, S.~{Ikeda},
  M.~{Inoue}, S.~{Issaoun}, D.J. {James}, B.T. {Jannuzi}, M.~{Janssen},
  B.~{Jeter}, W.~{Jiang}, M.D. {Johnson}, S.~{Jorstad}, T.~{Jung}, M.~{Karami},
  R.~{Karuppusamy}, T.~{Kawashima}, G.K. {Keating}, M.~{Kettenis}, J.Y. {Kim},
  J.~{Kim}, J.~{Kim}, M.~{Kino}, J.Y. {Koay}, P.M. {Koch}, S.~{Koyama},
  M.~{Kramer}, C.~{Kramer}, T.P. {Krichbaum}, C.Y. {Kuo}, T.R. {Lauer}, S.S.
  {Lee}, Y.R. {Li}, Z.~{Li}, M.~{Lindqvist}, K.~{Liu}, E.~{Liuzzo}, W.P. {Lo},
  A.P. {Lobanov}, L.~{Loinard}, C.~{Lonsdale}, R.S. {Lu}, N.R. {MacDonald},
  J.~{Mao}, S.~{Markoff}, D.P. {Marrone}, A.P. {Marscher},
  I.~{Mart{\'\i}-Vidal}, S.~{Matsushita}, L.D. {Matthews}, L.~{Medeiros}, K.M.
  {Menten}, Y.~{Mizuno}, I.~{Mizuno}, J.M. {Moran}, K.~{Moriyama},
  M.~{Moscibrodzka}, C.~{M{\"u}ller}, H.~{Nagai}, N.M. {Nagar}, M.~{Nakamura},
  R.~{Narayan}, G.~{Narayanan}, I.~{Natarajan}, R.~{Neri}, C.~{Ni},
  A.~{Noutsos}, H.~{Okino}, H.~{Olivares}, G.N. {Ortiz-Le{\'o}n}, T.~{Oyama},
  F.~{{\"O}zel}, D.C.M. {Palumbo}, N.~{Patel}, U.L. {Pen}, D.W. {Pesce},
  V.~{Pi{\'e}tu}, R.~{Plambeck}, A.~{PopStefanija}, O.~{Porth}, B.~{Prather},
  J.A. {Preciado-L{\'o}pez}, D.~{Psaltis}, H.Y. {Pu}, V.~{Ramakrishnan},
  R.~{Rao}, M.G. {Rawlings}, A.W. {Raymond}, L.~{Rezzolla}, B.~{Ripperda},
  F.~{Roelofs}, A.~{Rogers}, E.~{Ros}, M.~{Rose}, A.~{Roshanineshat},
  H.~{Rottmann}, A.L. {Roy}, C.~{Ruszczyk}, B.R. {Ryan}, K.L.J. {Rygl},
  S.~{S{\'a}nchez}, D.~{S{\'a}nchez-Arguelles}, M.~{Sasada}, T.~{Savolainen},
  F.P. {Schloerb}, K.F. {Schuster}, L.~{Shao}, Z.~{Shen}, D.~{Small}, B.W.
  {Sohn}, J.~{SooHoo}, F.~{Tazaki}, P.~{Tiede}, R.P.J. {Tilanus}, M.~{Titus},
  K.~{Toma}, P.~{Torne}, T.~{Trent}, S.~{Trippe}, S.~{Tsuda}, I.~{van Bemmel},
  H.J. {van Langevelde}, D.R. {van Rossum}, J.~{Wagner}, J.~{Wardle},
  J.~{Weintroub}, N.~{Wex}, R.~{Wharton}, M.~{Wielgus}, G.N. {Wong}, Q.~{Wu},
  K.~{Young}, A.~{Young}, Z.~{Younsi}, F.~{Yuan}, Y.F. {Yuan}, J.A. {Zensus},
  G.~{Zhao}, S.S. {Zhao}, Z.~{Zhu}, J.C. {Algaba}, A.~{Allardi}, R.~{Amestica},
  J.~{Anczarski}, U.~{Bach}, F.K. {Baganoff}, C.~{Beaudoin}, B.A. {Benson},
  R.~{Berthold}, J.M. {Blanchard}, R.~{Blundell}, S.~{Bustamente},
  R.~{Cappallo}, E.~{Castillo-Dom{\'\i}nguez}, C.C. {Chang}, S.H. {Chang}, S.C.
  {Chang}, C.C. {Chen}, R.~{Chilson}, T.C. {Chuter}, R.~{C{\'o}rdova Rosado},
  I.M. {Coulson}, T.M. {Crawford}, J.~{Crowley}, J.~{David}, M.~{Derome},
  M.~{Dexter}, S.~{Dornbusch}, K.A. {Dudevoir}, S.A. {Dzib}, A.~{Eckart},
  C.~{Eckert}, N.R. {Erickson}, W.B. {Everett}, A.~{Faber}, J.R. {Farah},
  V.~{Fath}, T.W. {Folkers}, D.C. {Forbes}, R.~{Freund}, A.I. {G{\'o}mez-Ruiz},
  D.M. {Gale}, F.~{Gao}, G.~{Geertsema}, D.A. {Graham}, C.H. {Greer},
  R.~{Grosslein}, F.~{Gueth}, D.~{Haggard}, N.W. {Halverson}, C.C. {Han}, K.C.
  {Han}, J.~{Hao}, Y.~{Hasegawa}, J.W. {Henning}, A.~{Hern{\'a}ndez-G{\'o}mez},
  R.~{Herrero-Illana}, S.~{Heyminck}, A.~{Hirota}, J.~{Hoge}, Y.D. {Huang},
  C.M.V. {Impellizzeri}, H.~{Jiang}, A.~{Kamble}, R.~{Keisler}, K.~{Kimura},
  Y.~{Kono}, D.~{Kubo}, J.~{Kuroda}, R.~{Lacasse}, R.A. {Laing}, E.M. {Leitch},
  C.T. {Li}, L.C.C. {Lin}, C.T. {Liu}, K.Y. {Liu}, L.M. {Lu}, R.G. {Marson},
  P.L. {Martin-Cocher}, K.D. {Massingill}, C.~{Matulonis}, M.P. {McColl}, S.R.
  {McWhirter}, H.~{Messias}, Z.~{Meyer-Zhao}, D.~{Michalik}, A.~{Monta{\~n}a},
  W.~{Montgomerie}, M.~{Mora-Klein}, D.~{Muders}, A.~{Nadolski}, S.~{Navarro},
  J.~{Neilsen}, C.H. {Nguyen}, H.~{Nishioka}, T.~{Norton}, M.A. {Nowak},
  G.~{Nystrom}, H.~{Ogawa}, P.~{Oshiro}, T.~{Oyama}, H.~{Parsons}, S.N.
  {Paine}, J.~{Pe{\~n}alver}, N.M. {Phillips}, M.~{Poirier}, N.~{Pradel}, R.A.
  {Primiani}, P.A. {Raffin}, A.S. {Rahlin}, G.~{Reiland}, C.~{Risacher},
  I.~{Ruiz}, A.F. {S{\'a}ez-Mada{\'\i}n}, R.~{Sassella}, P.~{Schellart},
  P.~{Shaw}, K.M. {Silva}, H.~{Shiokawa}, D.R. {Smith}, W.~{Snow},
  K.~{Souccar}, D.~{Sousa}, T.K. {Sridharan}, R.~{Srinivasan}, W.~{Stahm}, A.A.
  {Stark}, K.~{Story}, S.T. {Timmer}, L.~{Vertatschitsch}, C.~{Walther}, T.S.
  {Wei}, N.~{Whitehorn}, A.R. {Whitney}, D.P. {Woody}, J.G.A. {Wouterloot},
  M.~{Wright}, P.~{Yamaguchi}, C.Y. {Yu}, M.~{Zeballos}, S.~{Zhang},
  L.~{Ziurys}, Astrophys. J. Lett. \textbf{875}(1), L1 (2019).
\newblock \doi{10.3847/2041-8213/ab0ec7}

\bibitem{Eriguchi1983}
Y.~{Eriguchi}, I.~{Hachisu}, Progress of Theoretical Physics \textbf{70}(6),
  1534 (1983).
\newblock \doi{10.1143/PTP.70.1534}

\bibitem{Hachisu1984a}
I.~{Hachisu}, Y.~{Eriguchi}, Pub. Astron. Soc. Japan \textbf{36}(2), 259 (1984)

\bibitem{Eriguchi1985}
Y.~{Eriguchi}, I.~{Hachisu}, Astron. Astrophys. \textbf{142}(2), 256 (1985)

\bibitem{Hachisu86a}
I.~{Hachisu}, Astrophys. J., Supp. \textbf{62}, 461 (1986).
\newblock \doi{10.1086/191148}

\bibitem{Uryu98c}
K.~Uryu, Y.~Eriguchi, Mon. Not. Roy. Astron. Soc. \textbf{299}, 575 (1998).
\newblock \doi{10.1046/j.1365-8711.1998.01795.x}

\bibitem{Uryu98b}
K.~Uryu, Y.~Eriguchi, Mon. Not. Roy. Astron. Soc. \textbf{303}, 329 (1999).
\newblock \doi{10.1046/j.1365-8711.1999.02224.x}

\bibitem{Otani2009}
J.~{Otani}, R.~{Takahashi}, Y.~{Eriguchi}, Mon. Not. R. Astron. Soc.
  \textbf{396}(4), 2152 (2009).
\newblock \doi{10.1111/j.1365-2966.2009.14842.x}

\bibitem{Tomimura2005}
Y.~{Tomimura}, Y.~{Eriguchi}, Mon. Not. R. Astron. Soc. \textbf{359}, 1117
  (2005).
\newblock \doi{10.1111/j.1365-2966.2005.08967.x}

\bibitem{Yoshida2006}
S.~{Yoshida}, Y.~{Eriguchi}, Astrophys. J. Suppl. \textbf{164}, 156 (2006).
\newblock \doi{10.1086/501050}

\bibitem{Yoshida2006b}
S.~{Yoshida}, S.~{Yoshida}, Y.~{Eriguchi}, Astrophys. J. \textbf{651}(1), 462
  (2006).
\newblock \doi{10.1086/507513}

\bibitem{Yoshida2011}
S.~{Yoshida}, K.~{Fujisawa}, Y.~{Eriguchi}, S.~{Yoshida}, R.~{Takahashi}, in
  \emph{Advances in Plasma Astrophysics}, vol. 274, ed. by A.~{Bonanno}, E.~{de
  Gouveia Dal Pino}, A.G. {Kosovichev} (2011), vol. 274, pp. 437--440.
\newblock \doi{10.1017/S1743921311007435}

\bibitem{Fujisawa2012}
K.~{Fujisawa}, S.~{Yoshida}, Y.~{Eriguchi}, Mon. Not. R. Astron. Soc.
  \textbf{422}, 434 (2012).
\newblock \doi{10.1111/j.1365-2966.2012.20614.x}

\bibitem{Fujisawa2013a}
K.~{Fujisawa}, R.~{Takahashi}, S.~{Yoshida}, Y.~{Eriguchi}, Mon. Not. R.
  Astron. Soc. \textbf{431}(2), 1453 (2013).
\newblock \doi{10.1093/mnras/stt275}

\bibitem{Fujisawa2013}
K.~{Fujisawa}, Y.~{Eriguchi}, Mon. Not. R. Astron. Soc. \textbf{432}, 1245
  (2013).
\newblock \doi{10.1093/mnras/stt541}

\bibitem{Lander:2009}
S.K. {Lander}, D.I. {Jones}, Mon. Not. R. Astron. Soc. \textbf{395}, 2162
  (2009).
\newblock \doi{10.1111/j.1365-2966.2009.14667.x}

\bibitem{Lander2012a}
S.K. {Lander}, N.~{Andersson}, K.~{Glampedakis}, Mon. Not. R. Astron. Soc.
  \textbf{419}(1), 732 (2012).
\newblock \doi{10.1111/j.1365-2966.2011.19720.x}

\bibitem{Lander2013}
S.K. {Lander}, Phys. Rev. Lett. \textbf{110}(7), 071101 (2013).
\newblock \doi{10.1103/PhysRevLett.110.071101}

\bibitem{Palapanidis2015}
K.~{Palapanidis}, N.~{Stergioulas}, S.K. {Lander}, Mon. Not. R. Astron. Soc.
  \textbf{452}(3), 3246 (2015).
\newblock \doi{10.1093/mnras/stv1536}

\bibitem{Lander2021}
S.K. {Lander}, P.~{Haensel}, B.~{Haskell}, J.L. {Zdunik}, M.~{Fortin}, Mon.
  Not. R. Astron. Soc. \textbf{503}(1), 875 (2021).
\newblock \doi{10.1093/mnras/stab460}

\bibitem{Taniguchi2015}
K.~{Taniguchi}, M.~{Shibata}, A.~{Buonanno}, Phys. Rev. D \textbf{91}(2),
  024033 (2015).
\newblock \doi{10.1103/PhysRevD.91.024033}

\bibitem{Soldateschi2020}
J.~{Soldateschi}, N.~{Bucciantini}, L.~{Del Zanna}, Astron. Astrophys.
  \textbf{640}, A44 (2020).
\newblock \doi{10.1051/0004-6361/202037918}

\bibitem{Soldateschi2021}
J.~{Soldateschi}, N.~{Bucciantini}, L.~{Del Zanna}, Astron. Astrophys.
  \textbf{645}, A39 (2021).
\newblock \doi{10.1051/0004-6361/202038826}

\bibitem{Soldateschi2021a}
J.~{Soldateschi}, N.~{Bucciantini}, L.~{Del Zanna}, arXiv e-prints
  arXiv:2106.00603 (2021)

\end{thebibliography}


\end{document}